\documentclass[%
 reprint,
superscriptaddress, 
 amsmath,amssymb,
 aps,
]{revtex4-2}
\usepackage{comment}
\usepackage{graphicx}
\usepackage{dcolumn}
\usepackage{bm}
\usepackage{hyperref}
\hypersetup{
    colorlinks=true,
    linkcolor=blue,
    filecolor=magenta,      
    urlcolor=black,
    citecolor=blue, 
    pdftitle={SoLID White Paper 2022},
}
\usepackage[mathlines,modulo]{lineno}
\usepackage{booktabs}
\usepackage{sidecap}
\newcommand\lrpbf[1]{{ {#1}}}

\begin{document}

\preprint{APS/123-QED}

\title{The Solenoidal Large Intensity Device (SoLID) for JLab 12 GeV}



\author{J.~Arrington}
\affiliation{Lawrence Berkeley National Laboratory, Berkeley, CA, USA} 

\author{J.~Benesch}
\affiliation{Thomas Jefferson National Accelerator Facility, Newport News, VA, USA} 

\author{A.~Camsonne}
\affiliation{Thomas Jefferson National Accelerator Facility, Newport News, VA, USA}

\author{J.~Caylor}
\affiliation{Syracuse University, Syracuse, NY, USA}

\author{J.-P.~Chen}
\affiliation{Thomas Jefferson National Accelerator Facility, Newport News, VA, USA}

\author{S. Covrig Dusa}
\affiliation{Thomas Jefferson National Accelerator Facility, Newport News, VA, USA}

\author{A.~Emmert}
\affiliation{University of Virginia, Charlottesville, VA, USA}

\author{G. Evans}
\affiliation{Brigham Young University - Idaho
Rexburg, ID, USA}

\author{H.~Gao}
\affiliation{Duke University and Triangle Universities Nuclear Laboratory, Durham, NC, USA}

\author{J.-O.~Hansen}
\affiliation{Thomas Jefferson National Accelerator Facility, Newport News, VA, USA}

\author{G.~M.~Huber}
\affiliation{University of Regina, Regina, SK, Canada}

\author{S.~Joosten}
\affiliation{Argonne National Laboratory, Argonne, IL, USA}

\author{V.~Khachatryan}
\affiliation{Duke University and Triangle Universities Nuclear Laboratory, Durham, NC, USA}

\author{N.~Liyanage}
\affiliation{University of Virginia, Charlottesville, VA, USA}

\author{Z.-E.~Meziani}
\affiliation{Argonne National Laboratory, Argonne, IL, USA}

\author{M.~Nycz}
\affiliation{University of Virginia, Charlottesville, VA, USA}

\author{C.~Peng}
\affiliation{Argonne National Laboratory, Argonne, IL, USA}

\author{M.~Paolone}
\affiliation{New Mexico State University, Las Cruces, NM, USA}

\author{W.~Seay}
\affiliation{Thomas Jefferson National Accelerator Facility, Newport News, VA, USA}

\author{P.~A.~Souder}
\affiliation{Syracuse University, Syracuse, NY, USA}

\author{N.~Sparveris}
\affiliation{Temple University, Philadelphia, PA, USA}

\author{H.~Spiesberger}
\affiliation{PRISMA+ Cluster of Excellence, Institut f\"ur Physik, Johannes Gutenberg Universit\"at, 55099 Mainz, Germany}

\author{Y.~Tian}
\affiliation{Syracuse University, Syracuse, NY, USA}

\author{E.~Voutier}
\affiliation{Universit\'e Paris-Saclay, CNRS/IN2P3, IJCLab, Orsay, France}

\author{J.~Xie}
\affiliation{Argonne National Laboratory, Argonne, IL, USA}

\author{W.~Xiong}
\affiliation{Shandong University, Qingdao, China}

\author{Z.-Y.~Ye}
\affiliation{University of Illinois at Chicago, Chicago, IL, USA}

\author{Z.~Ye}
\affiliation{Tsinghua University, Beijing, China}

\author{J.~Zhang}
\affiliation{University of Virginia, Charlottesville, VA, USA}

\author{Z.-W.~Zhao}
\affiliation{Duke University and Triangle Universities Nuclear Laboratory, Durham, NC, USA}

\author{X.~Zheng}
\affiliation{University of Virginia, Charlottesville, VA, USA}


\collaboration{For the Jefferson Lab SoLID Collaboration}

\date{\today}

\begin{abstract}
The Solenoidal Large Intensity Device (SoLID) is a new experimental apparatus planned for Hall A at the Thomas Jefferson National Accelerator Facility (JLab). SoLID will combine large angular and momentum acceptance with the capability to handle very high data rates at high luminosity. With a slate of approved high-impact physics experiments, SoLID will push JLab to a new limit at the QCD intensity frontier that will exploit the full potential of its 12 GeV electron beam. In this paper, we present an overview of the rich physics program that can be realized with SoLID, which encompasses the tomography of the nucleon in 3-D momentum space from Semi-Inclusive Deep Inelastic Scattering (SIDIS), expanding the phase space in the search for new physics and novel hadronic effects in parity-violating DIS (PVDIS), a precision measurement of $J/\psi$ production at threshold that probes the gluon field and its contribution to the proton mass, tomography of the nucleon in combined coordinate and momentum space with deep exclusive reactions, and more. 
To meet the challenging requirements, the design of SoLID described here takes full advantage of recent progress in detector, data acquisition and computing technologies. 
In addition, we outline potential experiments beyond the currently approved program and discuss the physics that could be explored should upgrades of CEBAF become a reality in the future.
\end{abstract}

\maketitle

\tableofcontents

\section{Executive Summary}

To exploit the full potential of the 12 GeV energy upgrade of the Continuous Electron Beam Accelerator Facility (CEBAF) at Jefferson Lab (JLab), we have designed a new spectrometer, named the Solenoidal Large Intensity Device (SoLID)~\cite{Chen:2014psa,precdr:2019}. The main feature of SoLID is its large acceptance and the capacity to operate at the full CEBAF luminosity of up to $10^{39}$~cm$^{-2}$s$^{-1}$. A rich and diverse science program consisting of a set of high-impact physics experiments has been developed with SoLID. 
The SoLID proposal was submitted as a Major Item of Equipment (MIE) to the U.S. Department of Energy (DOE) and, after passing several Director's Reviews at JLab, received a successful Science Review from the DOE in March 2021. We are presently awaiting the full report describing the review outcome.

\begin{figure}[!ht]
\begin{center}
\includegraphics[width=\columnwidth]{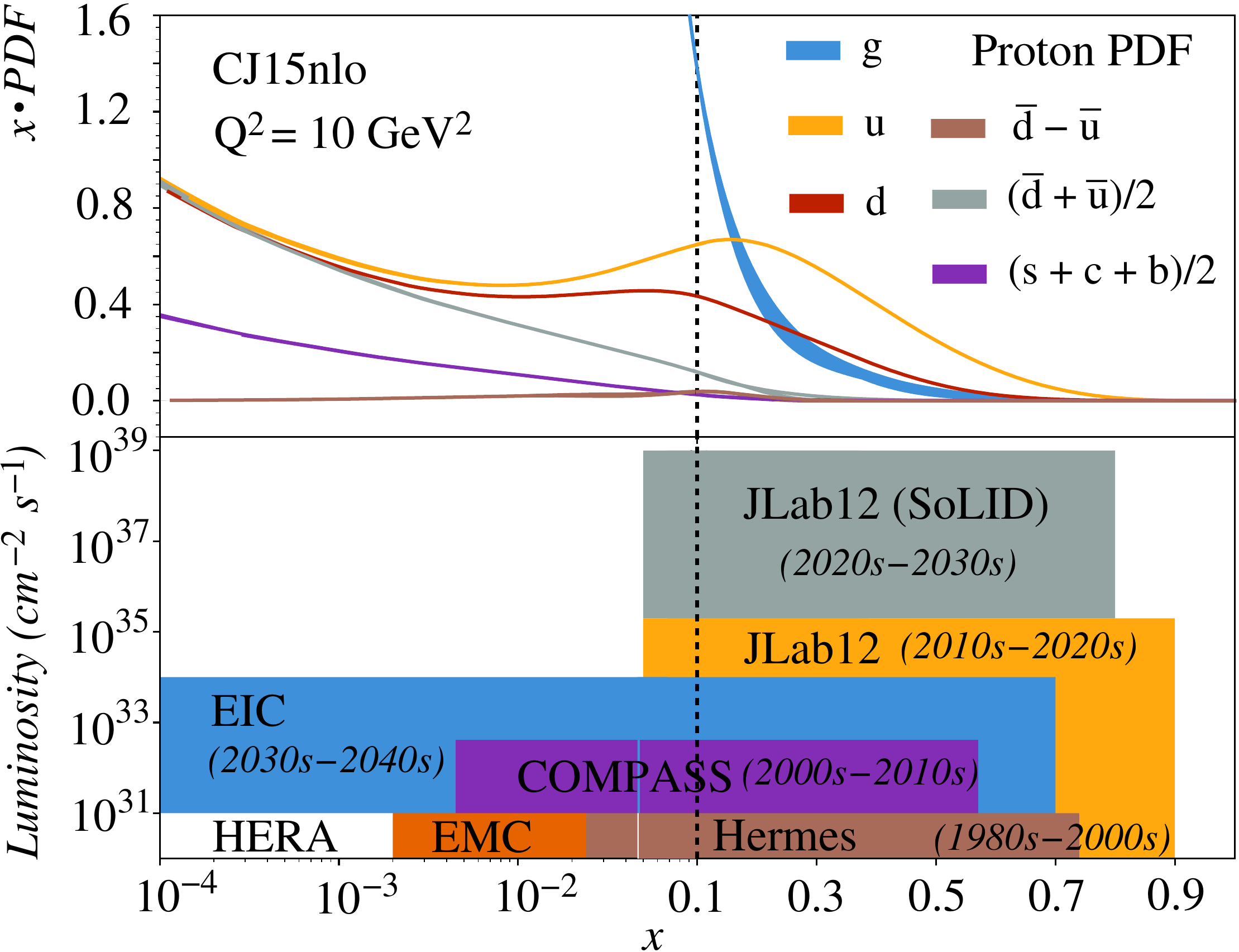}
\caption{Landscape of the QCD program.  SoLID expands the luminosity frontier in the large $x$ region whereas the EIC does the same for low $x$. Figure adapted from~\cite{Arrington:2021alx}.}
\label{fig:Cold}
\end{center}
\end{figure}

The SoLID spectrometer fills a critical void in the science reach of Quantum Chromodynamics (QCD) and fundamental symmetry studies. For illustration, Fig.~\ref{fig:Cold} shows the acceptance and luminosity range covered by JLab, the Electron-Ion Collider (EIC), and a number of 
lower-luminosity facilities that were designed to investigate the properties of quarks and gluons in the nucleon and their modified behavior in nuclei.  To maximize physics insight, it is essential to explore reactions over as large a range of $Q^2$ and Bjorken $x$ as possible. Together, JLab and the EIC will, over the next several decades, cover a broad and largely complementary kinematic range, with SoLID probing key physics and providing precision data primarily in the high-$x$ region.

SoLID can accommodate a variety of experimental configurations for a broad spectrum of physics.  Five primary experiments have been approved with the highest rating (``A'') by the JLab Program Advisory Committee (PAC). Three of these are measurements of Transverse-Momentum-Dependent Distributions (TMDs) describing the three-dimensional structure of the nucleon in momentum space via Semi-Inclusive Deep Inelastic Scattering (SIDIS) with polarized $^3$He and proton targets~\cite{JLabPR:E12-10-006,JLabPR:E12-11-007,JLabPR:E12-11-108}. The fourth aims at understanding the origin of the proton mass via measurements of near-threshold photo-production and electro-production of the $J/\psi$ meson~\cite{JLabPR:jpsi_solid}. The final one will test the electroweak sector of the Standard Model at low energy and study hadronic physics in the high-$x$ region~\cite{JLabPR:PVDIS_solid} through measurements of Parity-Violating Deep Inelastic Scattering (PVDIS).  In July 2022, two further experiments were approved, one to study the flavor dependence of the EMC effect using PVDIS with a $^{48}$Ca target~\cite{JLabPR:PVEMC} and the other to study hadronic physics with two-photon exchange via a measurement of the beam-normal single-spin asymmetry in DIS~\cite{JLabPR:bnssa_dis}. In addition, a series of approved experiments will run simultaneously with the main experiments. These include Deep Exclusive Meson Production (DEMP)~\cite{JLabPR:demp} and Time-like Compton Scattering (TCS)~\cite{JLabPR:E12-12-006A}, which access the Generalized Parton Distributions (GPDs) and improve our knowledge of the spatial three-dimensional structure of the nucleon.

The SoLID spectrometer can operate at such high luminosity in large part due to recent developments in detector, data acquisition, and computing technologies. High-rate tracking detector such as Gas Electron Multipliers (GEMs), Cherenkov counters with advanced photon detectors, and fast Multi-gap Resistive Plate Chambers (MRPCs) for time-of-flight measurements are key examples.  Fast electronics developed at JLab will handle the high trigger and background rates.  The large data volume can be handled by the advanced computing facility at JLab. These technological advancements, not available in the initial planning stages of the 12 GeV program, have become a reality and allow us to fully exploit the available accelerator capabilities to advance the frontiers of 
QCD studies.

\section{Introduction}
Since commencing operation in 1995, CEBAF has been the medium-energy electron scattering facility with the worldwide highest luminosity for conducting experiments with fixed proton and nuclear targets. Initially delivering electron beams with energies of up to 6 GeV, CEBAF was successfully upgraded in 2017, raising the beam energy to 12 GeV. Along with the energy upgrade, another experimental hall, Hall D, was added to the facility, and detectors in the other experimental halls were improved. At the same time, JLab's physics program has evolved to match the progress in our understanding of the structure of the nucleon within the theory of the strong interaction, known as QCD, and to push for higher precision in measurements of fundamental symmetries. Progress on both these frontiers requires first and foremost higher statistics. QCD studies aim to describe nucleon structure in three dimensions in both momentum and coordinate space using SIDIS and deeply virtual exclusive processes. Obtaining the desired 3-D mapping involves dividing the experimental data into many multi-dimensional bins, which is only meaningful if the total data set contains a very large number of events. Meanwhile, decades of experience in improving systematic uncertainties of parity-violating electron scattering (PVES) experiments allow us to measure spin-dependent asymmetries in DIS with a precision of better than parts per million (ppm), which calls for event counts of order $10^{12}$. Similarly, $J/\psi$ production on the proton requires high luminosity so that a sufficient number of events can be accumulated near the production threshold, where the cross section falls rapidly.

\begin{figure}[!ht]
 \includegraphics[width=0.5\textwidth]{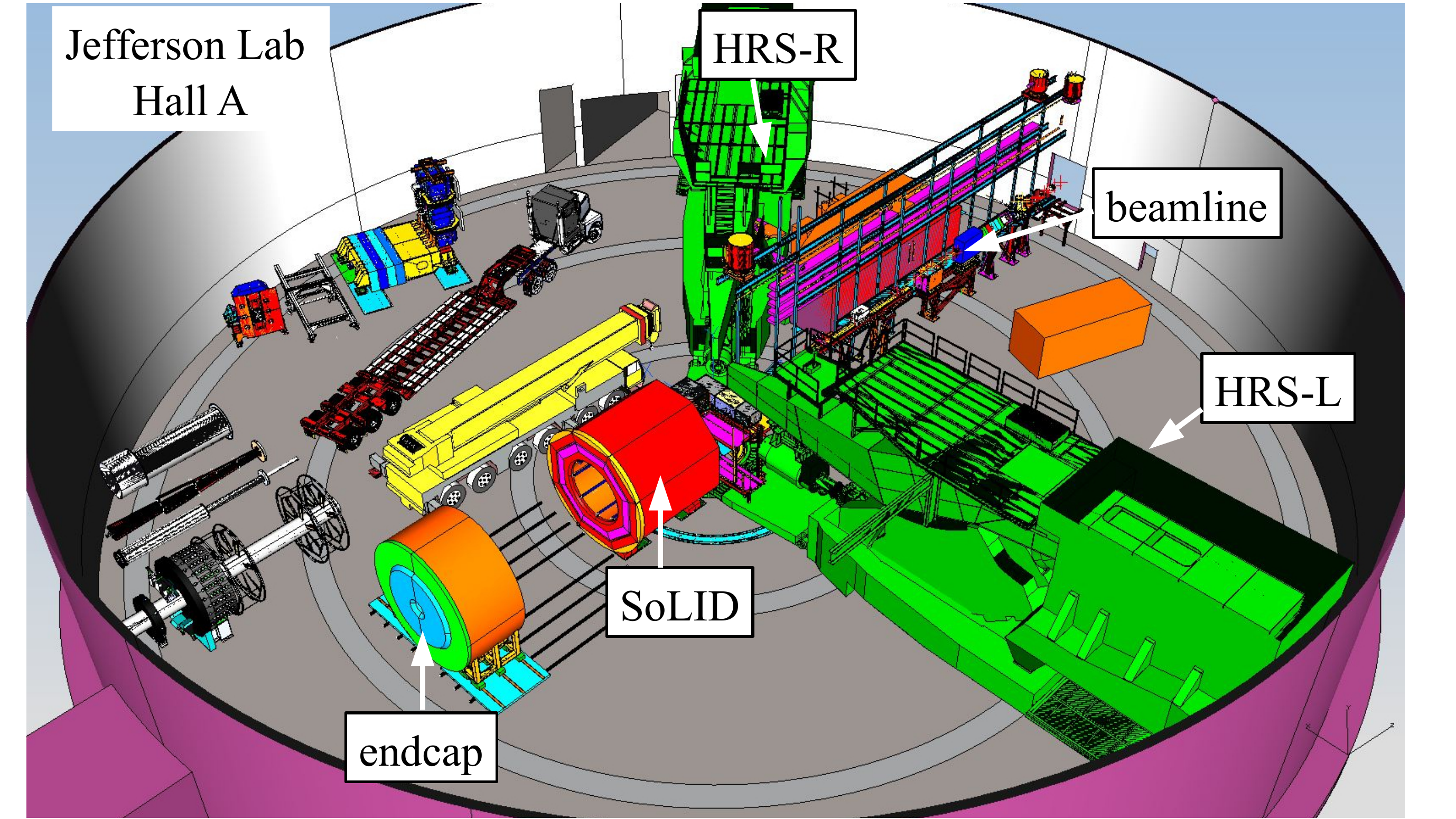}
 \caption{Schematic layout of SoLID in Hall A, with the endcap pulled downstream to allow detector installation and reconfiguration. The two high resolution spectrometers (HRS-L and HRS-R, not in use) are parked at backward angles.}\label{fig:solid_in_halla}
\end{figure}

SoLID is designed to fulfill these needs. By combining a 1.4 T solenoid magnet and a large-acceptance detector that covers $2\pi$ azimuthal angle, SoLID is particularly suitable to collect data with high statistics from DIS, SIDIS and Deep-Virtual exclusive processes. 
In addition, the SoLID design fully incorporates the ability to reconfigure all detector systems in order to optimize detection capabilities for SIDIS and $J/\psi$ meson production on the one hand, and for the PVDIS program on the other. SoLID is intended to be installed in experimental Hall A, as shown in Fig.~\ref{fig:solid_in_halla}.

In SIDIS, both a hadron and the scattered primary electron are detected in the final state.  The SIDIS process probes the distributions of quarks as a function of their transverse momentum and transverse spin.  These distributions are the transverse-momentum-dependent parton distributions (TMDs).  
At leading twist, there are eight independent TMDs. All of them can be extracted from SIDIS experiments with different combinations of target/beam polarization and angular modulations.  
There are three approved SoLID SIDIS experiments, including two on a $^3$He target: one with a longitudinal polarization~\cite{JLabPR:E12-11-007}, the other with a transverse polarization~\cite{JLabPR:E12-10-006}. The third experiment will use a transversely polarized proton target~\cite{JLabPR:E12-11-108}.

For SIDIS, the relevant variables are $P_T$, the hadron tranverse momentum, $z$, the fraction of longitudinal momentum carried by the hadron, and $Q^2$ and $x$ which are characteristic of the inclusive DIS process. Thus the SIDIS process to extract TMDs has multi-variable dependence, and a large data set is required to attain good statistics without integrating over one or more of the variables.  This is the main reason the high luminosity of the SoLID spectrometer is required.  

The main goal of the proposed SoLID $J/\psi$ measurement is to study the gluonic field contributions to the proton structure and proton mass. Gluons play an essential role in the structure of the proton, which is evident from the difference between the proton's total mass and its constituents' current quark masses. Most of the proton information carried by the gluons can be encoded in three gravitational form factors dubbed $A_g$, $B_g$ and $C_g$ that are part of the matrix element of the QCD energy momentum tensor. 
A compelling way to access these form factors is through virtual heavy-meson photo- and electro-production over the widest possible range of photon-nucleon invariant mass. Following recent studies, the region near the $J/\psi$ production threshold seems to be a very promising kinematic sector, not only for obtaining these form factors and thus determining the mass and scalar radius (dominated by gluons) of the proton, but also for exploring the trace anomaly that underlies the origin of the proton mass.

As a consequence, extensive data are required very close to threshold, where the cross section is very small.  The large acceptance of SoLID and its ability to handle high luminosity make it the ideal detector to study this physics with threshold $J/\psi$ production~\cite{JLabPR:jpsi_solid}. The EIC will provide complementary information through the production of the higher-mass $\Upsilon$ particle.  Since extraction of the gluonic form factors are model independent, measurements at both facilities agreeing with lattice QCD will give strong confidence in the interpretation. 

The goal of the SoLID PVDIS program~\cite{JLabPR:PVDIS_solid} is to measure the cross section asymmetry, $A_{PV}$, between right- and left-handed beam electrons with high precision. This asymmetry originates from parity non-conservation in weak interactions. At JLab energies, it can be determined from the interference between photon and $Z^0$ exchange processes in DIS. SoLID will provide data on $A_{PV}$ with sub-percent relative precision over a wide $(x,Q^2$) range. Measured on a deuteron target, the $A_{PV}^{(d)}$ data can be used to determine parameters of the electroweak Standard Model and to set limits on new physics up to an energy scale that is comparable to the reach of the LHC. The SoLID PVDIS deuteron measurement is unique in that it measures the strength of a particular contact interaction, the effective electron-quark $VA$ couplings, that cannot be isolated by any other experiments at present. Measured on a proton target, $A_{PV}^{(p)}$ can help determine the Parton Distribution Function (PDF) ratio $d/u$ at large $x$ without nuclear effects. Lastly, PVDIS asymmetries can probe specific hadronic physics effects such as charge symmetry violation (CSV). CSV at the quark level would be reflected in a specific kinematic dependence of the deuteron asymmetry, while effects of CSV at the nuclear level can be studied by measuring PVDIS asymmetries on a nuclear target such as $^{48}$Ca~\cite{JLabPR:PVEMC}. 

With SoLID being a versatile spectrometer, many other processes can be measured. The full azimuthal coverage of SoLID allows for the determination of the beam-normal single-spin asymmetry to high precision in DIS~\cite{JLabPR:bnssa_dis}, providing a new observable for studying two-photon-exchange effects. A number of run-group experiments will collect data at the same time as the SIDIS and $J/\psi$ experiments, including some that aim at studying Generalized Parton Distributions (GPDs)~\cite{JLabPR:demp,JLabPR:E12-12-006A}. 

This paper is organized as follows: the SIDIS, PVDIS, and $J/\psi$ programs are described in Sections~\ref{sec:sidis}, \ref{sec:pvdis}, and \ref{sec:jpsi}, respectively. In Section~\ref{sec:gpd} we expand on the GPD program (both approved run-group experiments and key measurements under study) with SoLID, and in Section~\ref{sec:others} all other run-group experiments, the beam normal single-spin-asymmetry (BNSSA) experiment, and an idea to measure PVDIS asymmetry using a polarized target. The SoLID instrumentation is detailed in Section~\ref{sec:instrum}. Finally, in Section~\ref{sec:upgrades} we discuss unique measurements that will become possible should a positron beam or an energy upgrade of CEBAF be realized in the future.

\section{Semi-inclusive Deep Inelastic Scattering}\label{sec:sidis}
\subsection{\label{sec:intro} 
The Three-dimensional Momentum Structure of the Nucleon}

A substantial amount of our knowledge of the internal structure of nucleons and nuclei in terms of quarks and gluons, the
 fundamental degrees of freedom of QCD, has been obtained though experimental and theoretical studies of the Parton Distribution Functions (PDFs) \cite{Ethier:2020way} and Fragmentation Functions (FFs) \cite{Metz:2016swz}. Within the 
collinear factorization scheme of deep inelastic lepton-nucleon scattering (DIS), leading-twist integrated  PDFs are defined as
probability densities for finding an unpolarized or longitudinally polarized parton in a fast-moving unpolarized or longitudinally 
polarized nucleon (``longitudinal'' is defined as along the nucleon moving direction). These PDFs depend on two variables 4-momentum transfer-squared, $Q^2$, and longitudinal momentum fraction, $x$. The dependence on $Q^2$ is due to gluon radiations and is governed by the QCD evolution equation.   
These PDFs are considered to be one-dimensional, {\it i.e.}\ they depend only on the longitudinal momentum, and are well-investigated. On the other hand, over the past 
two decades, the frontier of studies has moved forward to include three-dimensional PDFs, which describe the partonic motion and spatial distributions in the transverse direction, {\it i.e.}\ perpendicular to the nucleon's momentum. 

Semi-Inclusive Deep Inelastic Scattering (SIDIS) of leptons off nucleons, in which the scattered 
lepton and a leading hadron are detected in the final state, is a powerful tool to probe the transverse momentum and spin structure of the nucleon in addition to the longitudinal structure. Through this process, one can extract the transverse-momentum-dependent parton distribution functions 
(TMD-PDFs or just TMDs), which permit a three-dimensional tomography of the nucleon in momentum space.
Through exclusive processes such as deeply 
virtual Compton scattering, one can extract different views of nucleon through generalized parton distribution 
functions (GPDs), where the three dimensions are the longitudinal momentum and the two spatial coordinates in the transverse plane. All 
the information on TMDs and GPDs is contained in the five-dimensional Wigner distribution functions
\cite{Belitsky:2003nz,Lorce:2011kd}. The study of TMDs on the partonic structure of the nucleon in three-dimensional momentum space probes rich non-perturbative QCD dynamics and phenomena, and provides essential information on partonic orbital motion and spin-orbit correlations inside the nucleon. In addition, TMDs cast light on multi-parton correlations at leading twist, which helps uncover the dynamics of the nucleon's quark-gluon structure.

\subsection{\label{sec:TMDs} TMDs and Spin Asymmetries}

\begin{figure}[hbt!]
\centering
\includegraphics[width=0.48\textwidth]{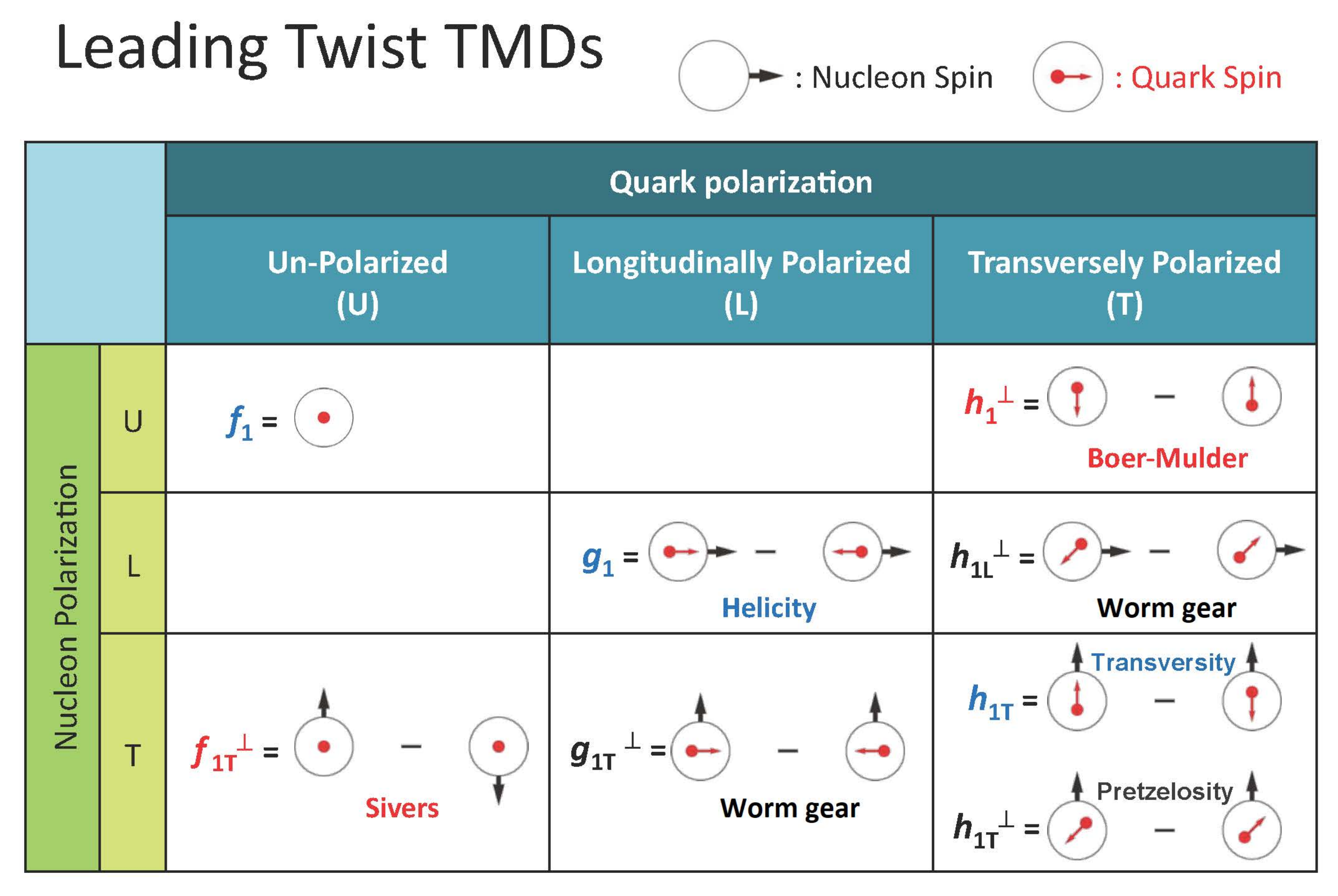}
\vspace*{-0.2cm}
\caption{Eight leading-twist TMDs arranged according to the quark ${\rm (f, g, h)}$ and nucleon ${\rm (U, L, T)}$ polarizations. 
Figure from Ref.~\cite{precdr:2019}.} 
\label{fig:TMD-table}
\end{figure}

Most TMDs stem from the coupling of the quark transverse momentum to the spin of the nucleon and quark. Hence, one can
study spin-orbit correlations in QCD with TMDs. At leading twist, if one integrates over the quark 
transverse momenta inside the nucleon, the surviving TMDs are the unpolarized parton distribution $f_{1}$,  
the longitudinally polarized parton distribution $g_{1}$ (Helicity), and the transversely polarized quark distribution function 
$h_{1T}$ (Transversity) \cite{Gao:2010av}. In addition to $f_{1}$, $g_{1}$, and $h_{1T}$, there are five additional leading-twist 
TMDs \cite{Mulders:1995dh,Boer:1997nt}, some of which vanish in the absence of quark orbital angular momentum (OAM). 
Figure~\ref{fig:TMD-table} tabulates these eight TMDs according to quark and nucleon polarization, where $U$, $L$,
and $T$ denote unpolarized, longitudinal, transverse polarization, respectively. All are functions of the longitudinal 
momentum fraction $x$ (Bjorken $x$) and the quark transverse momentum $\mathbf{k_{\perp}}$.

Let us focus on the following TMDs shown in Fig.~\ref{fig:TMD-table}: transversity, pretzelosity, 
Sivers, and worm-gear. Also given are the dependence on nucleon spin $\mathbf{S_{T}}$, quark spin $\mathbf{s_{q}}$, and virtual photon three-momentum
$\mathbf{P}$, which defines the longitudinal, $z$, direction.

\begin{itemize}
\item[(i)] \textbf{Transversity TMD}, depending on $\mathbf{S_{T}} \!\cdot\! \mathbf{s_{q}}$: in the parton model, provides information 
on the probability of finding transversely polarized quarks (anti-quarks) in a transversely polarized nucleon. Due to relativistic effect, the transversity TMD 
behaves differently from the helicity TMD, which provides information on the probability of finding longitudinally polarized quarks (anti-quarks) in a longitudinally polarized nucleon. The integral of transversity over $x$ yields the tensor charge \cite{Jaffe:1991kp,Barone:2001sp,DAlesio:2020vtw}, which is an important property of the nucleon that has 
been calculated precisely by lattice QCD. Precise measurements of the tensor charges of the proton and neutron will allow for their quark flavor separation and provide direct comparisons to lattice QCD predictions. Quark tensor charges are coefficients for the quark electric dipole moments (EDMs) to connect to nucleon EDMs if nucleon EDMs originate from quark EDMs,
making them important for tests of the Standard Model (SM) and searches for new physics beyond the SM. 
\item[(ii)] \textbf{Pretzelosity TMD}, depending on $\mathbf{S_{T}} \!\cdot\! \left[\mathbf{k_{\perp}\,k_{\perp}}\right] \!\cdot\! \mathbf{s_{qT}}$, 
describes the correlation among the transverse spin of the nucleon, transverse spin of the quark, as well as the transverse 
motion of the quark inside the nucleon. The pretzelosity distribution reflects the difference between helicity and transversity TMDs. This difference is due to relativistic effects. In various quark and QCD inspired models, pretzelosity TMD has been shown to provide quantitative 
information about the orbital angular momentum of the partons inside the nucleon.
\item[(iii)] \textbf{Sivers TMD}, depending on $\mathbf{S_{T}} \!\cdot\! \mathbf{k_{\perp}} \!\times\! \mathbf{P}$, describes a correlation between 
the nucleon transverse spin and the quark orbital motion. 
The Sivers TMD would vanish if there were no parton Orbital Angular Momentum (OAM).
Hence, studies of Sivers TMD help understand the
contribution of the quark OAM to the nucleon spin. Another interesting aspect is the predicted sign change between the Sivers function extracted from SIDIS process versus that from Drell-Yan process based on QCD. The experimental test of such a sign change 
has been another important motivation for the study of the Sivers TMD.
\item[(iv)] \textbf{Worm-gear TMDs}, $g_{1T}$ and $h_{1L}^{\perp}$, are twist-2 TMD PDFs related to the transverse motion of quark, nucleon spin, and quark spin. 
They are also known as ``worm-gear'' functions since they link perpendicular spin configurations between the nucleon and quarks.
More specifically, $g_{1T}$ describes the distribution of a longitudinally polarized quark inside a transversely polarized nucleon, while $h_{1L}^{\perp}$
describes the distribution of a transversely polarized quark inside a longitudinally polarized nucleon.
Interestingly, the worm-gear functions can not be generated dynamically from coordinate space densities by final-state interactions, and thus have no analogous terms in impact parameter (coordinate) space described by GPDs.
Their appearance is a sign of intrinsic transverse motion of quarks.
\end{itemize}

\begin{figure}[hbt!]
\centering
\includegraphics[width=0.45\textwidth]{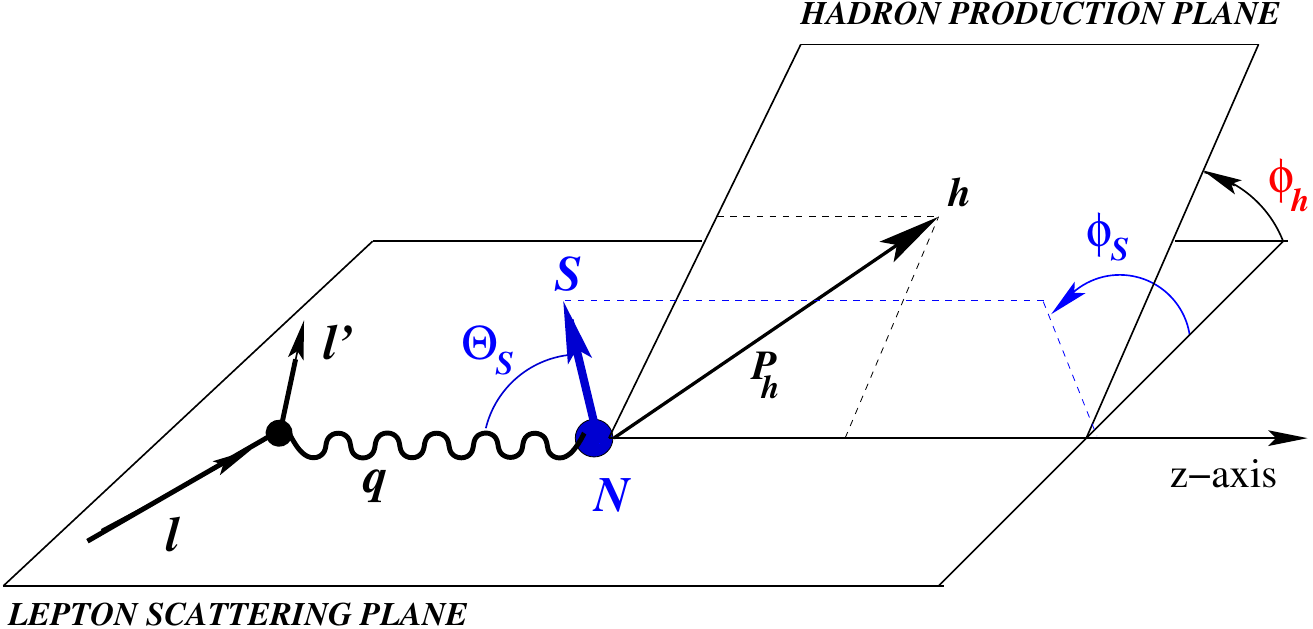}
\caption{Kinematics of SIDIS in the one-photon exchange approximation. 
This figure is from Ref.~\cite{Bastami:2018xqd}.}
\label{fig:fig_SIDIS}
\end{figure}

Figure~\ref{fig:fig_SIDIS} illustrates the SIDIS process in terms of the azimuthal angles defined with respect to the lepton scattering plane. 
$\phi_{h}$ is the angle between the lepton scattering plane and the hadron production plane, while $\phi_{S}$ is the angle 
between the lepton scattering plane and that defined by the polarization vector of the target’s spin and the virtual photon three-momentum vector.
In SIDIS of unpolarized leptons from transversely polarized nucleons, the target single-spin asymmetries 
(SSAs) allow one to experimentally explore the three aforementioned TMDs---Transversity, Pretzelosity, and Sivers---through various azimuthal angular dependencies.

In the leading twist formalism, the SSAs can be written with these three leading twist terms as:
\begin{eqnarray}
A_{UT} & = & A_{UT}^{\rm Collins} \sin\!{(\phi_{h} + \phi_{S})} +
\nonumber \\
& + &  A_{UT}^{\rm Pretzelosity} \sin\!{(3\phi_{h} - \phi_{S})} 
\nonumber \\
& + &
A_{UT}^{\rm Sivers} \sin\!{(\phi_{h} - \phi_{S})} .
\label{eq:eqn_SSA}
\end{eqnarray}
Here in $A_{UT} $, the first subscript U (or L) refers to the unpolarized beam (or longitudinally polarized beam). The second 
subscript  T (or U, or L) refers to the target, which is transversely polarized (or unpolarized, or longitudinally polarized) with respect to the virtual photon three-momentum vector. The SSAs in Eq.~(\ref{eq:eqn_SSA}) 
are represented as follows, assuming TMD factorization holds:
\begin{eqnarray}
\!\!\!\!\!\!\!\!\!\!
\mbox{(i)}~~~
A_{UT}^{\rm Collins} & \propto &
\nonumber \\
& & \!\!\!\!\!\!\!\!\!\!\!\!
\langle \sin\!{(\phi_{h} + \phi_{S})} \rangle_{UT} \propto h_{1T} \otimes H_{1}^{\perp} ,
\label{eq:eqn_Collins}
\end{eqnarray}
where, $H_{1}^{\perp}$ is the Collins fragmentation function \cite{Collins:1992kk}, extracted from charged pion pair production 
based on $e^{+}e^{-}$ annihilation \cite{Belle:2005dmx}, $\otimes $ represents a convolution.
\begin{eqnarray}
\!\!\!\!\!\!\!\!\!\!
\mbox{(ii)}~~~
A_{UT}^{\rm Pretzelosity} & \propto &
\nonumber \\
& & \!\!\!\!\!\!\!\!\!\!\!\!\!\!\!\!\!\!\!\!\!\!
\langle \sin\!{(3\phi_{h} - \phi_{S})}  \rangle_{UT} \propto h_{1T}^{\perp} \otimes H_{1}^{\perp} ,
\label{eq:eqn_Pretzelosity}
\end{eqnarray}
where $h_{1T}^{\perp}$ is the pretzelosity TMD, and the same Collins fragmentation function appears. Models show that non-zero 
pretzelosity requires interference between the nucleon wave function components differing by two units of OAM of the quarks 
(e.g., the interference of the $p$-$p$ or $s$-$d$ OAM states). The Pretzelosity asymmetry stems from quarks that are polarized 
perpendicularly to the nucleon spin direction, in the transverse plane within a 
transversely polarized nucleon. 
\begin{eqnarray}
\!\!\!\!\!\!\!\!\!\!
\mbox{(iii)}~~~
A_{UT}^{\rm Sivers} & \propto &
\nonumber \\
& & \!\!\!\!\!\!\!\!\!\!
\langle \sin\!{(\phi_{h} - \phi_{S})} \rangle_{UT} \propto f_{1T}^{\perp} \otimes D_{1} ,
\label{eq:eqn_Sivers}
\end{eqnarray}
where $f_{1T}^{\perp}$ is the Sivers function, describing the probability density of finding unpolarized quarks inside a 
transversely polarized nucleon, and $D_{1}$ is the unpolarized fragmentation function.

These three asymmetries are part of the total 18 terms in the SIDIS differential cross section expression
\cite{Bastami:2018xqd}. There are five terms depending on target transverse spin direction $S_T$. Two of the five terms are higher-twist terms. At leading twist, only the following three terms are relevant for the target transverse single spin asymmetry:
\begin{eqnarray}
& &
\frac{d\sigma_{\rm SIDIS}}{dx\,dy\,dz\,dP_{h\perp}^{2}\,d\phi_{h}\,d\phi_{{\!}_{S}}} =
\nonumber \\
& & = \frac{\alpha^{2}}{x y Q^{2}}
\left( 1 - y + \frac{1}{2}y^{2} \right) F_{UU}(x,y,P_{h\perp}^{2}) \times
\nonumber\\
& &
~~~ \times \Bigg\{ 1 + ... +  S_{T}\,\sin(\phi_{h}+\phi_{S})\,p_{1}\,A_{UT}^{\rm Collins} + 
\nonumber\\
& &
~~~~~~~~ + S_{T}\,\sin(3\phi_{h}-\phi_{S})\,p_{1}\,A_{UT}^{\rm Pretzelosity} + 
\nonumber\\
& &
~~~~~~~~ + S_{T}\,\sin(\phi_{h}-\phi_{S})\,A_{UT}^{\rm Sivers} + ... \Bigg\} .
\label{eqn:eq_SIDIS_kin}
\end{eqnarray}
For definitions of the 
kinematic variables and the coefficient $p_{1}$, see Eq.~(2.1) and Eq.~(2.3) in \cite{Bastami:2018xqd}.

These SIDIS SSAs depend on four kinematic variables that are $(x, P_{hT}, z, Q^{2})$, and such asymmetries are 
typically small and kinematic dependent. Therefore, high-precision measurements of these asymmetries in such a 4-D kinematic 
space will require a {\it large acceptance} + {\it high luminosity} device (such as SoLID) with a full azimuthal angular 
range to disentangle various azimuthal angular dependencies.  

The experimental SSA for a detector such as SoLID with a full 2$\pi$ azimuthal angular acceptance is defined as 
\cite{JLabPR:E12-10-006}
\begin{eqnarray}
\!\!\!\!\!\!\!\!\!\!
A_{UT}(\phi_{h}, \phi_{S}) & = & \frac{2}{P_{T}^{1} + P_{T}^{2}} \times
\nonumber \\
& & \!\!\!\!\!\!\!
\times \frac{ \sqrt{N_{1}\!\uparrow\,N_{2}\!\downarrow} -  \sqrt{N_{1}\!\downarrow\,N_{2}\!\uparrow} }
{ \sqrt{N_{1}\!\uparrow\,N_{2}\!\downarrow} +  \sqrt{N_{1}\!\downarrow\,N_{2}\!\uparrow} } .
\label{eq:eqn_ExpAsym}
\end{eqnarray}
In this formula, the given number of counts $N_{1}\!\uparrow \equiv N_{1}(\phi_{h},\phi_{S})$ and 
$N_{1}\!\downarrow \equiv N_{1}(\phi_{h},\phi_{S} + \pi)$ are taken at the same time while the target 
polarization is $P_{T}^{1}$. $N_{2}\!\uparrow \equiv N_{2}(\phi_{h},\phi_{S})$ and 
$N_{2}\!\downarrow \equiv N_{2}(\phi_{h},\phi_{S} + \pi)$ are taken at the same time with the target 
polarization being $P_{T}^{2}$, when the target spin is flipped by $180^{\circ}$.

The JLab PAC50 in July 2022 reviewed all SoLID SIDIS experiments and reaffirmed their importance and re-approved all SIDIS experiments with the highest scientific rating of ``A". 
SoLID's full $2\pi$ azimuthal angular coverage has a unique advantage in reducing systematic uncertainties associated with flipping the target spin direction 
apart from those associated with luminosity and detection efficiencies.

While we use these three SSAs to illustrate how one can access information concerning certain TMDs from SIDIS processes, 
we point out that all eight leading-twist TMDs can be accessed through various lepton and nucleon polarization combinations 
from SIDIS processes.
For example, the aforementioned worm-gear function, $g_{1T}$, can be accessed through the beam-target double spin asymmetry (DSA) of $A_{LT}$
with an azimuthal angular modulation of $\cos\!{(\phi_{h} - \phi_{S})}$. Such DSA measurements require a longitudinally polarized lepton beam and a transversely polarized 
target, as was used in \cite{JeffersonLabHallA:2011vwy}.
The other worm-gear piece, $h_{1L}^{\perp}$, and helicity $g_{1L}$ can be accessed with a longitudinally polarized target
through SSA and DSA measurements of $A_{UL}$ (with an angular modulation of $\sin2\phi_h$) and $A_{LL}$, respectively.
For details, we refer to a recent review article \cite{Anselmino:2020vlp}.

\subsection{\label{sec:SIDIS} The SoLID SIDIS program} 

The 12-GeV physics era at JLab opens a great new window to accomplish precision studies of the transverse spin and 
TMD structure of the nucleon in the valence quark region. The experimental program on TMDs is one of the science 
pillars of the 12-GeV program at JLab. The SoLID SIDIS program aims to address the following questions. \\
$\bullet$ Is it possible to precisely determine the tensor charge of the u and d quarks, which are fundamental properties of the nucleon?

The tensor charge of the $u$ and $d$ quarks can be determined to a high precision from the SoLID SIDIS program. They will provide a precision test of lattice QCD calculations.

$\bullet$ How do we quantify the quark transverse motion inside the nucleon and observe spin-orbit correlations?

The Sivers TMD has been predicted in a variety of models to have the sensitivity to spin-orbit correlations and can provide quantitative information about the transverse motion of the quarks inside the nucleon. With the kinematic reach of SoLID at 12 GeV and the precision SoLID measurements will have, the SoLID SIDIS program will answer the above question and also determine whether the confined motion in the transverse plane is dependent on Bjorken $x$ in the valence quark region. 

$\bullet$ Is it possible to provide quantitative information on the quark OAM?

Based on the previous discussion, both Sivers and pretzelosity TMDs are able to provide quantitative information on the quark OAM.

$\bullet$ Are there clear signatures for relativistic effect inside the nucleon and can we observe them? 

Both transversity and prezelosity TMDs will provide clear and quantitative information about relativistic effects inside the nucleon.
The transverse TMD would be the same as that of the helicity TMD if it were not for the relativistic effect. The relation among the helicity, the transversity and the pretzelosity TMDs provides another signature for the relativistic effect inside the nucleon. Again with the high-precision SoLID will achieve, this question will be answered.


In summary, these questions will be answered by three ``A''-rated SoLID experiments approved by the JLab PAC \cite{JLabPR:E12-10-006,JLabPR:E12-11-007,JLabPR:E12-11-108}, 
along with two run group experiments \cite{JLabPR:sidis_dihadron,JLabPR:sidis_kaon}. 
Recently the JLab PAC50 in July 2022 reviewed these experiments and reaffirmed their importance.  
These new experiments will employ a superconducting solenoid magnet, 
a detector system consisting of forward-angle and large-angle sub-detectors, as well as a high-pressure transversely/longitudinally 
polarized ${}^{3}$He (neutron) target and a transversely polarized NH${}_{3}$ (proton) target, positioned upstream of the magnet. 
In order to extract TMDs with precision from SSA and DSA measurements, the SoLID detection system will have a capability of 
handling large luminosities with a large acceptance, a full azimuthal angular coverage, good kinematic coverage in terms 
of the $x$, $P_{hT}$, $z$, $Q^{2}$ variables for SIDIS, and good particle identification for electrons and charged pions and kaons. \\

\noindent\underline{The three approved SIDIS experiments} \\

\begin{figure*}[hbt!]
\hspace*{-0.2in}
\centering
\begin{tabular}{cc}
\includegraphics[width=0.5\textwidth]{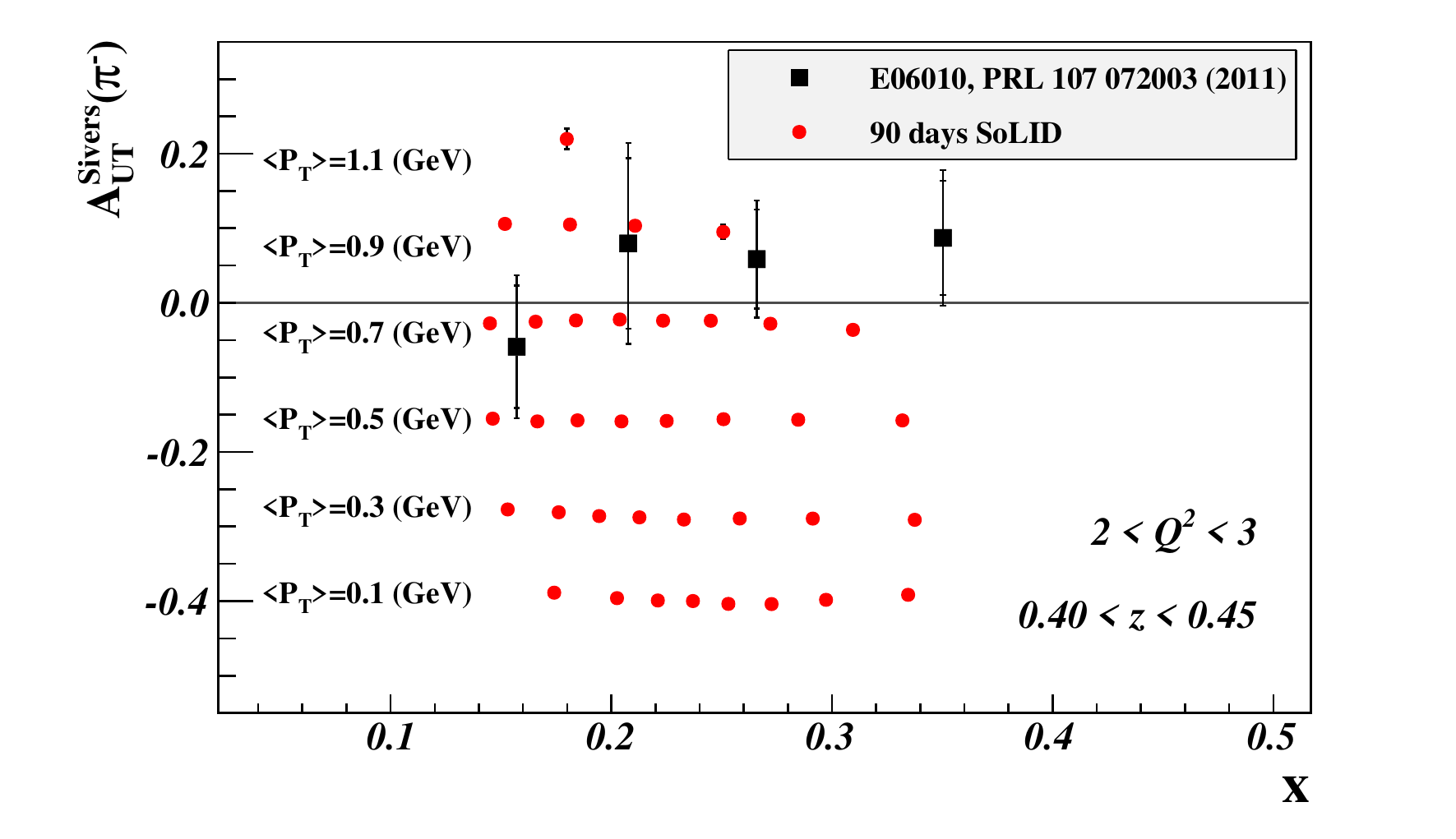}
&
\hspace*{-0.3in}
\includegraphics[width=0.5\textwidth]{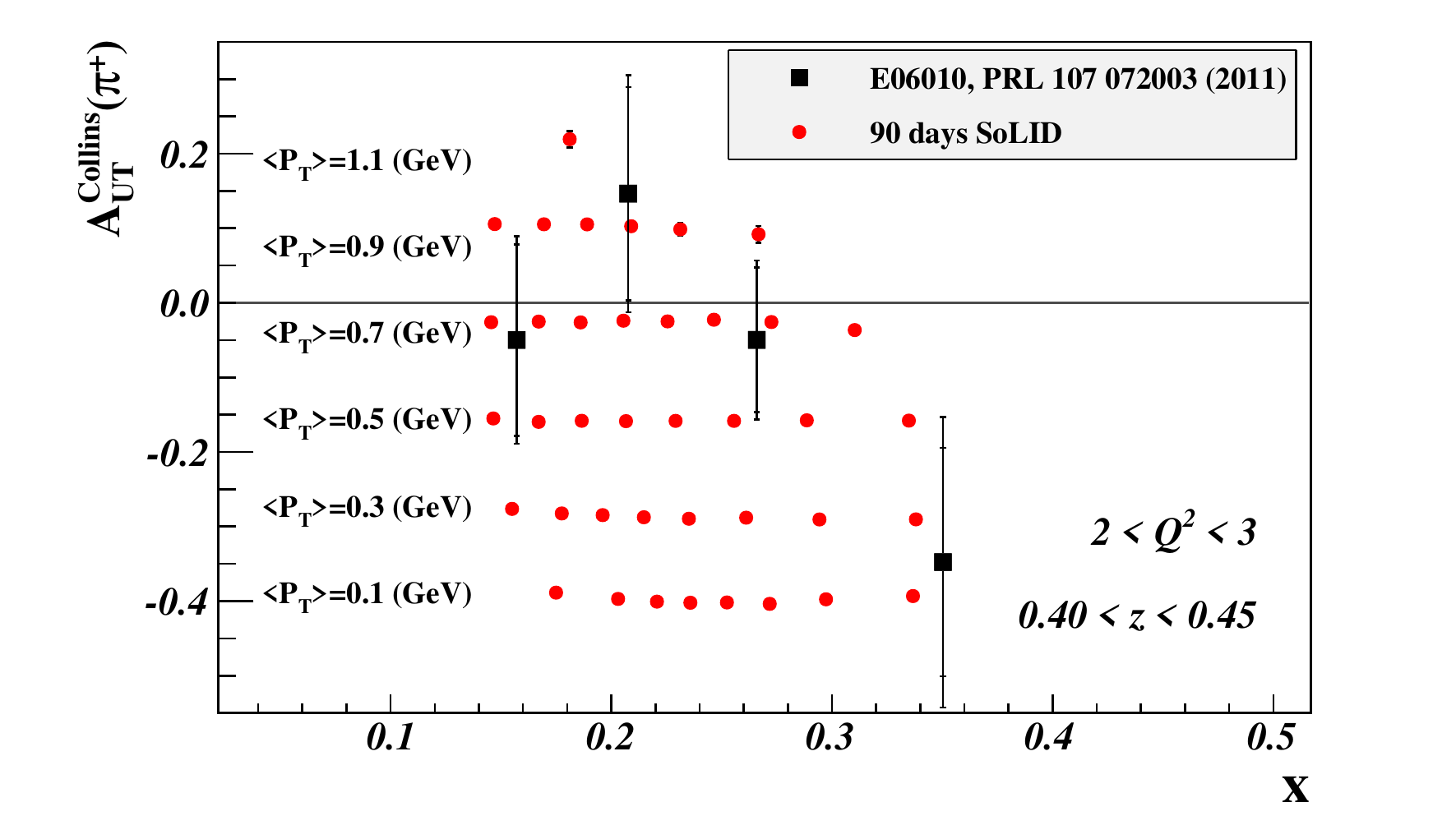} 
\end{tabular}
\linespread{0.5}
\caption{
{} The left panel shows the projected Sivers asymmetry measurement for $\pi^{-}$ for a typical $z$ and $Q^2$ bin ($0.40 < z < 0.45$, 
$2~{\rm GeV}^{2} < Q^{2} < 3~{\rm GeV}^{2}$) as a function of $x$, with different ranges of the hadron $P_{hT}$ labeled. In the 
plots $P_{hT} = P_{T}$. The right panel shows the projected Collins asymmetry measurement for $\pi^{+}$ in the same binnig. Also 
shown are the results from the 6-GeV experiment E06-010~\cite{JeffersonLabHallA:2011ayy}. Both plots are from \cite{Chen:2014psa}.
}
\label{fig:E12-10-006}
\end{figure*}

\paragraph{Experiment E12-10-006 {\rm \cite{JLabPR:E12-10-006} with a transversely polarized ${}^{3}$He target:}}
The experiment E12-10-006 has been approved for 90 days of total beam time with 15 $\mu$A, 11/8.8 GeV electron beams on a 40-cm long,
10 amgs (1 amg = 1 atm at 0$^o$C) transversely polarized $^3$He target. 
The projected data from E12-10-006 are binned in 
$(x, P_{hT}, z, Q^{2})$ space, and only SoLID allows for such 4-D binning with excellent precision for each bin. As examples, for a typical $z$ and $Q^2$ bin ($0.40 < z < 0.45$, $2~{\rm GeV}^{2} < Q^{2} < 3~{\rm GeV}^{2})$, 
data projections for the Sivers and Collins asymmetry measurements are shown in Fig.~\ref{fig:E12-10-006} with the left 
panel for the Sivers $\pi^-$ and right panel for the Collins $\pi^+$ asymmetries. For the complete projections, which consist over
1400 data points, we refer to the proposal~\cite{JLabPR:E12-10-006}.

\begin{figure*}[hbt!]
\vspace*{-0.1in}
\centering
\begin{tabular}{cc}
\includegraphics[width=0.475\textwidth]{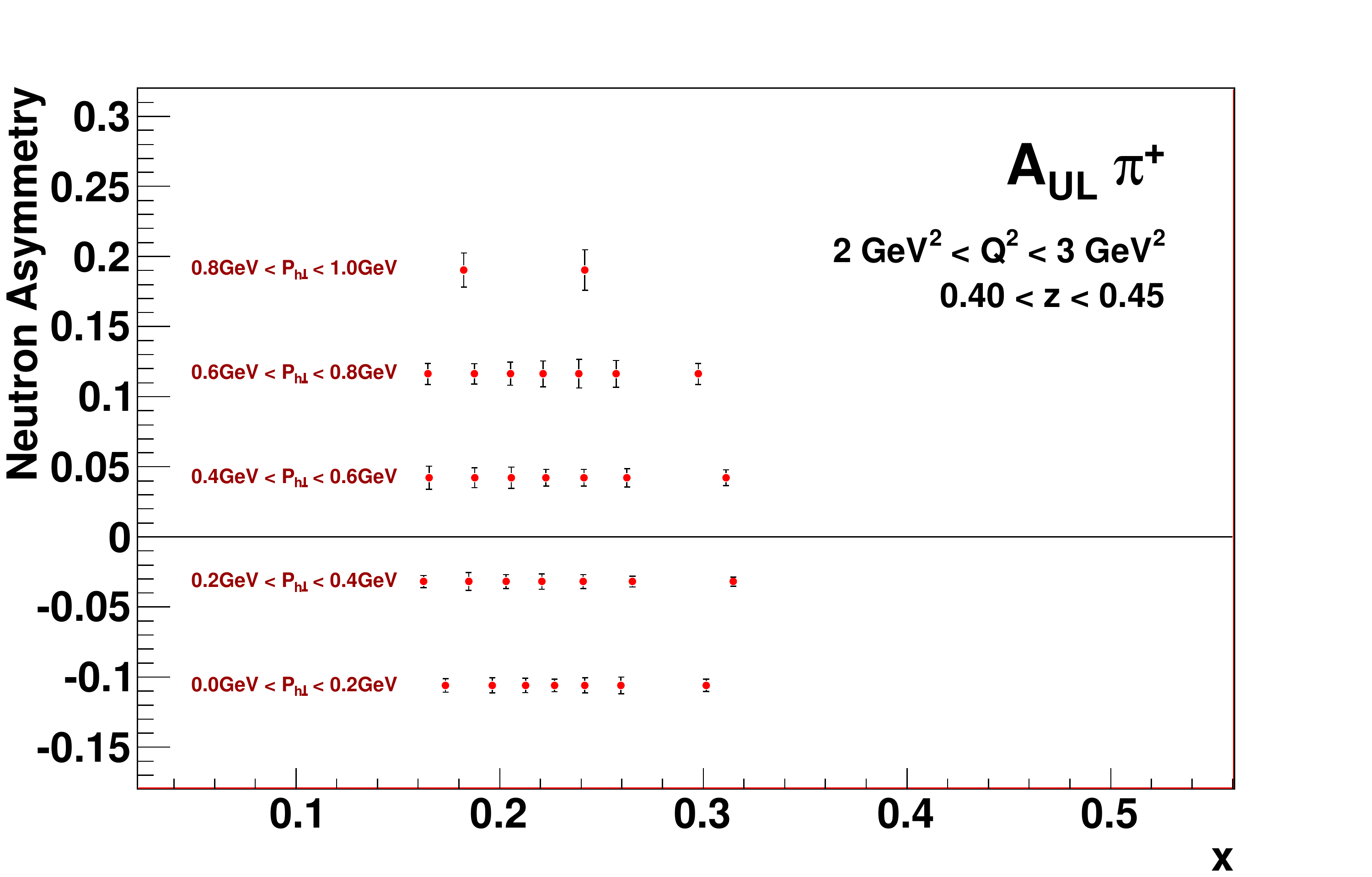}
&
\includegraphics[width=0.475\textwidth]{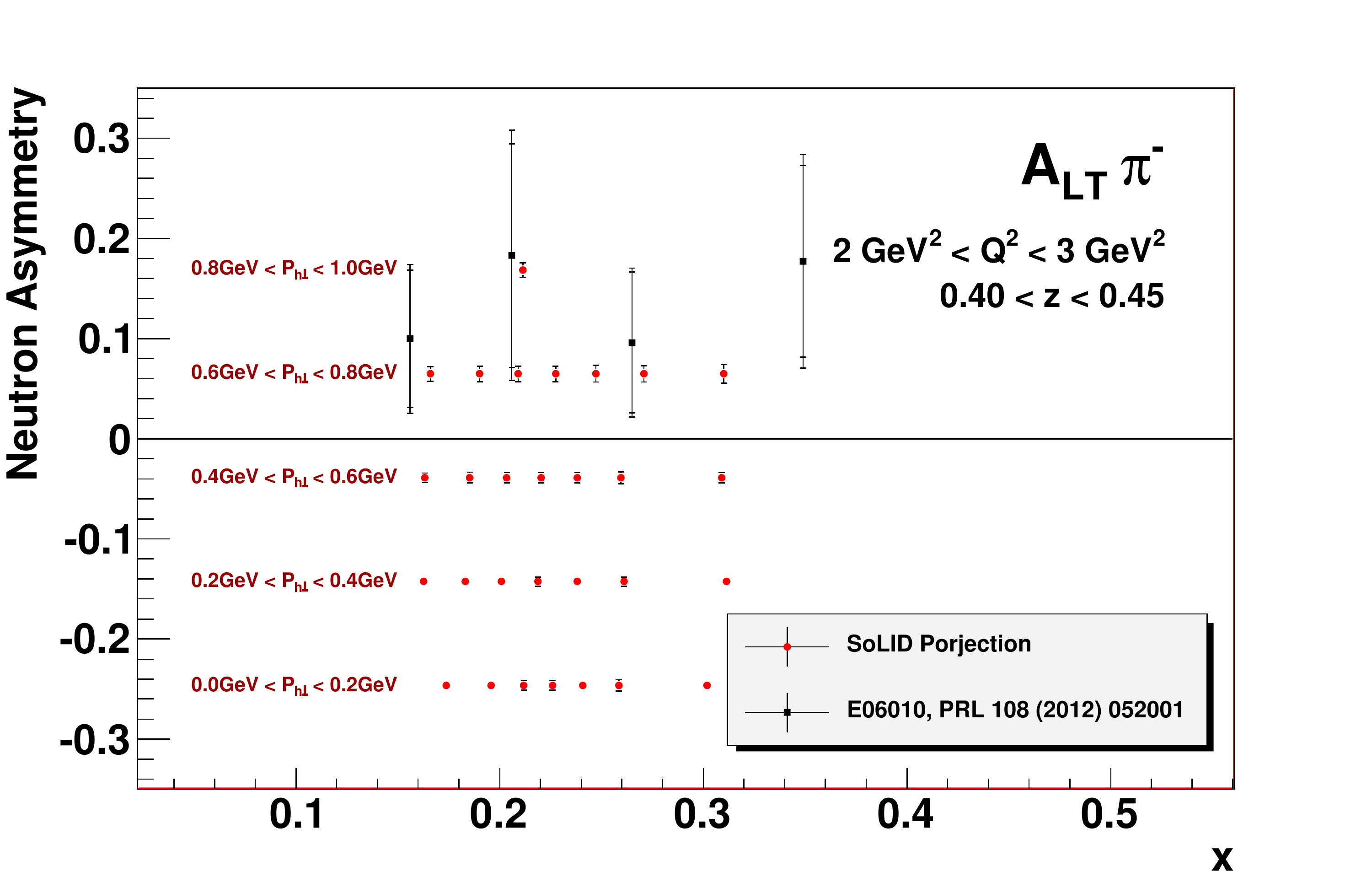} 
\end{tabular}
\linespread{0.5}
\caption{
{} The left panel shows the projection for a typical $z$ and $Q^2$ bin ($0.40 < z < 0.45$, $2~{\rm GeV}^{2} < Q^{2} < 3~{\rm GeV}^{2}$) 
for the $\pi^{+}$ single-target spin asymmetry $A^{sin(2\Phi_{h})}_{UL}$ measurement as a function of $x$, with different ranges of the 
hadron $P_{hT}$ labeled. In the plots $P_{hT} = P_{T}$. The right panel shows the projection for the corresponding $z$-$Q^2$ bin for the 
$\pi^{-}$ double-target spin asymmetry $A^{cos(\Phi_{h}-\Phi_{S})}_{LT}$ measurement. Also shown are the results from the 6-GeV 
experiment E06-010~\cite{JeffersonLabHallA:2011vwy}.
} 
\label{fig:E12-11-007}
\end{figure*}

\paragraph{Experiment E12-11-007 {\rm \cite{JLabPR:E12-11-007} with a longitudinally polarized ${}^{3}$He target:}}
The experiment E12-11-007 has been approved for 35 days of total beam time with 15 $\mu$A, 11/8.8 GeV electron beams on a 40-cm long, 
10 amgs longitudinally polarized $^3$He target to match about 50\% statistics of the experiment E12-10-006. 
The projected data are binned into $(x, P_{hT}, z, Q^{2})$ bins. For a typical $z$ and $Q^2$ bin ($0.40 < z < 0.45$, 
$2~{\rm GeV}^{2} < Q^{2} < 3~{\rm GeV}^{2}$, one of the total 48 $z$-$Q^2$  slices), data projections are shown in 
Fig.~\ref{fig:E12-11-007} as examples. For the complete projections, we refer to the proposal~\cite{JLabPR:E12-11-007}.

\paragraph{Experiment E12-11-108 {\rm \cite{JLabPR:E12-11-108} with a transversely polarized NH${}_{3}$ target:}}

\begin{figure*}[hbt!]
\vspace*{-0.1in}
\centering
\begin{tabular}{cc}
\includegraphics[width=0.475\textwidth]{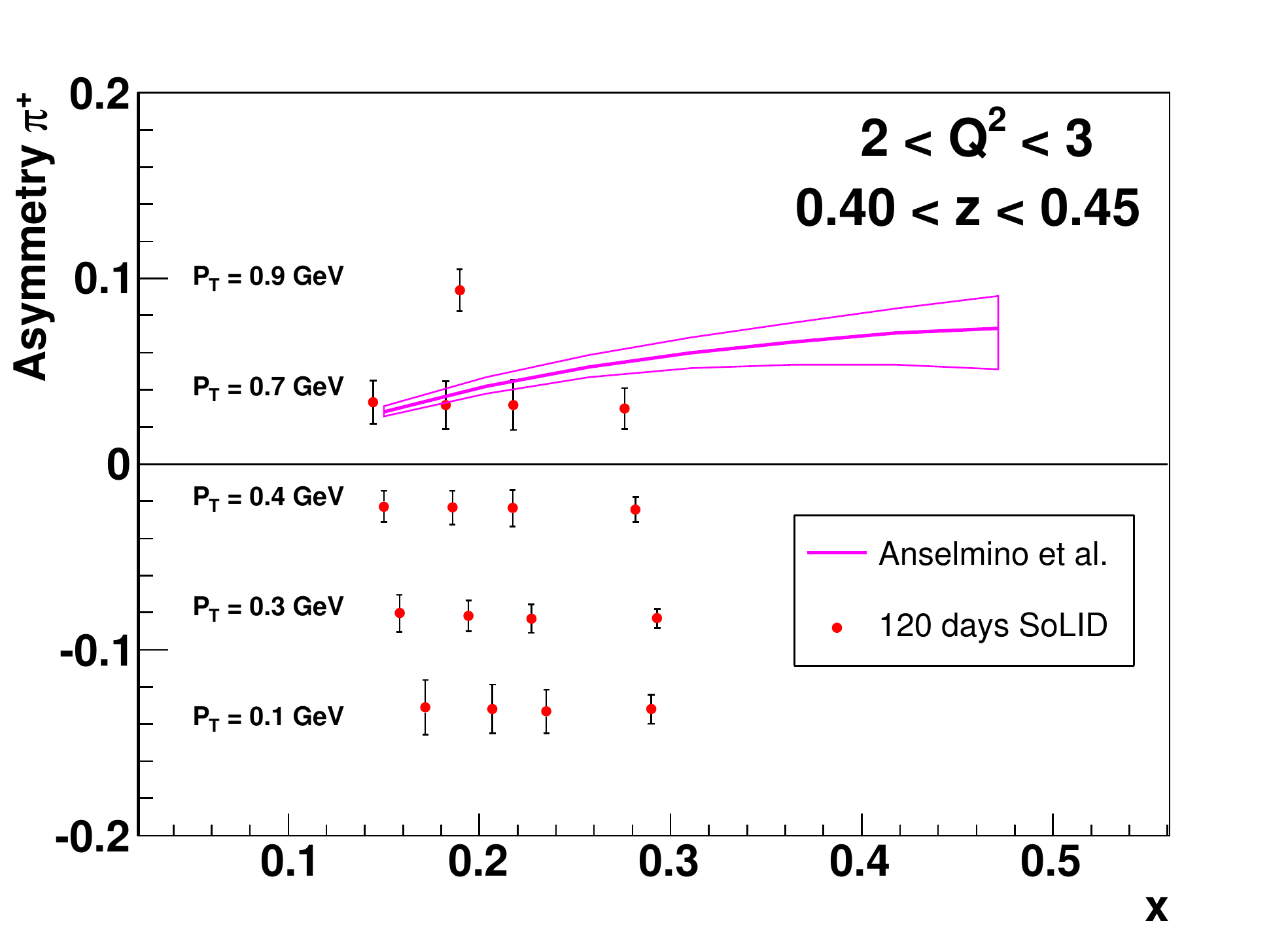}
&
\includegraphics[width=0.475\textwidth]{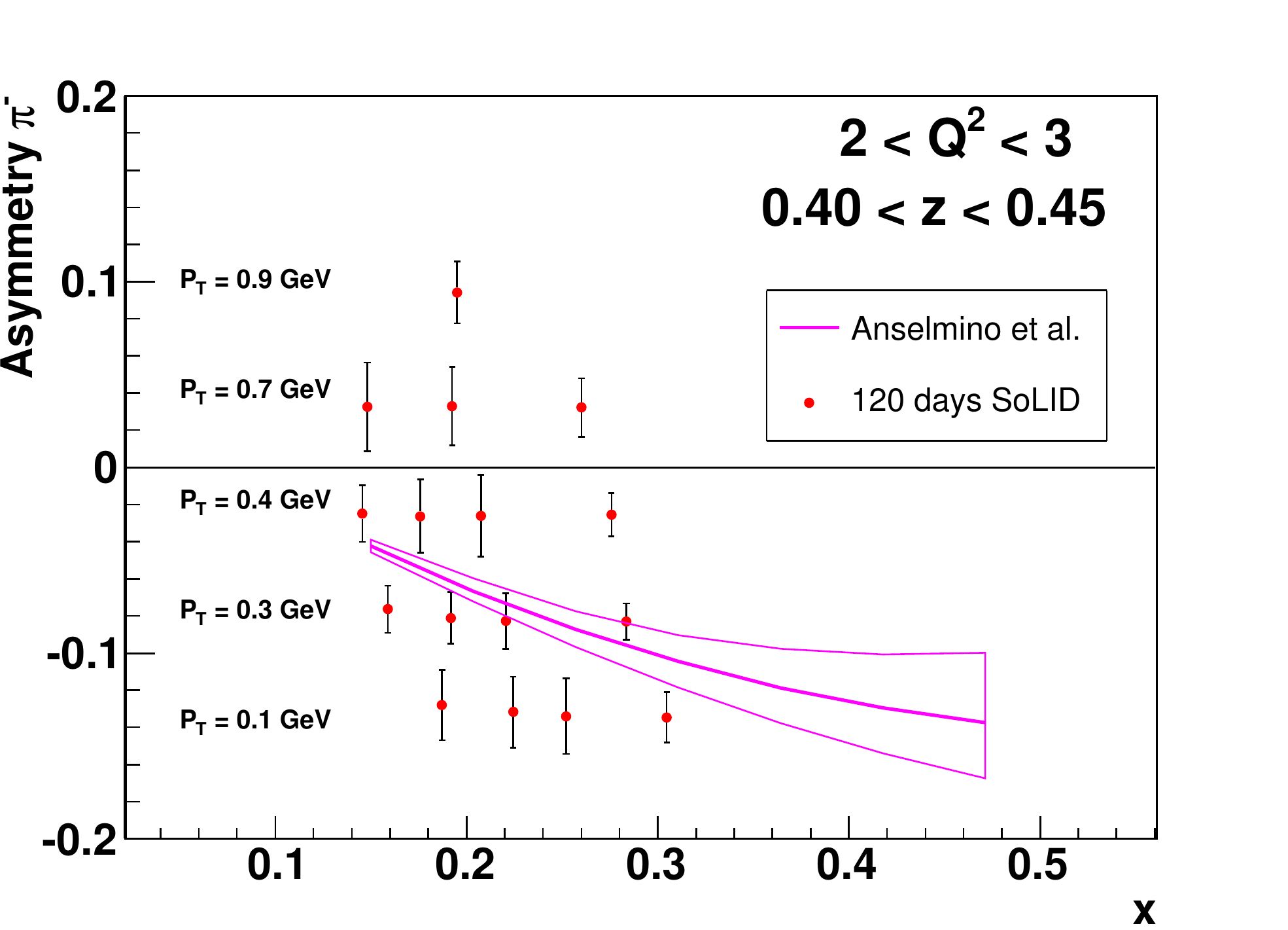} 
\end{tabular}
\linespread{0.5}
\caption{
{} The left panel shows the projection for a typical $z$ and $Q^2$ bin ($0.40 < z < 0.45$, $2~{\rm GeV}^{2} < Q^{2} < 3~{\rm GeV}^{2}$) 
for the $\pi^{+}$ Collins asymmetry measurement as a function of $x$, with different ranges of the hadron $P_{hT}$ labeled. In the plots 
$P_{hT} = P_{T}$. The right panel shows the projection for the corresponding $z$-$Q^2$ bin for the $\pi^{-}$ measurement. Also shown are 
the predictions of the Collins asymmetry from Anselmino {\it et al.}~\cite{Anselmino:pv} with model uncertainties.} 
\label{fig:E12-11-108}
\end{figure*}

The experiment E12-11-108 has been approved for 120 days 
with 100 nA, 11/8.8 GeV electron beams on a 3-cm long, polarized 
NH$_3$ target. 
The 8.8 GeV beam energy will provide precision data for radiative corrections along with the increased $Q^2$ coverage. 
The projected data from E12-11-108 are binned into $(x, P_{hT}, z, Q^{2})$ bins. As an example, for a typical $z$ and $Q^2$ bin 
($0.40 < z < 0.45$, $2~{\rm GeV}^{2} < Q^{2} < 3~{\rm GeV}^{2}$), data projections for the Collins asymmetry measurements are 
shown in Fig.~\ref{fig:E12-11-108} with the left panel for $\pi^+$ and right panel for $\pi^-$. For the complete projections of 
E12-11-108, we refer to the proposal~\cite{JLabPR:E12-11-108}. \\

In July 2022 these three SoLID SIDIS proposals were presented to the JLab PAC50 as part of the JLab jeopardy review process. 
The PAC reaffirmed the importance of the program and all three experiments remain active with ``A" rating. The PAC evaluation summary for each of the three SIDIS experiments is quoted here
``This experiment will provide data of unprecedented quality on SIDIS in JLab-12 GeV kinematics. The theory and phenomenology developments in the last decade make this experiment yet more compelling and highlight the impact of SoLID program."\\

\noindent\underline{The SIDIS run group experiments} \\

\paragraph{Dihadron Electroproduction in DIS with Transversely Polarized ${}^{3}$He Target at 11 and 8.8 GeV {\rm \cite{JLabPR:sidis_dihadron}:}} 
A study of transversity parton distribution using measurements of semi-inclusive electroproduction of two charged pions in the DIS 
region will be carried out. The data will provide input to extract the $u$ and $d$ transversity distributions in a model independent way.
This experiment will be run in parallel with the approved experiment E12-10-006.

\paragraph{$K^{\pm}$ Production in Semi-Inclusive Deep Inelastic Scattering using Transversely Polarized Targets and the SoLID Spectrometer 
{\rm \cite{JLabPR:sidis_kaon}:}} A study of measurements of $K^{\pm}$ production in SIDIS using both the transversely polarized ${}^{3}$He and 
NH${}_{3}$ targets will be performed, to extract the $K^{\pm}$ Collins, Sivers and other TMD asymmetries. The data will provide input 
to determine the $u$, $d$ and sea quarks' TMDs. This experiment will be run in parallel with the approved experiments E12-10-006 and 
E12-11-108.

More details on these two run group experiments will be given in Section VII.

\subsection{\label{sec:tensor} Transversity, Tensor Charge, and EDM}

\begin{figure}[hbt!]
\centering
\includegraphics[width=0.35\textwidth]{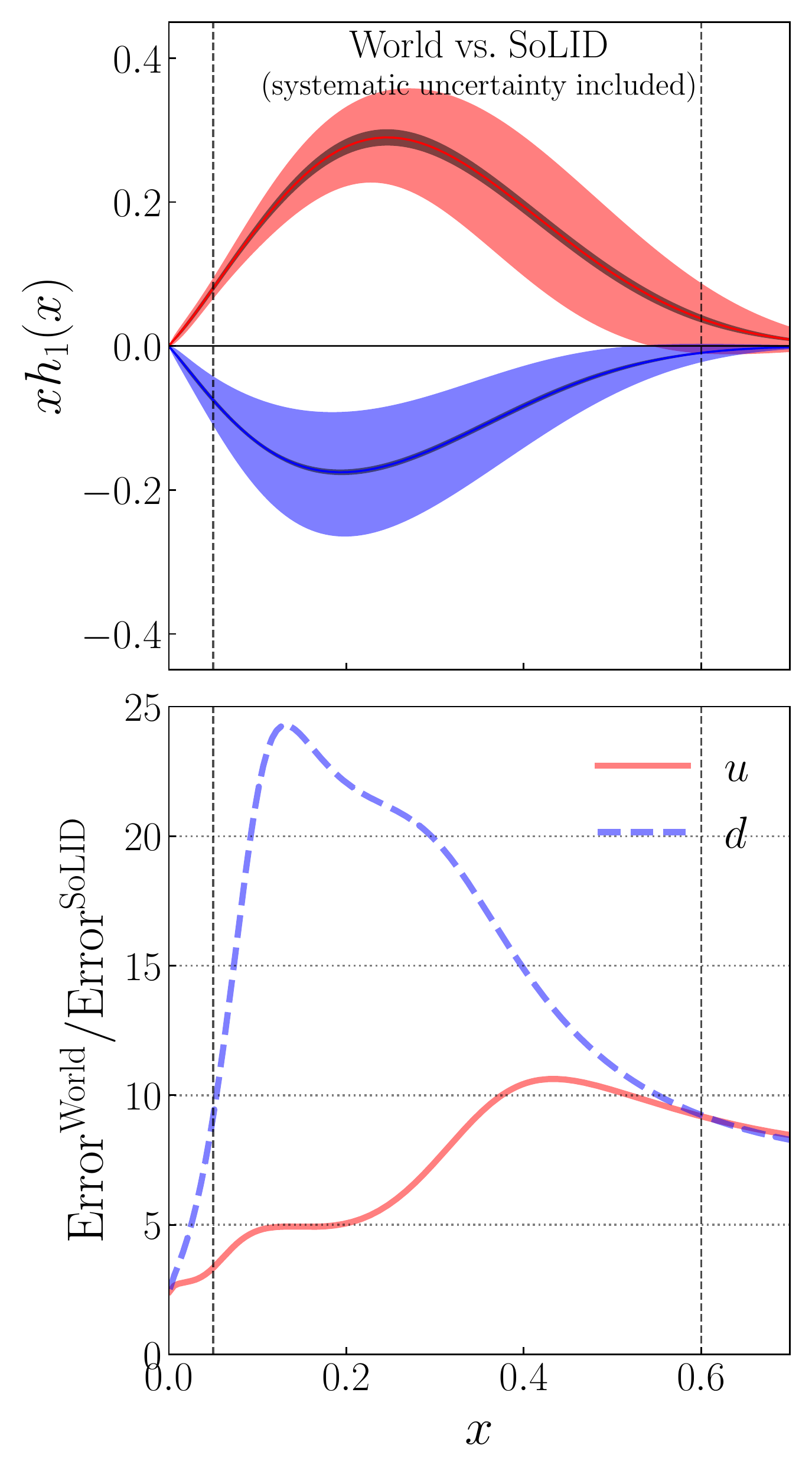}
\caption{The impact on the $u$ and $d$ quarks' transversity distributions by the SoLID SIDIS program. In the top panel, 
the wide uncertainty bands show our current knowledge from the world data global analysis, whereas the narrow uncertainty 
bands show the SoLID projections. The bottom panel shows the improvements, manifested as the ratios between the current 
and projected uncertainties. 
} 
\label{fig:transversity_solid}
\end{figure}

The combination of the SIDIS experiments discussed above will give an opportunity for accessing essential information on TMDs from the neutron and the proton in the valence quark region, and for flavor separation of TMDs (e.g., Transversity, Pretzelosity, Sivers, and $g_{1T}$) 
for $u$ and $d$ quarks. Fig.~\ref{fig:transversity_solid} shows 
the projected SoLID transversity distributions for the $u$ and $d$ quarks at a typical value of $Q^2=2.4$~GeV$^2$  obtained with our up-to-date knowledge of evolution of TMDs and FFs, including both systematic and statistical uncertainties.
The $x$-range between the two vertical dashed lines is directly measurable by SoLID. 
The precision data in the valence quark region
will make a major improvement in our knowledge of the transversity distribution. The program will also allow us to study the $k_T$ dependence and the $Q^2$ evolution of TMDs.

\begin{figure}[hbt!]
\vspace*{-0.15in}
\hspace*{-0.28in}
\centering
\includegraphics[width=0.48\textwidth]{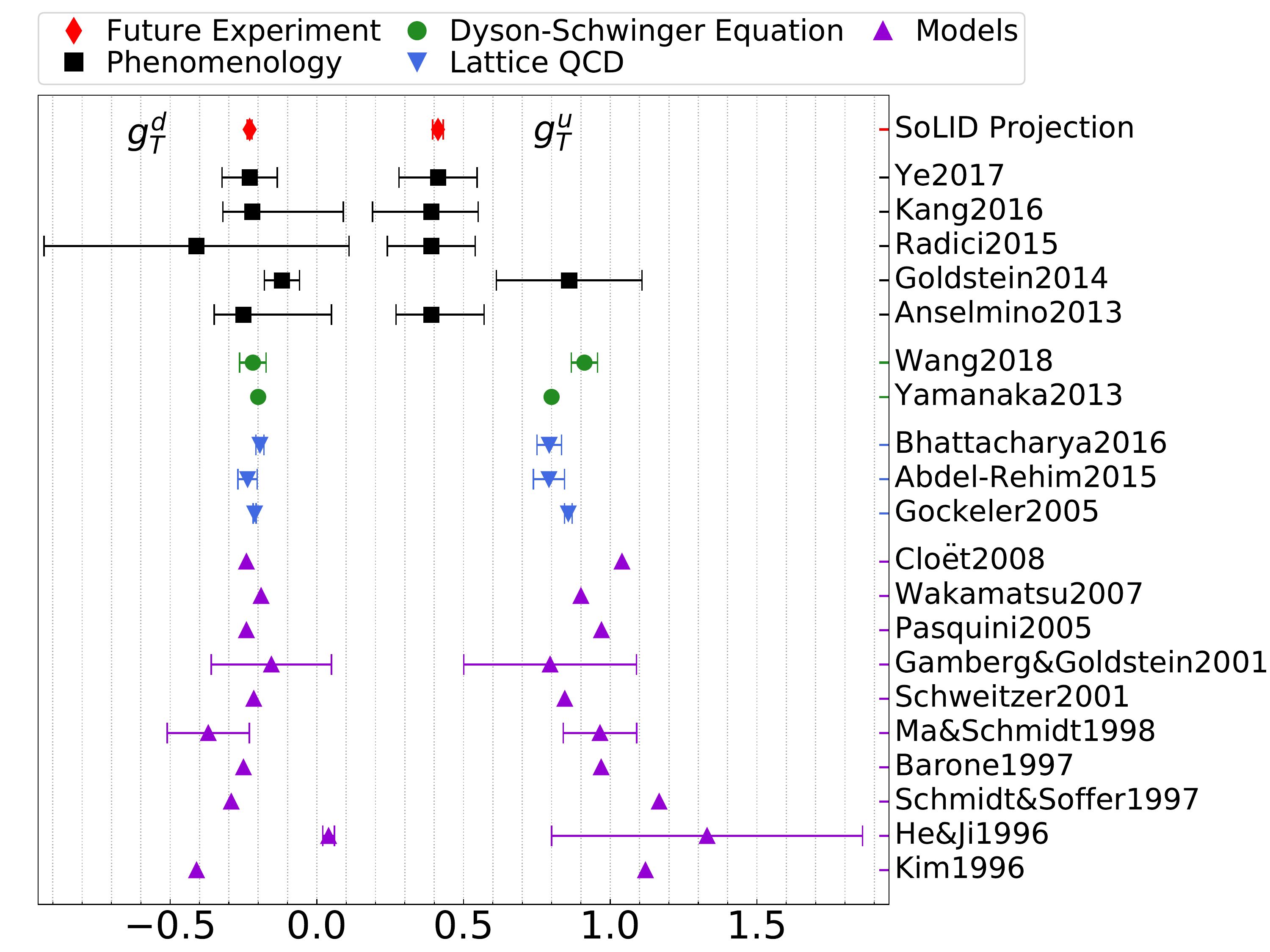}
\caption{The impact of the projected SoLID measurement of the down (left) and up (right) quark tensor charges (see Eq.~(\ref{eq:eqn_gT})) together with the 
current knowledge from various models, Dyson-Schwinger equations, global analyses, and lattice QCD simulations. 
} 
\label{fig:tensorcharge}
\end{figure}

\begin{figure}[hbt!]
\vspace*{-0.15in}
\hspace*{-0.28in}
\centering
\includegraphics[width=0.48\textwidth]{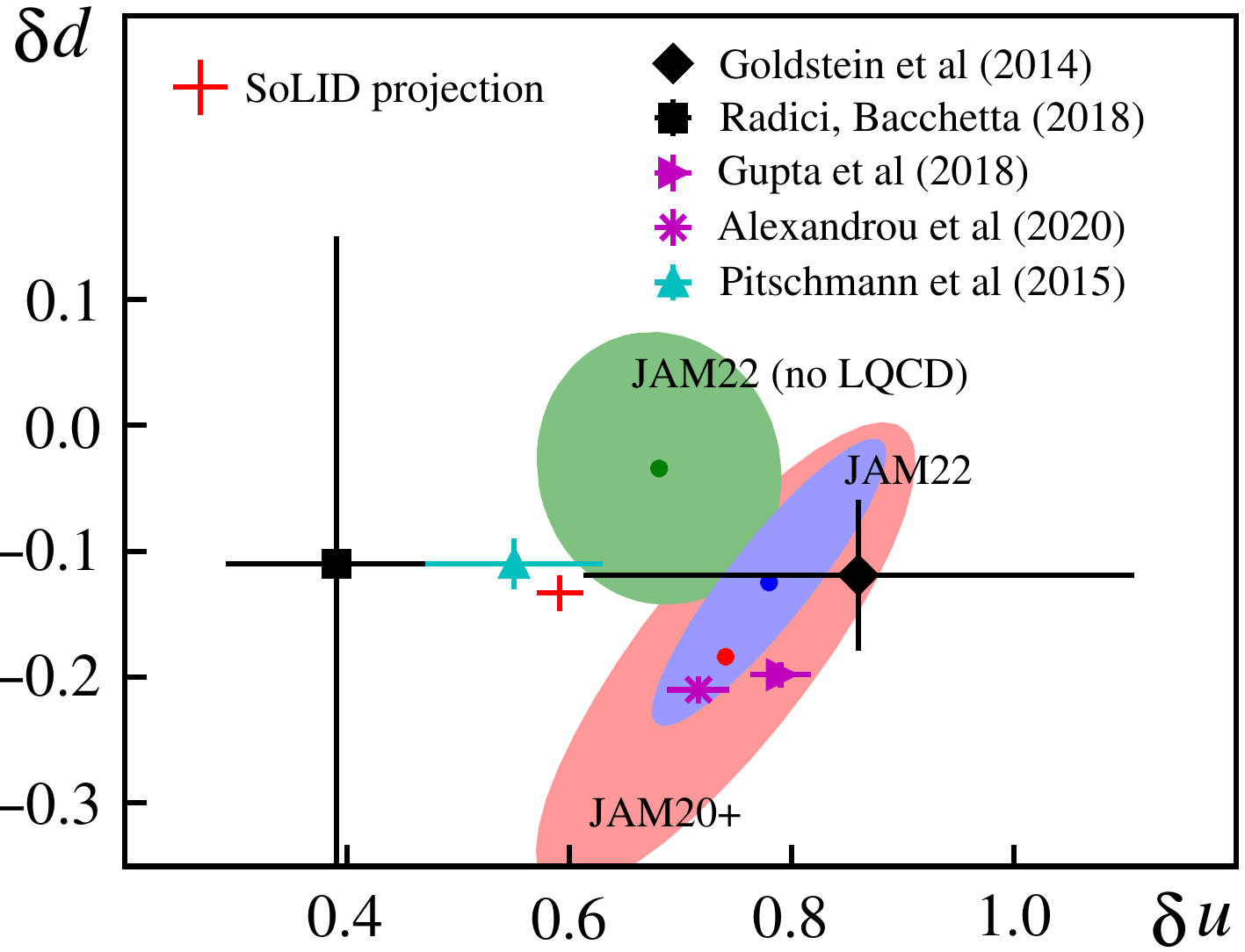}
\caption{The impact of the projected SoLID measurement of the tensor charge added to the left panel of Fig. 3 of ~\cite{Gamberg_2022} with the horizontal axis showing the u-quark tensor charge and the vertical axis showing the d-quark tensor charge. 
} 
\label{fig:tensorcharge2d}
\end{figure}

Moreover, we will obtain precise information on the quark tensor charge defined as
\begin{equation}
g_{T}^{q} = \int_{0}^{1} \left[ h_{1}^{q}(x) - h_{1}^{\bar{q}}(x) \right] dx .
\label{eq:eqn_gT}
\end{equation}
The nucleon (quark) tensor charge is as important as its electric charge, mass and the spin. It has been calculated by lattice QCD and the prediction is becoming increasingly precise. It is also an input for the tests of the Standard Model (see below).
A quantitative study in \cite{Ye:2016prn} shows that the SoLID SIDIS program will improve the accuracy of the tensor charge determination 
by one order of magnitude, allowing for a benchmark test of lattice QCD predictions. The high impact of the SoLID projections on the extraction 
of the tensor charge of the $u$ and $d$ quarks is demonstrated in Fig.~\ref{fig:tensorcharge} and Fig.~\ref{fig:tensorcharge2d}. 
 The projected SoLID $u$ and $d$ quark tensor charges are 
$g_{T}^{u} = 0.547 \pm 0.021$, $g_{T}^{d} = -0.376 \pm 0.014$. They represent less than 4\% relative uncertainty 
for the SoLID extraction of the $u$ and $d$ quark tensor charge, and should be compared to the 2019 FLAG review~\cite{FlavourLatticeAveragingGroup:2019iem} of the 
Lattice QCD calculations where the corresponding numbers are 4\% and 7\% for $u$ and $d$ quark, respectively. 
The Lattice QCD calculations\cite{PNDME18,ETMC20} of isovector tensor charge reached precision of 3\%.   
Therefore, these results 
will provide a benchmark test of precise lattice calculations. 

More recent global analysis and extraction of the tensor charge by the JAM collaboration included world data and Lattice QCD results~\cite{Gamberg_2022,PhysRevD.102.054002}. Their extracted tensor charges of the u and d quarks are shown in 2-d plot Fig.\ref{fig:tensorcharge2d}, along with the SoLID projection. The JAM collaboration also made projections~\cite{GAMBERG2021136255} for both SoLID and EIC, concluded that combing both would lead to the most precise extractions of the tensor charge.

The tensor charge is also connected to 
the neutron and proton electric dipole moments (EDMs), giving us a unique opportunity to test the Standard Model (SM) and to search for new physics beyond SM. The nucleon EDM is related to the quark EDM as~\cite{Ellis:1996dg,Bhattacharya:2012bf,Pitschmann:2014jxa,Xu:2015kta}:
\begin{equation}
d_{p} = g_{T}^{u}d_{u} + g_{T}^{d}d_{d} + g_{T}^{s}d_{s}
\end{equation}
\begin{equation}
d_{n} = g_{T}^{d}d_{u} + g_{T}^{u}d_{d} + g_{T}^{s}d_{s}, 
\label{eq:eqn_edm}
\end{equation}
where quark tensor charges appear as the coefficients in front of the corresponding quark EDMs. In these two equations, the heavy flavor contributions are neglected, and isospin symmetry is applied in Eq.~(\ref{eq:eqn_edm}).
Notably, a phenomenological study in~\cite{Liu:2017olr} puts experimental constraints on quark EDMs by combining nucleon EDM measurements 
with tensor charge extractions. With the current projected sensitivity of the neutron/proton EDM experiments and the existing precision of 
tensor charge extractions (based on the study from \cite{Ye:2016prn}), the upper limit on quark EDMs is
$1.27 \times 10^{-24}\,e \cdot {\rm cm}$ for the $u$ quark, and $1.17 \times 10^{-24}\,e \cdot {\rm cm}$ for the $d$ quark, where a 
10\% uncertainty from the isospin symmetry breaking is included. Both are determined at the scale of $4~{\rm GeV^{2}}$. 
Future precise measurements of the tensor charge from the SoLID SIDIS program and the nucleon EDMs will reduce the upper limit on quark EDMs by about three orders 
of magnitude, {\it i.e.} to the level of $10^{-27}\,e\cdot\rm cm$~\cite{Liu:2017olr}. 
Based on the result from a dimensional analysis that gives the quark EDM as $d_{q} \sim {\frac{em_q}{4\pi{\Lambda}^2}}$~\cite{POSPELOV2005119},
 we estimate the new physics scale probed 
by the current quark EDM limit to be about $1\,\rm TeV$. With the quark EDM limit improved by three orders of magnitude from future experiments, 
it can probe new physics up to $30$--$40$\,TeV~\cite{Liu:2017olr}, beyond the LHC energy.

\section{Parity Violating Deep Inelastic Scattering}\label{sec:pvdis}
The main observable to be measured by the PVDIS program of SoLID~\cite{JLabPR:PVDIS_solid} is the Parity-Violating (PV) cross section asymmetry, defined as
\begin{eqnarray}
  A_{RL}\equiv \frac{\sigma_R-\sigma_L}{\sigma_R+\sigma_L}~,
\end{eqnarray}
where $\sigma_{R,L}$ are differential cross sections of right- and left-handed incoming electron, respectively. 
The first Parity Violating Electron Scattering (PVES) experiment,  SLAC E122~\cite{Prescott:1978tm,Prescott:1979dh}, played a pivotal role in establish the Standard Model of electroweak physics. During the 6 GeV era of JLab, PVES has provided data on the strangeness content of the nucleon (see {\it e.g.}\ G0 experiment~\cite{G0:2005chy,G0:2009wvv}), the excess of the neutron distribution in heavy nuclei and its connection to neutron star physics~\cite{PREX:2021umo}, and  determination of the proton weak charge~\cite{Androic:2013rhu,Androic:2018kni} as a test of the SM~\cite{Young:2007zs}. Furthermore, through measurement of $A_{RL}$ in DIS, a measurement of the effective electron-quark neutral current couplings $g_{VA}^{eq}$~\cite{Wang:2014bba,Wang:2014guo} was completed that improved the precision of SLAC E122 by an order of magnitude.

In the DIS region, the asymmetry can be written as
\begin{eqnarray}
 A_{PV} &=& -\frac{G_FQ^2}{4\sqrt{2}\pi\alpha}
         \left[a_1+a_3Y\right]~,\label{eq:Apvdis1}
\end{eqnarray}
where $G_F=1.166\times 10^{-5}$~GeV$^{-2}$ is the Fermi constant, $\alpha$ is the fine structure constant, and
\begin{equation}
  a_1(x)=2 g_A^e\frac{F_1^{\gamma Z}}{F_1^\gamma}~,~~
  a_3(x)=g_V^e\frac{F_3^{\gamma Z}}{F_1^\gamma}~.  \label{eq:a31}
\end{equation}
The structure functions $F_{1,3}^{\gamma, \gamma Z}$ can be written in the parton model in terms of PDFs $q_i(x,Q^2)$ and $\bar q_i(x,Q^2)$ of the target:
\begin{eqnarray}
  F_1^\gamma(x,Q^2)&=&\frac{1}{2} \sum{Q_{q_i}^2 \left[q_i(x,Q^2)+\bar q_i(x,Q^2
)\right]}~,\label{eq:F1gqpm}\\
  F_1^{\gamma Z}(x,Q^2)&=&\sum{Q_{q_i} g_V^i\left[q(x,Q^2)+\bar q_i(x,Q^2)\right
]}~,\label{eq:F1gzqpm}\\
  F_3^{\gamma Z}(x,Q^2)&=&2\sum{Q_{q_i} g_A^i \left[q_i(x,Q^2)-\bar q_i(x,Q^2)\right]}~.\label{eq:F3gzqpm}
\end{eqnarray}
Here, $Q_{q_i}$ denotes the quark's electric charge and 
the summation is over the quark flavors $i=u,d,s\cdots$. The $g_{V,A}^{e,i}$ are the vector and axial coupling of the electron or quark of flavor $i$, and in the SM are determined by the weak mixing angle, and the electric and weak hypercharge of the particle. 
The variable $Y$ is a kinematic factor given approximately by 
\begin{eqnarray}
  Y&=& \frac{1-(1-y)^2}
   {1+(1-y)^2}~.\label{eq:y13_prd}
\end{eqnarray}
Detailed expressions of $Y$ that include target-mass effect and the longitudinal structure function $F_L$ can be found in Ref.~\cite{Wang:2014guo}. 

Equation~(\ref{eq:Apvdis1}) shows that by measuring the PVDIS asymmetry on the proton or nuclei, different physics topics can be explored. 
The PVDIS program of SoLID includes three components: the PVDIS deuteron program that is aimed at the precision determination of electroweak parameters and a search for Beyond-the-Standard Model (BSM) physics; the PVDIS proton program that will provide the PDF ratio $d/u$ in the valence quark region free of nuclear model dependence; and the PVEMC program that will study isospin dependence of the EMC effect by the use of neutron-rich isotopes. 
With SoLID fully exploring the high luminosity potential of CEBAF, we expect to improve the precision of PVDIS measurement by a factor of ten compared with 6 GeV.

\subsection{PVDIS Deuteron Measurement}
\subsubsection{SoLID as a EW/BSM Facility}
The SM is a theoretical framework that explains successfully nearly all existing phenomena of particle physics. On the other hand, it is often referred to as an effective theory at the electroweak scale, and believed to be only part of a theory that would ultimately encompass all three (or four) interactions of nature. Given that current evidence of new physics, such as dark matter and neutrino mass, allows many possibilities to extend the SM to higher energy scales, it is imperative that we carry out as many high-precision measurements as possible to test the SM and to shed light on where BSM physics might occur.

The high intensity beam of CEBAF provides a unique opportunity for SM and BSM study. The reach of experiments that search for BSM physics, if focused on new heavy particles, can be approximately characterized by the product $s\sqrt{\mathcal L} $ where $\mathcal L$ is the luminosity and $s$ is the center-of-mass energy of the lepton-nucleon scattering process.  Even with the electron ion collider (EIC) coming online in the near future, the BSM reach of fixed-target experiments at JLab is still at least one order of magnitude higher than the EIC if the intensity of CEBAF's 11 GeV beam is matched by the use of a large acceptance spectrometer, placing SoLID at a unique position to provide an impact on the landscape of EW/BSM physics study for the next decade(s). 

\subsubsection{Determination of EW Parameters}
To access EW paramters, we measure the PVDIS asymmetry on a deuteron target, for which the SM expression simplifies to: 
\begin{eqnarray}
  A_{PV,(d)}^{SM}= \frac{3G_F Q^2}{10\sqrt{2}\pi\alpha}\left[(2g_{AV}^{eu}-g_{AV}^{ed})+R_V Y(2g_{VA}^{eu}-g_{VA}^{ed})\right]~,\nonumber\\
\end{eqnarray}
where $R_V(x) \equiv ({u_V+d_V})/({u^+ + d^+})$ 
with $q^+\equiv q(x)+\bar q(x)$ and $q_V\equiv q(x)-\bar q(x)$. Using the appropriate electric charge and the weak isospin of quarks, they are related to the weak mixing angle $\theta_W$. We define the low energy electron-quark effective couplings, given in the SM at the tree level as: 
\begin{eqnarray}
 g^{eu}_{AV} &=& 2g^e_A g^u_V
          = -\frac{1}{2} + \frac{4}{3} \sin^2\theta_{W}~,\label{eq:c1ufactorized}\\
  g^{eu}_{VA} &=& 2g^e_V g^u_A
          = - \frac{1}{2} + 2 \sin^2\theta_{W}~,\label{eq:c1dfactorized}\\
 g^{ed}_{AV} &=& 2g^e_A g^d_V
          = \frac{1}{2} - \frac{2}{3} \sin^2\theta_{W}~,\label{eq:c2ufactorized}\\
  g^{eu}_{VA}&=& 2g^e_V g^d_A 
          =  \frac{1}{2} - 2 \sin^2\theta_{W}~.\label{eq:c2dfactorized}
\end{eqnarray} 
Note that in BSM physics extensions, the couplings may no longer be expressed as products of electron and quark couplings.

\begin{figure}[!ht]
\includegraphics[width=0.5\textwidth]{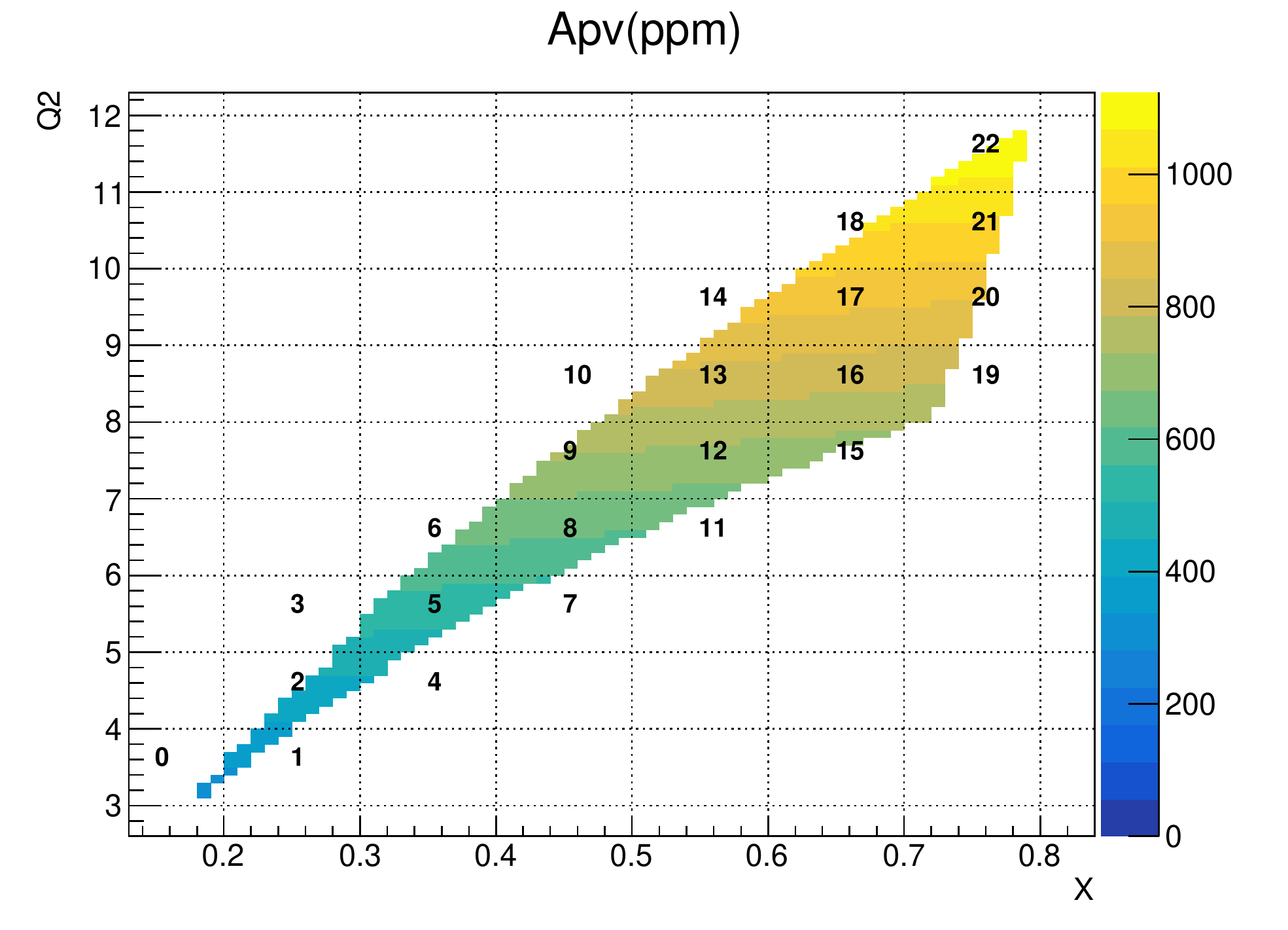}
\caption{Illustration of PVDIS asymmetry on a deuteron target in ppm on the $(x,Q^2)$ plane. The data are divided into evenly spaced grid with the bin number shown. The expected statistical uncertainty is less than 1\% in most of the bins. }
\label{fig:Apv_ed}
\end{figure}

Using 120~days of 50~$\mu$A electron beam with 85\% polarization incident on a 40-cm long liquid deuterium target, we can measure the PVDIS asymmetry to sub-percent-level precision within a wide $(x,Q^2)$ range, see Fig.~\ref{fig:Apv_ed}. 
The dominant uncertainties will be from experimental systematics including beam polarimetry (0.4\%) and $Q^2$ determination (0.2\%), assumed to be fully correlated among all bins, and radiative corrections (0.2\%) and event reconstruction (0.2\%), assumed to be fully uncorrelated in the present projection study.  The treatment of both radiative corrections and event reconstruction as uncorrelated is a simplifying assumption and will be studied in full in the actual data analysis.

\begin{figure}[!ht]
\begin{center}
\includegraphics[height=0.5\textwidth,angle=270]{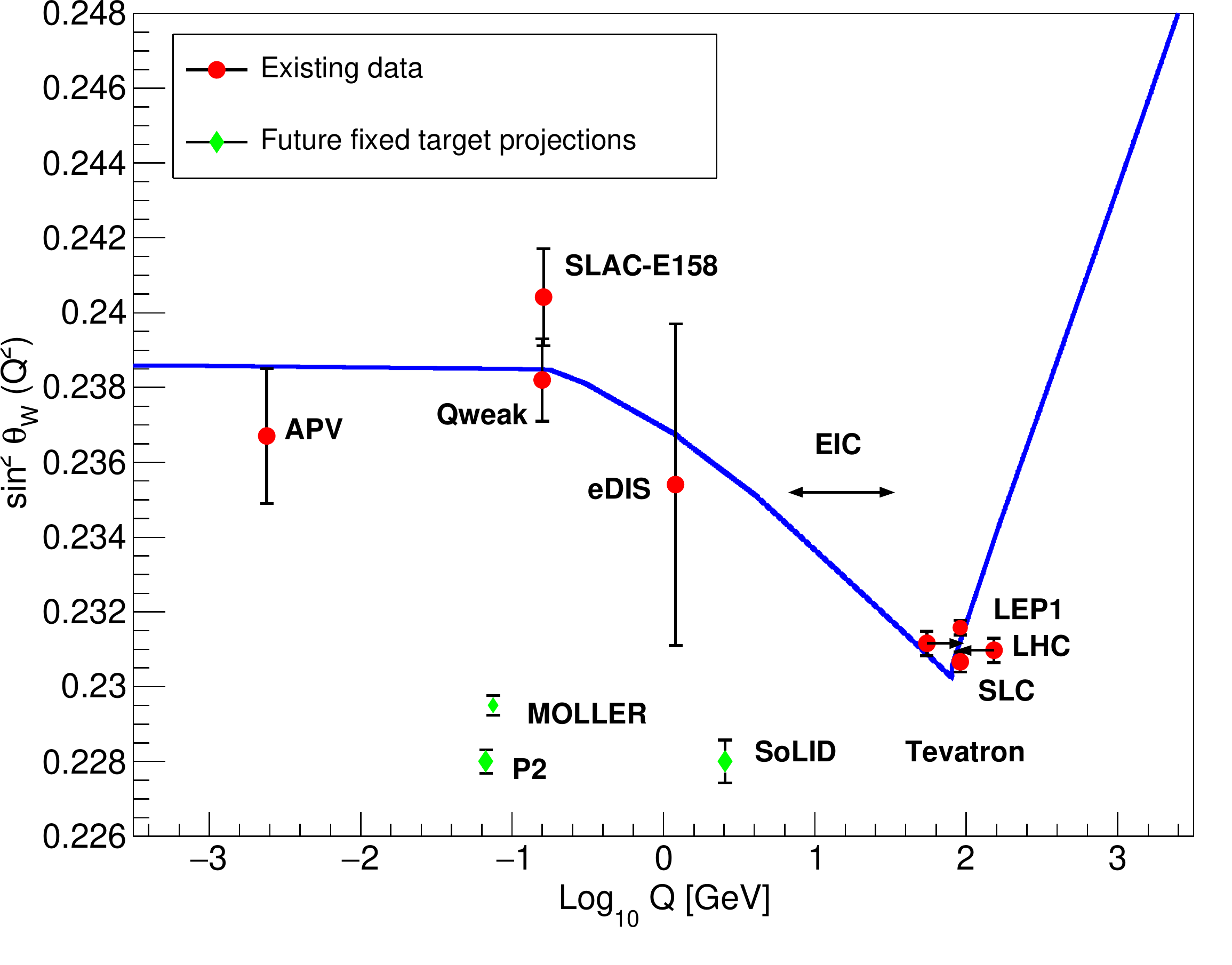}
\end{center}
\caption{Experimental determination of the weak mixing angle $\sin^2\theta_W$. Data points for Tevatron and LHC are shifted horizontally for clarity. 
}\label{fig:thetaW}
\end{figure}

Fitting projected $A_{PV}$ data using the function: 
\begin{eqnarray}
  A_{PV}^{\rm data} = A_{PV,(d)}^{\rm SM}\left(1+\frac{\beta_{\rm HT}}{(1-x)^3 Q^2}+\beta_{\rm CSV} x^2\right)~,\label{eq:Apvdis_fit}
\end{eqnarray}
where $A_{PV,(d)}^{\rm SM}$ is expressed in terms of $\sin^2\theta_W$ and accounting for all correlated and uncorrelated systematic effects, we arrive at the uncertainty projection shown in Fig.~\ref{fig:thetaW}. In Eq.~(\ref{eq:Apvdis_fit}), the use of the two $\beta$ parameters is to account for possible hadronic effects, to be discussed in Section~\ref{sec:pvdis_hadronic_physics}.

The SoLID deuteron PVDIS measurement, along with the upcoming MOLLER~\cite{Benesch:2014bas} at JLab and the P2 experiment~\cite{Becker:2018ggl} at the MESA facility in Mainz, will provide three new cornerstone measurements of the weak mixing angle $\sin^2\theta_W$ in the low to intermediate energy region. 
Regarding relevant BSM physics, one possible extension is for a dark boson ($Z_d$) that would induce an apparent deviation of $\sin^2\theta_W$ from the SM prediction at low $Q^2$. In this scenario, a comparison of all three experiments will help to determine the mass of the $Z_d$. Here, SoLID PVDIS is unique in that its $Q^2$ range will help to distinguish between $Z_d$ of light masses $50-200$~MeV~\cite{Davoudiasl:2014kua} and those of intermediate masses (10-35)~GeV~\cite{Davoudiasl:2015bua}. 
Another possible extension involves a heavy dark photon~\cite{Thomas:2022qhj}, whose parameter space can be constrained by PVES, Atomic PV, and the recent results on the $W$ mass~\cite{CDF:2022hxs,Thomas:2022gib}.

\begin{figure}[t]
\includegraphics[width=0.47\textwidth]{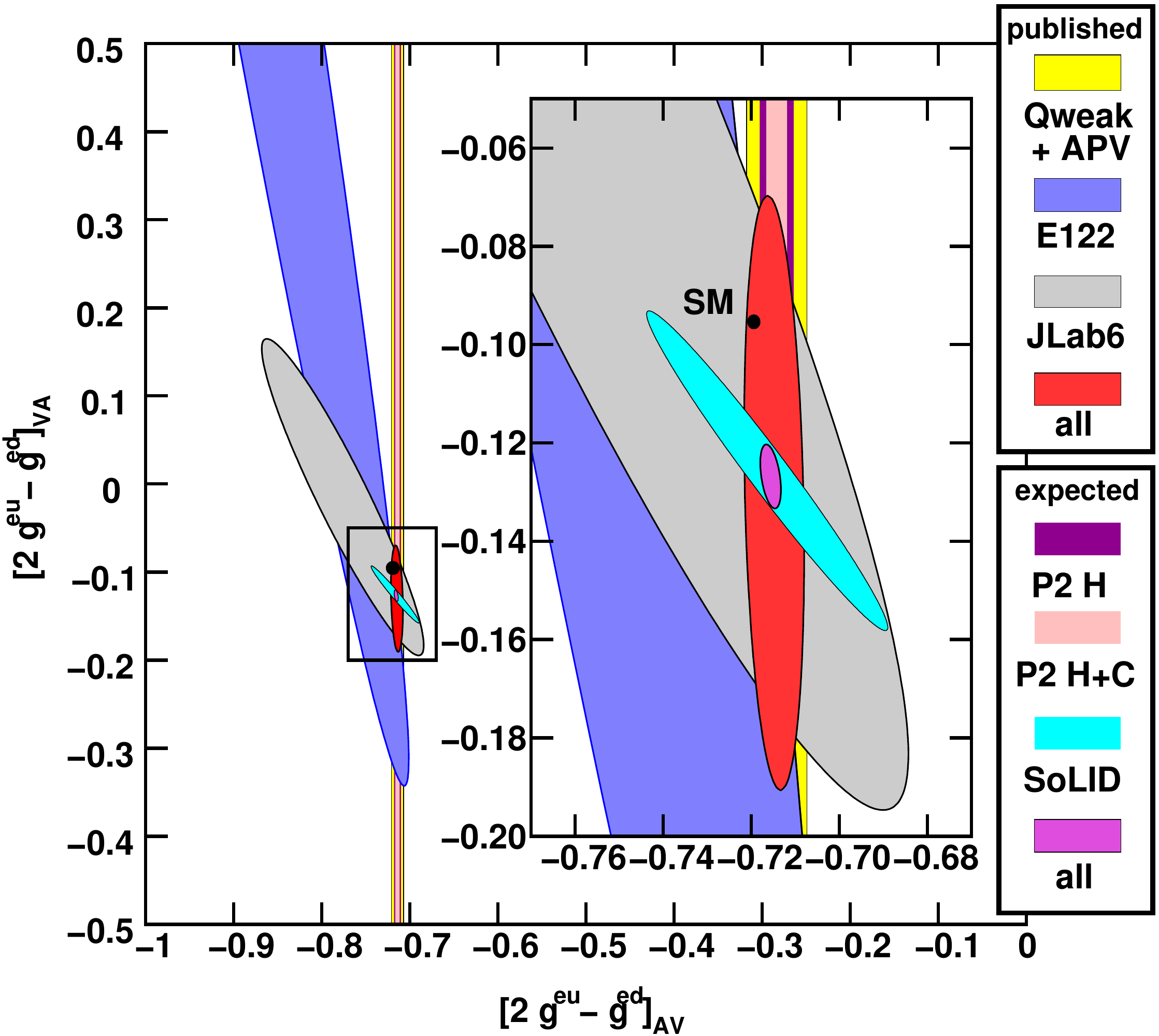}
\caption{Adapted from Ref.~\cite{Zheng:2021hcf}: Current experimental knowledge of the couplings $g_{VA}^{eq}$ (vertical axis). The latest world data constraint (red ellipse) is provided by combining the 6 GeV Qweak~\cite{Androic:2018kni} on $g_{AV}^{eq}$ (yellow vertical band) and the JLab 6 GeV PVDIS~\cite{Wang:2014bba,Wang:2014guo} experiments (grey ellipse). The SoLID projected result is shown as the cyan ellipse. Also shown are expected results from P2 (purple and pink vertical bands) and the combined projection using SoLID, P2, and all existing world data (magenta ellipse), centered at the current best fit values. 
}
\label{fig:c2}
\end{figure}

Furthermore, to fully explore BSM physics, one must study as many individual components of lepton-lepton or lepton-quark interactions as precisely as possible, in addition to the weak mixing angle. 
The upcoming MOLLER, P2, and the SoLID PVDIS deuteron measurements will provide precision measurements of the low-energy effective couplings $g_{VA}^{ee}$, $g_{VA}^{eq}$, and $g_{AV}^{eq}$, respectively.  For PVDIS, we do so by expressing $A_{PV,(d)}^{\rm SM}$ in Eq.~(\ref{eq:Apvdis_fit}) as functions of the electron-quark effective couplings and perform a simultaneous fit of the combinations $(2g_{AV}^{eu}-g_{AV}^{ed})$ and $(2g_{VA}^{eu}-g_{VA}^{ed})$, shown as the cyan-colored ellipse in Fig.~\ref{fig:c2}. The PVDIS projection can be further combined with that from P2 to provide a global fit, represented by the magenta-colored ellipse. 
Due to the small value of $g_{VA}^{eq}$'s in the SM, they could be particularly sensitive to BSM physics. One model that the $g_{VA}^{eq}$'s are sensitive to involves the leptophobic $Z'$s~\cite{Babu:1996vt}, corresponding to additional neutral gauge bosons ($Z'$) with negligible couplings to leptons, and thus would cause only sizable axial couplings to quarks while leaving the $g_{AV}^{eq}$ unaffected.

\subsubsection{BSM Reach of PVDIS with SoLID}
The potential of BSM searches can be generally characterized by the mass scale $\Lambda$, quantified as modifications of the SM Lagrangian by replacing
\begin{eqnarray}
\frac{G_F}{\sqrt{2}} g_{ij} \rightarrow \frac{G_F}{\sqrt{2}} g_{ij} + \eta_{ij}^{q}\frac{4\pi}{(\Lambda_{ij}^{q})^2}\ ,
\label{eq:ciqmodified}
\end{eqnarray}
where $ij=AV,VA$. We assume that the new physics is strongly coupled with a coupling $g$ given by $g^2 = 4\pi$, and $\eta_{ij}^{q}=\pm 1$ corresponds to cases where the new physics increases ($+1$, constructive) or decreases ($-1$, destructive) the couplings. Once combined with the expected results from the P2 experiment~\cite{Becker:2018ggl}, the 90\% C.L. mass limit that can be reached by the SoLID PVDIS deuteron measurement is 
\begin{eqnarray}
\Lambda_{VA
}^{eq} &=& g \sqrt{\frac{\sqrt{2}\sqrt{5}}{G_F 1.96\Delta \left(2g_{VA}^{eu}-g_{VA}^{ed}\right)}}=17.6~\mathrm{TeV}~,
\end{eqnarray}
where the $\sqrt{5}$ represents the ``best case scenario'' where BSM physics affects maximally the quark flavor combination being measured~\cite{Erler:2014fqa}. 
Such BSM limits are complementary to those from high energy facilities. As an example: the LHC Drell-Yan cross section data determine certain combinations of both the parity-violating and parity-conserving electron quark couplings, defined by the observable measured. As a result, the LHC constraints appear to have a flat direction in the BSM parameter space~\cite{Boughezal:2021kla}.
In this context, PVDIS observables are sensitive to different combinations of the couplings, 
resolving the ambiguity in the determination of BSM parameters. 

SoLID will undoubtedly push forward the EW/BSM physics study in the low to medium energy regime. On the other hand, a variety of challenges exist. First, one must carry out both electromagnetic and electroweak radiative corrections to high precision.  Significant progress has been made on this topic:  The event generator {\tt Djangoh}~\cite{Charchula:1994kf}, originally developed for HERA cross section analysis, has been adapted to fixed-target experiments and to nuclear targets. Modifications were made such that it can be used to calculate parity violating asymmetries to high precision, immune to the statistical limit of a Monte-Carlo program. 
While there is still detailed work to be done, we anticipate that the 0.2\% uncertainty projected on the radiative corrections can be reached. Such progress will also be useful for the similar program at the EIC.

\subsection{PVDIS Proton Measurement and Hadronic Physics Study }\label{sec:pvdis_hadronic_physics}
In Eq.~(\ref{eq:Apvdis_fit}), the use of the two $\beta$ parameters is to account for possible hadronic effects: $\beta_{HT}$ for 
higher twist (HT) and $\beta_{CSV}$ for charge symmetry violation (CSV)~\cite{Rodionov:1994cg} at the quark level, both expected to have distinct $x$ and $Q^2$ dependencies, especially at high $x$ values. 
The PVDIS deuteron measurement has the special property that most HT effects cancel in the asymmetry, and thus any sizable HT contribution will indicate the significance of quark-quark correlations~\cite{Mantry:2010ki}.
The CSV effect refers to the possibility that the up quark PDF in the proton and down quark PDF in the neutron are different. 
Together, these hadronic physics effects may support certain explanations of the apparent inconsistency of the NuTeV experiment~\cite{Zeller:2001hh} with the SM~\cite{Londergan:2003ij,Londergan:2009kj}. 

In addition to the deuteron measurement, PVDIS asymmetries on the proton target will allow us to determine PDF ratio $d(x)/u(x)$ at high $x$ based on the dependence of the structure functions in Eq.~(\ref{eq:Apvdis1}). 
The standard determination of the $d/u$ ratio relies on fully inclusive DIS on a proton target compared to a deuteron target.  In the large $x$ region, nuclear corrections in the deuteron target lead to large uncertainties in the $d/u$ ratio. 
However, they can be completely eliminated if the $d/u$ ratio is obtained from the proton target alone. 
For a proton target in the parton model and omitting sea quark distributions~\cite{Zheng:2021hcf}, the PVDIS asymmetry is given by: 
\begin{eqnarray}
  A_{PV,(p)} =\frac{3 G_F Q^2}{2\sqrt{2}\pi\alpha}\frac{(2g_{AV}^{eu}-\frac{d}{u}g_{AV}^{ed})+Y[{2g_{VA}^{eu}-\frac{d}{u}g_{VA}^{ed}}]}
  {{4+\frac{d}{u}}}~.\nonumber\\
  \label{eq:Apvdis_proton}
\end{eqnarray}
which provides a direct access to $d/u$ without any nuclear physics effects. 
\begin{figure}[!h]
    \centering
    \includegraphics[width=0.5\textwidth]{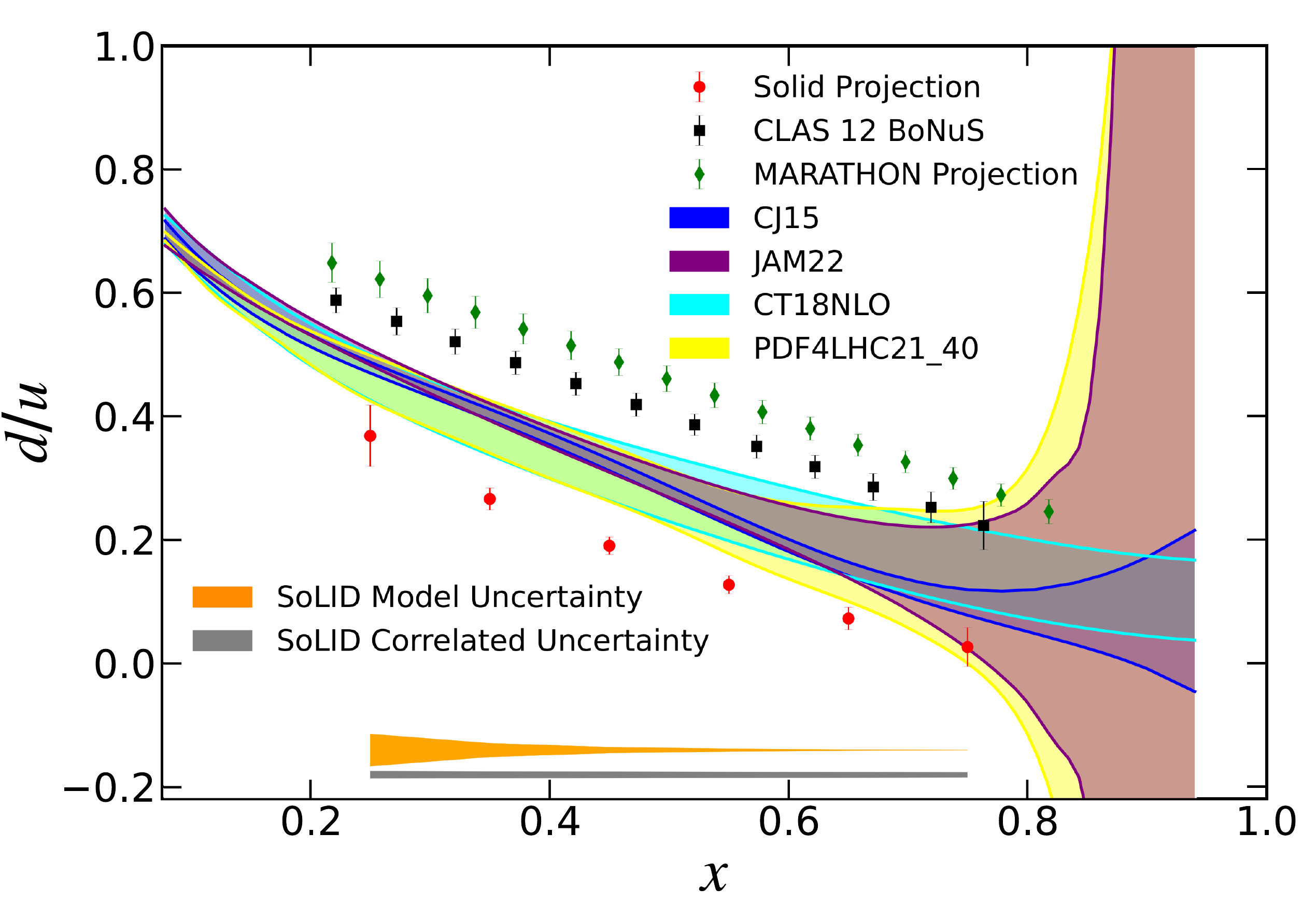}
    \caption{Projected results on the PDF ratio $d/u$ from the PVDIS proton measurement (red points) compared with the current world fits from a number of PDF groups and their uncertainties. The error bars of the SoLID projection indicate the uncertainty in the extracted $d/u$ from statistical uncertainties, while uncorrelated systematic uncertainties are negligible. The two horizontal shaded bands show the uncertainty in $d/u$ due to omitting sea quarks in Eq.~(\ref{eq:Apvdis_proton}) (model uncertainty, orange-colored band), and from correlated systematic uncertainties (dark grey band). Projections on MARATHON and CLAS12 BoNuS are from their respective experimental proposals~\cite{JLabPR:MARATHON,JLabPR:BONUS12}.
    }
    \label{fig:pvdis_du} 
\end{figure}
In this way, SoLID is complementary to the recent MARATHON experiment at low $Q^2$ as well as $W$ production data from Fermilab at high $Q^2$~\cite{Accardi:2016qay}. 
The MARATHON data have been interpreted in different ways~\cite{JeffersonLabHallATritium:2021usd,Cocuzza:2021rfn,Cui:2021gzg}, highlighting the importance of the PVDIS proton measurement that will provide information both directly on $d/u$ and on nuclear physics models relevant for future inclusive scattering measurement involving the deuteron or heavier nuclear targets. Using 90~days of 50~$\mu$A electron beam with 85\% polarization incident on a 40-cm long liquid hydrogen target, the projection on $d/u$ is shown in Fig.~\ref{fig:pvdis_du}. 

SoLID in its PVDIS configuration can be used to study more hadronic physics topics. For example, data on PV asymmetry for nucleon resonances will be collected simultaneously with PVDIS running. Resonance $A_{PV}$ data will help to test how well we model the nucleon, explore quark-hadron duality in the electroweak sector, and help constrain inputs for radiative corrections of PVDIS. Measurements of the beam-normal single-spin asymmetry $A_n$ provides information on two-photon-exchange physics, see Section~\ref{sec:bnssa}. 

\subsection{Flavor dependence of the EMC effect}

Just as PVDIS can be used to study the $d/u$ ratio in the valence quark region when measured for the proton, it can also be used to study the flavor structure of PDFs if a nuclear target is used.  For an isoscalar target with mass number $A$, where charge symmetry provides the expectation $u_A(x)=d_A(x)$, the PVDIS asymmetry is independent of the EMC effect as long as all PDFs are modified in the same way.  In an isoscalar nucleus such a deuteron or $^{40}$Ca, it can be used to look for charge-symmetry violation, although the expectation is that this would yield a small effect (as discussed in section~\ref{sec:pvdis_hadronic_physics}): While the EMC effect modifies the PDFs in these nuclei, it is assumed that the modification of the up- and down-quarks is identical, and as such, will cancel exactly in the ratio of $F_1^{\gamma Z}/F_1^\gamma$ and $F_3^{\gamma Z}/F_1^\gamma$, making the asymmetry completely insensitive to the conventional (flavor-independent) EMC effect. 

If the EMC effect yields different nuclear modification for the up-quark and down-quark PDFs, this modifies $A_{PV}$ making it sensitive to the flavor dependence of the EMC effect. In non-isoscalar nuclei, the flavor dependence that arises from the difference in Fermi smearing for protons and neutrons is expected to be extremely small, except for $x > 0.7$-$0.8$, as conventional smearing and binding effects are a small part of the EMC effect~\cite{Miller:2001tg, Smith:2002ci}. Over the past decade there have been several indications that the EMC effect may have a significant flavor dependence in non-isoscalar nuclei, as seen in calculations of the EMC effect using different coupling for up- and down-quarks to the QCD scalar and vector potentials~\cite{Cloet:2012td}, and PDF analyses~\cite{Schienbein:2009kk, Kovarik:2010uv} which explains the tension between neutrino charged-current scattering and DIS plus Drell-Yan data by allowing for a flavor-dependent EMC effect. In addition, a range of models~\cite{Arrington:2015wja} inspired by the observed correlation between the EMC effect and short-range correlations~\cite{Weinstein:2010rt, Arrington:2012ax} also predict a flavor dependence of the EMC effect associated with the isospin structure of short-distance or high-momentum pairs of nucleons.  In all cases, these models, calculations, and fits predict an increase in the EMC effect for protons inside of neutron-rich nuclei.

An experiment has been conditionally approved by JLab PAC50~\cite{JLabPR:PVEMC} to measure PVDIS on $^{48}$Ca. The experiment, called PVEMC, uses the exact same configuration as the PVDIS measurements on hydrogen and deuterium, except with a 2.4-g/cm$^2$ $^{48}$Ca target. The $^{48}$Ca was chosen to provide a nucleus with a significant EMC effect and a large neutron excess, while avoiding very high-$Z$ material which would yield significantly more radiation for the same target thickness.  
The kinematic coverage is similar to that shown in Fig.~\ref{fig:Apv_ed}, though the data will be binned only in $x$ and with a statistical precision at about 1\% or less within each $x$ bin accumulated with 68 days of data taking. The experimental systematic uncertainties are expected to be also similar to the deuteron measurement. 

From the measured $A_{PV}$, we can extract the dominant $a_1$ contribution (Eq.~\ref{eq:Apvdis1}) which is sensitive to the $d/u$ ratio of the nuclear structure function. This sensitivity is clear if one evaluates $a_1$ under the assumption that only light quark distributions $u_A(x)$ and $d_A(x)$ contribute and express $a_1$ as: 
\begin{equation}
    a_1 \simeq \frac{9}{5} - 4\sin^2\theta_W - \frac{12}{25} \frac{u_A^+ - d_A^+}{u_A^+ + d_A^+}
\end{equation}
with the convention that $q^\pm = q(x) \pm \bar{q}(x)$. This expression is a good approximation at large $x$, where the sea quarks do not contribute significantly, and shows that the PVDIS asymmetries are directly sensitive to flavor dependence of the EMC effect that modifies $u^+_A$ and $d^+_A$ differently.

\begin{figure}[htb]
\includegraphics[angle=90,width=0.45\textwidth,trim=20mm 25mm 20mm 30mm,clip]{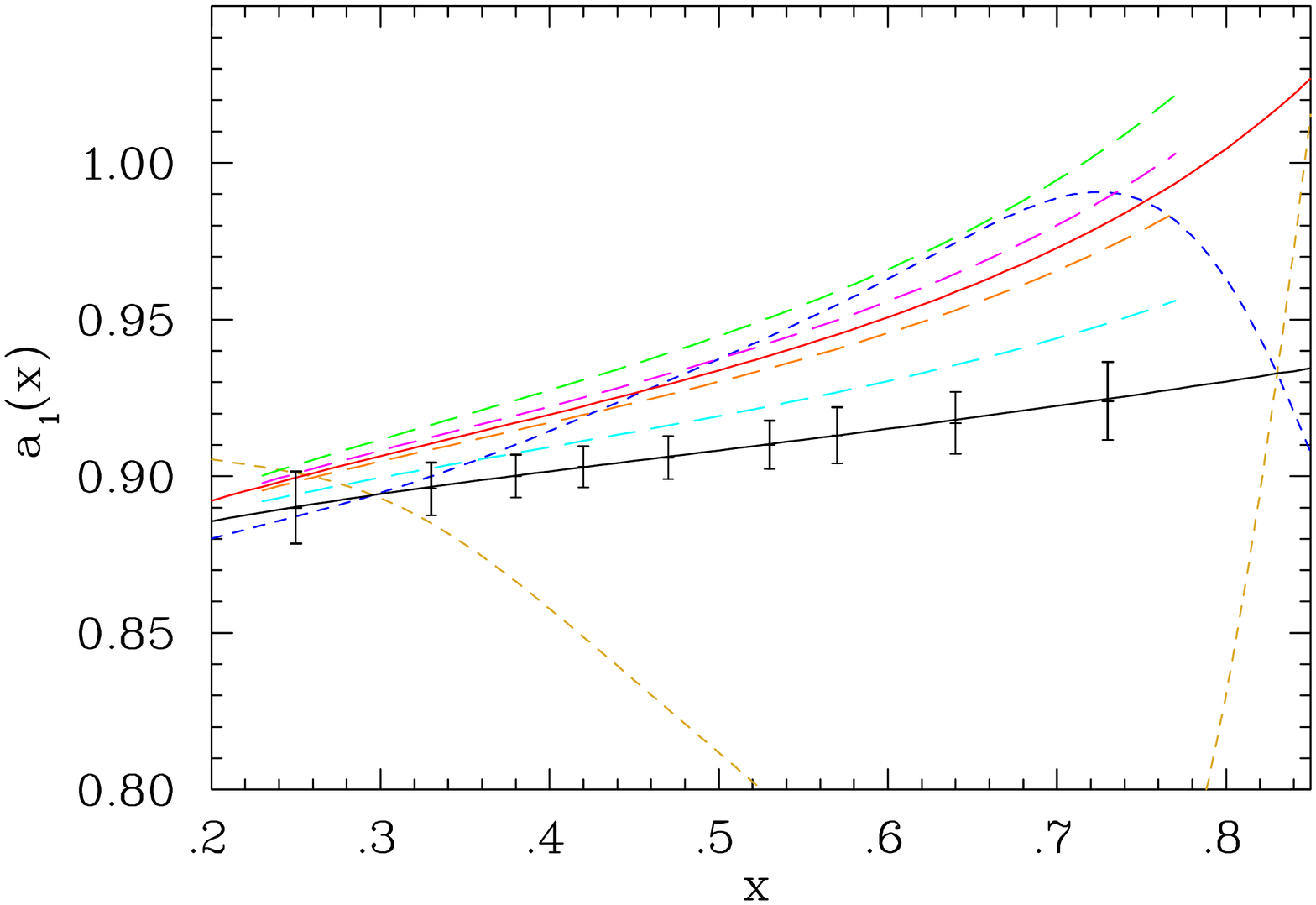}
\caption{
Projections for the extracted $a_1(x)$ for the PVEMC proposal~\cite{JLabPR:PVEMC} (black points), including statistical, systematic, and normalization (0.4\%) uncertainties. The black curve represents the prediction for the flavor-independent EMC effect, the red curve is the CBT model~\cite{Cloet:2012td}, the long-dashed curves represent the predictions from simple scaling models mentioned above~\cite{Arrington:2015wja}, and the short-dashed curves represent extreme cases where the EMC effect is caused entirely by up-quark (blue) or down-quark (brown) modification.}
\label{fig:pvemc-projection}
\end{figure}

Figure~\ref{fig:pvemc-projection} shows the projected uncertainties for the proposed measurement, including statistical and systematic uncertainties, as well as the estimated uncertainty on the baseline prediction in the absence of a flavor-dependent EMC effect. The points are shown on the flavor-independent prediction, and the various curves represent projections based on calculations or simple models of the EMC effect, as described in the caption. The projected results give $7\sigma$ sensitivity to the CBT model prediction~\cite{Cloet:2012td} and $>$3$\sigma$ 
sensitivity to all but the smallest effect among the models evaluated. Thus, the data will provide a search for a non-zero flavor dependence in the EMC effect, be able to differentiate between `large' and `small' effects, and set stringent limits on such a flavor-dependent effect should the results be consistent with a flavor-independent effect.

The presence and size of a flavor-dependent modification of the nuclear PDFs has wide-ranging implications.  First, the size of the flavor dependence is sensitive to the underlying physics behind the EMC effect. In addition, observing a flavor-dependent EMC effect would imply that the PDFs used for non-isoscalar nuclei are incorrect, modifying the expectation for high energy lepton-nucleus scattering such as $e-A$ or for $A-A$ collisions. This could be significant for heavy nuclei which have a large neutron excess, as well as for measurements utilizing polarized $^3$He as an effective neutron target.

\section{Near-Threshold \texorpdfstring{$J/\psi$}{J/psi} Production 
}\label{sec:jpsi}
The proton's fundamental properties, like its electric charge, mass, and spin, are the hallmarks of our knowledge of the visible universe. More than 60 years ago, through a novel experimental investigation of its charge using electron scattering, we learned that the proton is not a point-like particle but has a finite volume with primary constituents. In the following 20 years, these constituents dubbed ``partons" were identified through electron and muon deep inelastic scattering studies as being the quarks and gluons we know today. In tandem, the theory of strong interactions, a non-abelian field theory known as Quantum Chromodynamics (QCD)~\cite{Shifman:1978bx,Shifman:1978by,Shifman:1978zn} was developed and brought our understanding and knowledge of the proton's interior to a new level. In practice, the theory was intractable analytically but could be approximated and tested such as in DIS experiments. Our naive three valence quark picture providing the total spin 1/2 of the proton was challenged, and experimental studies in the last 40 years have described the proton's spin in terms of its partonic structure front and center. Today we know that both constituents quarks gluons and their angular momentum play a role in providing the proton's total spin 1/2. Furthermore, the spin of the proton provided a laboratory to test and better understand QCD with various controlled approximations in corners of its full phase space.

Many studies have focused on the proton electric charge and spin. The proton mass, however, has received less attention. Although the proton's total mass is measured and calculated in QCD with high precision\cite{Durr:2008zz,Borsanyi:2014jba}, its origin, gravitational density distribution, among its partonic constituents and the trace anomaly are yet to be investigated and fully understood through direct measurements. A few facts are crucial to know why further studies are needed to get a deeper insight into the constituents' role in providing the proton's total mass. First, it is well known that the Higgs mechanism provides for the mass of the quark constituents and breaks chiral symmetry in the QCD Hamiltonian. However, this is only a small fraction of the proton's total mass, about 10\%. Second, we also know that scale symmetry is broken in QCD, and this violation is responsible for most of the proton mass. This is reflected by contributions from the gluons' energy, self-interactions, and interactions with the moving quarks.

Recent measurements at JLab~\cite{Ali:2019lzf,Duran:2022xag}, motivated by the LHCb charm pentaquarks discovery\cite{Aaij:2015tga,Aaij:2019vzc}, have given new impetus to using the $J/\psi$ particle, a small color dipole, not only to search for these pentaquarks but also to probe the gluonic gravitational mass density in the proton and determine the mass radius and scalar radius. These two radii encode information contained in the gluonic gravitational form factors (GFFs) known as $A_g(k)$ and $C_g(k)$ form factors, where $A_g(k)$ is the response to a graviton-like tensor glueball (2$^{++}$) probe and $C_g(k)$ is a response to a scalar (0$^{++}$) probe. Because the production of the $J/\psi$ particle at JLab  occurs at photon energies near threshold, the region of the measurement is highly non-perturbative. Different theoretical approaches with various approximations have been explored in this non-perturbative region of production to extract these gravitational form factors~\cite{Kharzeev:2021qkd,Hatta:2018ina,Hatta:2019lxo,Mamo:2022eui,Mamo:2019mka,Guo:2021ibg,Sun:2021gmi}. Recent lattice QCD calculations~\cite{Pefkou:2021fni,Shanahan:2018pib} of these gravitational form factors, albeit at a large pion mass of 450 MeV, will enable comparisons with the various extraction methods of the GFFs. 

\lrpbf{Close to threshold, the smallness of the electro- and photoproduction cross sections requires a dedicated experiment with a well-designed detector to exploit the full potential of the beam luminosity and capture the full phase space of this process in a measurement of key observables. SoLID provides all the necessary tools to realize the highest statistics exclusive measurements of $J/\psi$ through both the $e^+e^-$ and $\mu^+\mu^-$ channels while cross checking these two complementary channels and controlling the systematic errors.}

\subsection{The SoLID \texorpdfstring{$J/\psi$}{J/psi} Experiment}

\begin{figure}[ht]
    \centering
    \includegraphics[width=0.45\textwidth]{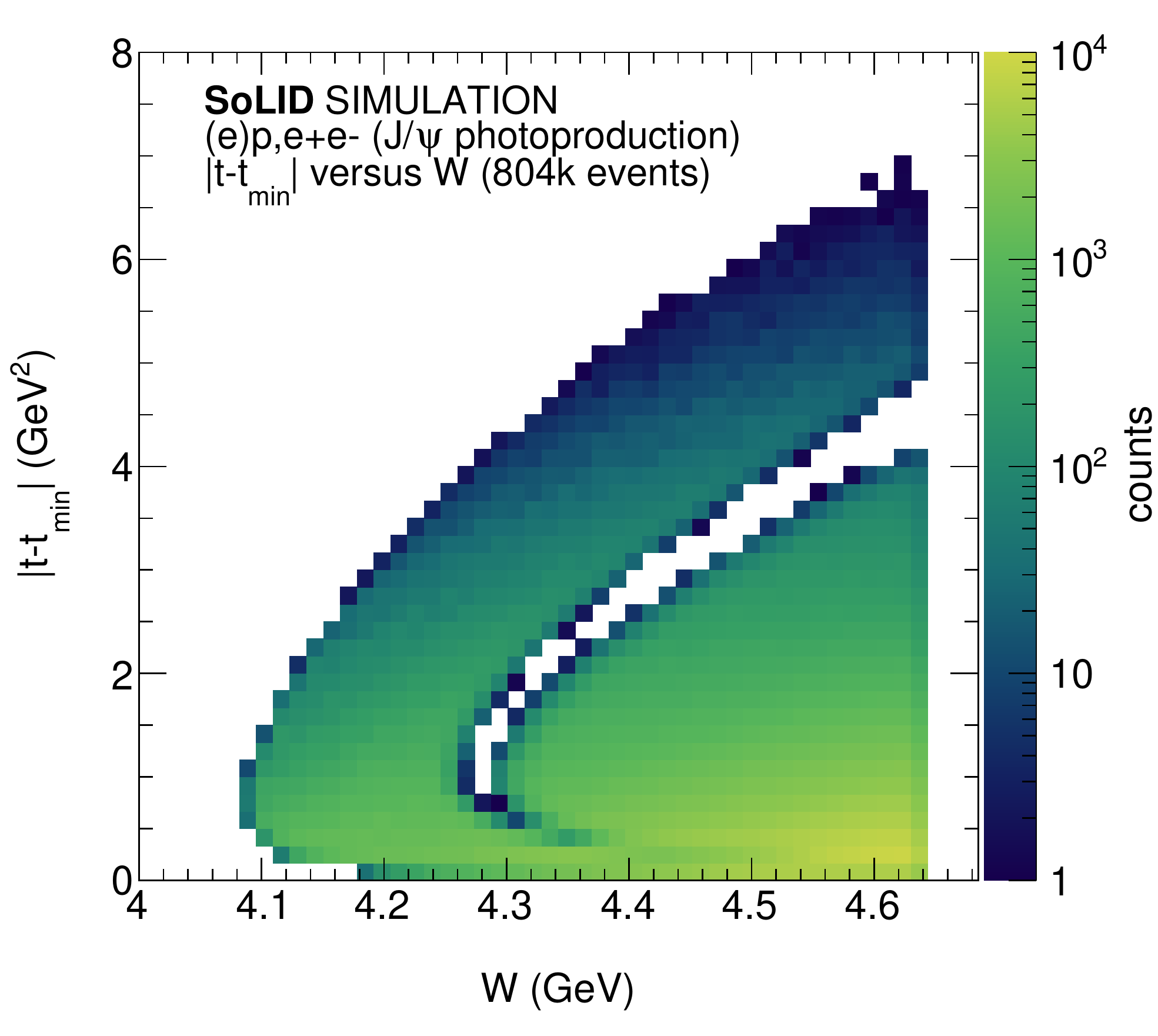}\\
    \includegraphics[width=0.45\textwidth]{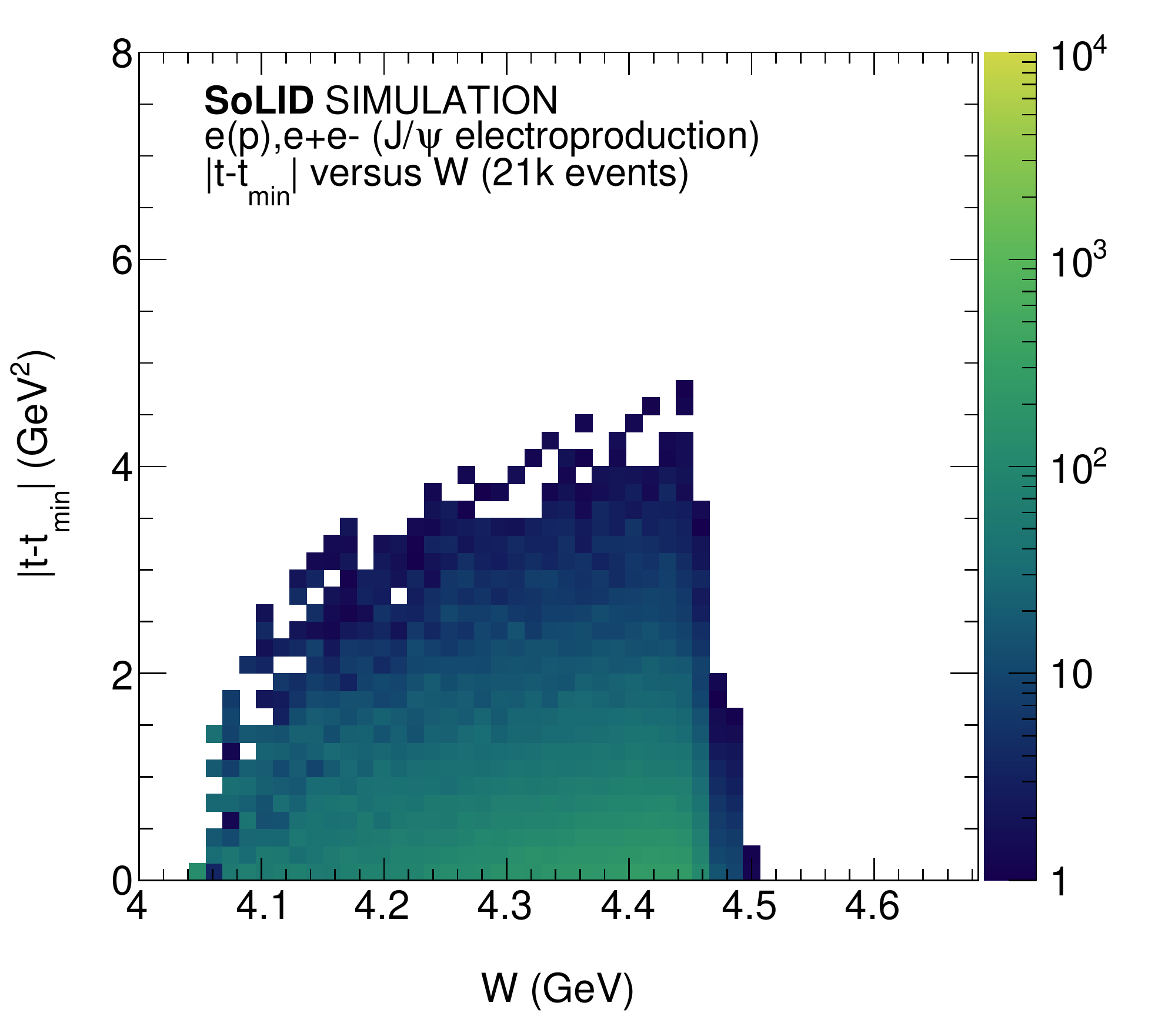}
    \caption{Mandelstam variable $|t-t_\text{min}|$ versus the invariant mass of the final state $W$ for exclusive photo-(top) and electroproduction (bottom) of $J/\psi$ near threshold. The high statistics of the photoproduction channel allow for a precise measurement of the $t$-dependence at larger values of $t$, important for constraining gravitational form factors. The electroproduction measurement complements the photoproduction measurement through improved acceptance near threshold.}
    \label{fig:jpsi:phasespace}
\end{figure}

The detector setup for SoLID-$J/\psi$~\cite{JLabPR:jpsi_solid} is similar to SoLID-SIDIS, except for the unpolarized liquid hydrogen target.
The nominal luminosity for SoLID-$J/\psi$ is 50 days at $10^{37}$~{cm}$^{-2}${s}$^{-1}$.
In Fig.~\ref{fig:jpsi:phasespace} we show the kinematic phase space reachable for the electroproduction and photoproduction channels for the nominal luminosity. 

\begin{figure}[ht]
    \includegraphics[width=0.45\textwidth]{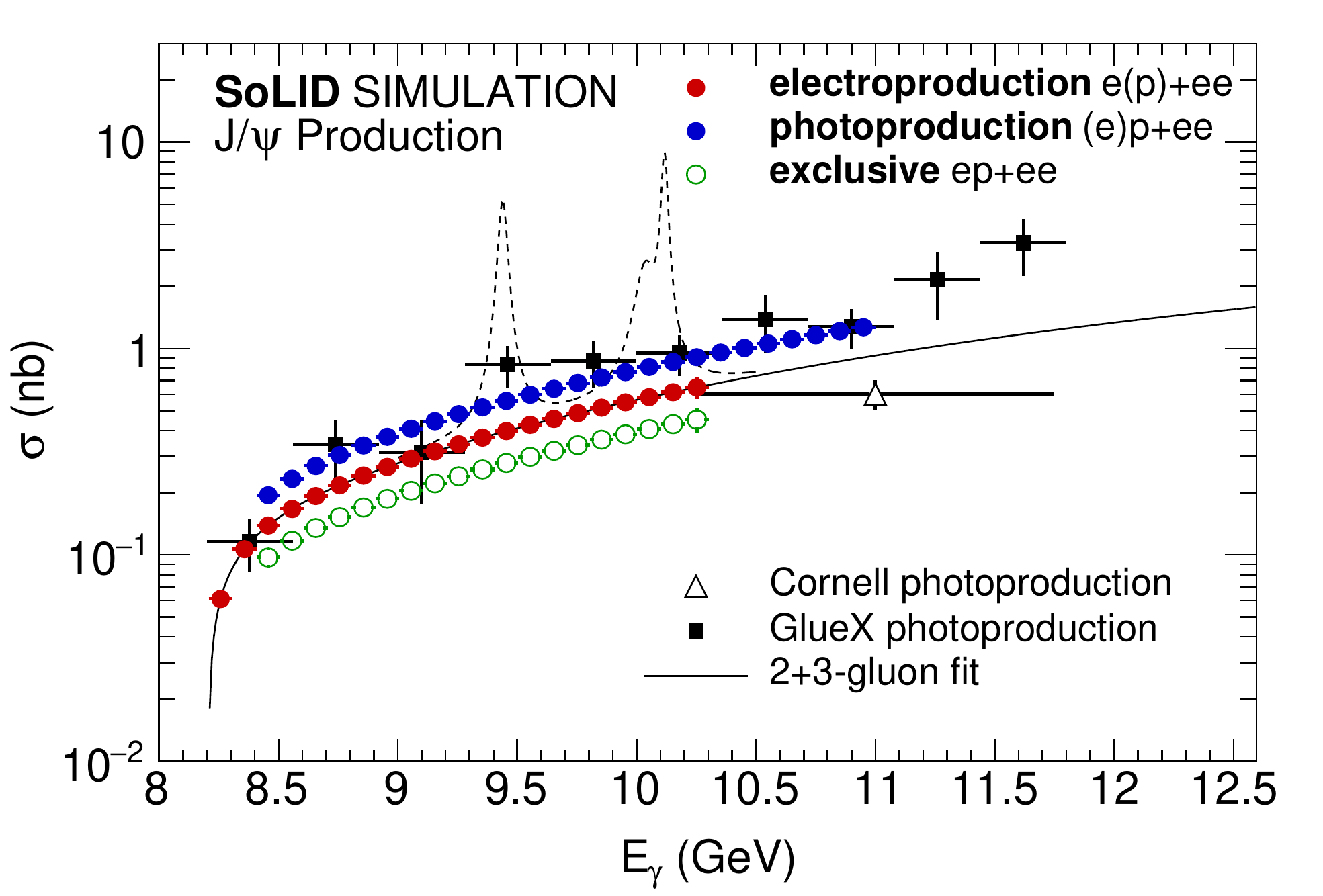}
    \caption{Projected 1-D $J/\psi$ cross section results as a function of photon energy $E_\gamma$ compared with the available world data. The blue disks show the photoproduction results, while the red disks show the electroproduction results, and the green circles show the results for a fully exclusive electroproduction measurement. Each of the measurements on this figure has a corresponding high-precision measurement of the $t$-dependent differential cross section.}
    \label{fig:jpsi:1d}
\end{figure}
\begin{figure}[ht]
    \includegraphics[width=0.45\textwidth]{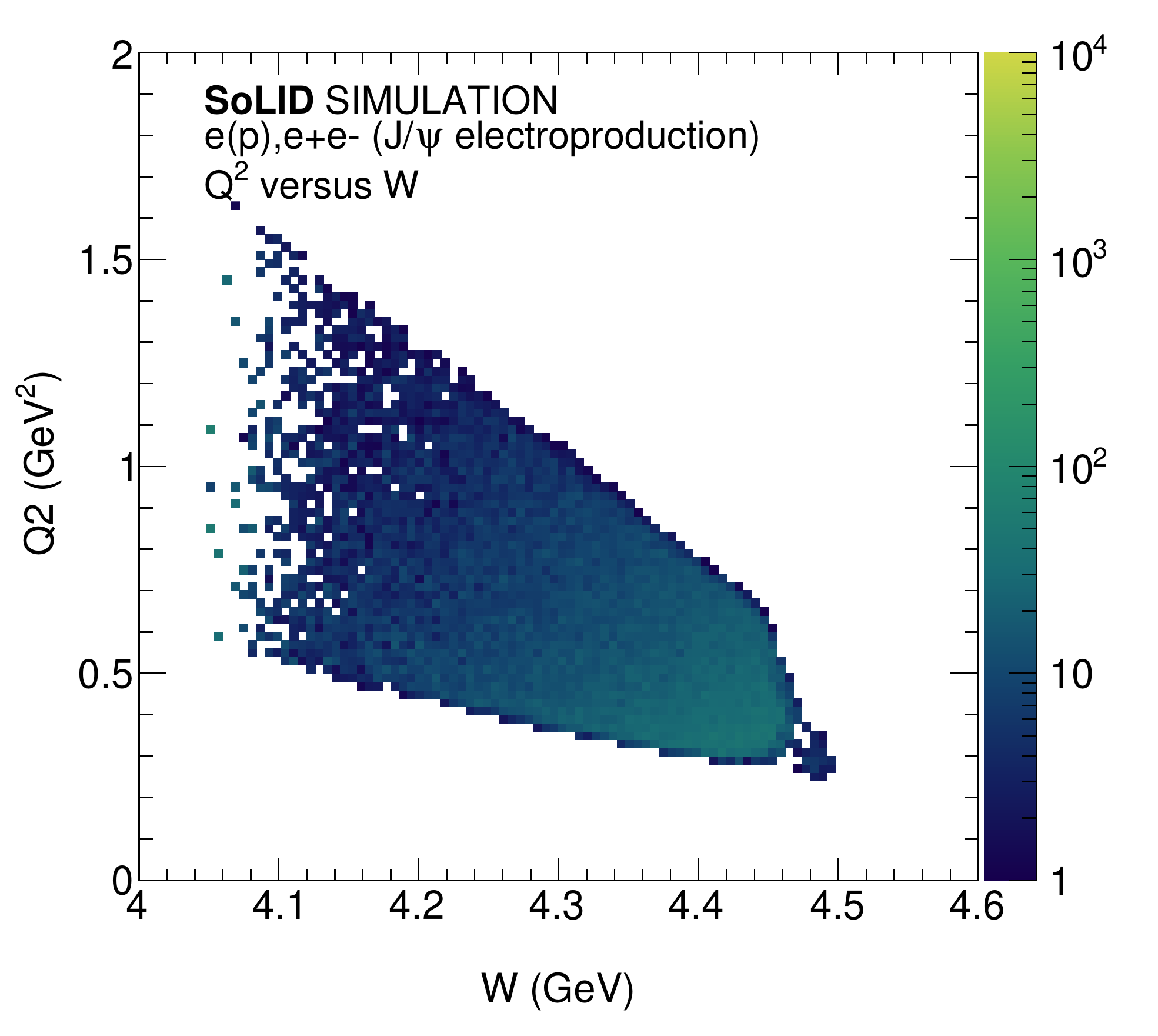}
    \caption{Photon virtuality $Q^2$ versus the invariant mass of the final state $W$ for exclusive and electroproduction (bottom) of $J/\psi$ near threshold. At threshold, there is a modest lever arm in $Q^2$, with an average virtuality of about 1 GeV$^2$.}
    \label{fig:jpsi:leverarm}
\end{figure}

The photoproduction channel receives approximately equal contributions from quasi-real electroproduction events and direct photoproduction events due to bremsstrahlung in the extended target. The photoproduction channel maximizes the statistical impact the SoLID-$J/\psi$ experiment can achieve.
We measure these events by requiring a coincidence between the $J/\psi$ decay electron-positron pair and the recoil proton.

To measure the electroproduction events, we measure the scattered electron in coincidence with the $J/\psi$ decay electron-positron pair. For a subset of events, we also detect the recoil proton for a full exclusive measurement. This redundant measurement is important for understanding the physics and detector backgrounds necessary to precisely determine the absolute cross section.

The projected 1D cross section results for the nominal luminosity is shown in Fig.~\ref{fig:jpsi:1d}.
The photoproduction and electroproduction channels are truly complementary to each other: the photoproduction channel has superior statistics and $t$-reach at higher $W$, while electroproduction has superior reach in the region very close to the threshold.
The relation between the photon virtuality $Q^2$ and $W$ are shown in Fig.~\ref{fig:jpsi:leverarm}. The average $Q^2$ at threshold is about 1 GeV$^2$, dropping as a function of $W$. Combining the electroproduction results with the photoproduction results yields a modest but important lever arm in $Q^2$.

\subsection{Gluonic Gravitational Form Factors and SoLID}
\begin{figure}[ht]
    \includegraphics[width=0.23\textwidth]{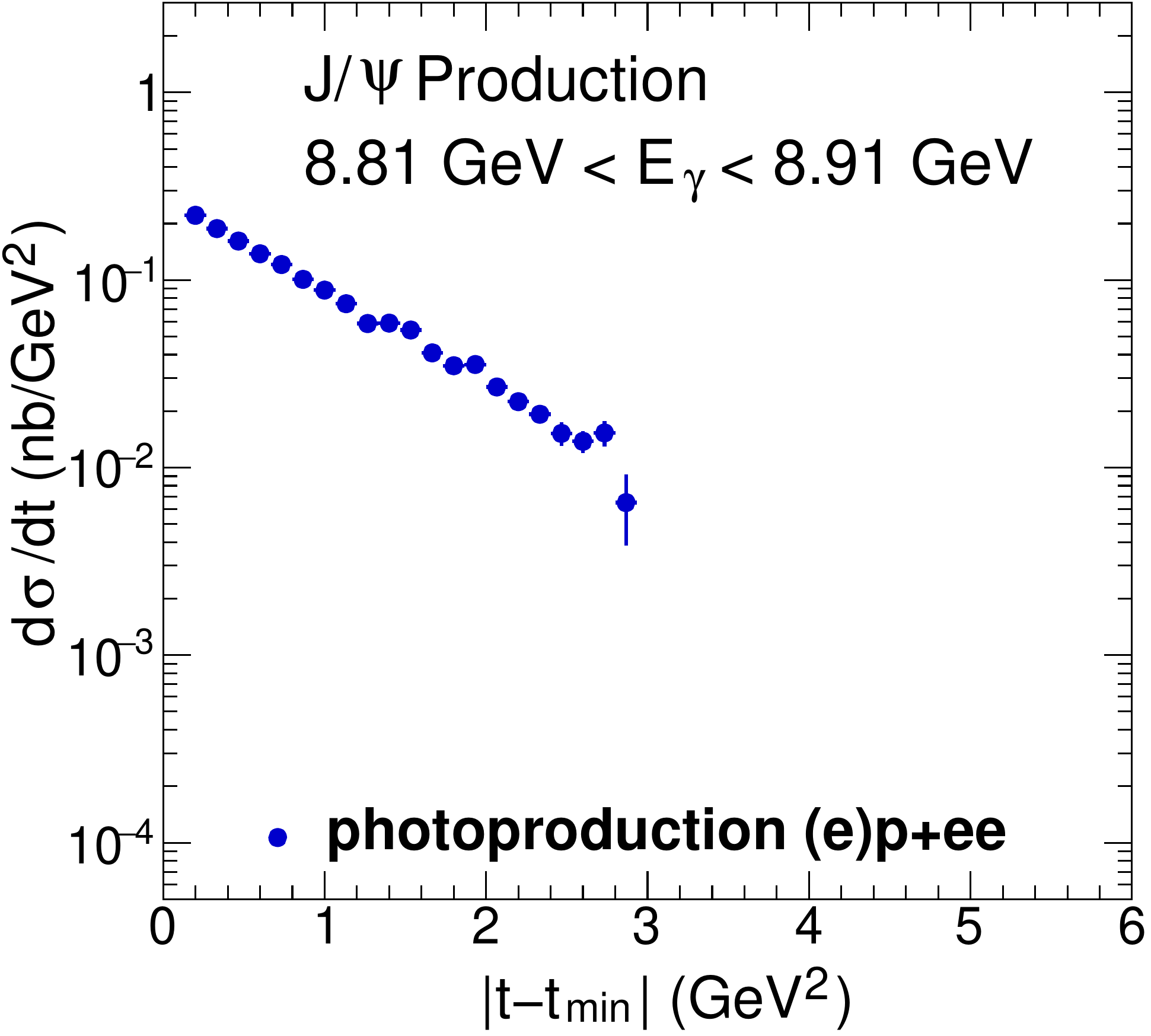}
    \includegraphics[width=0.23\textwidth]{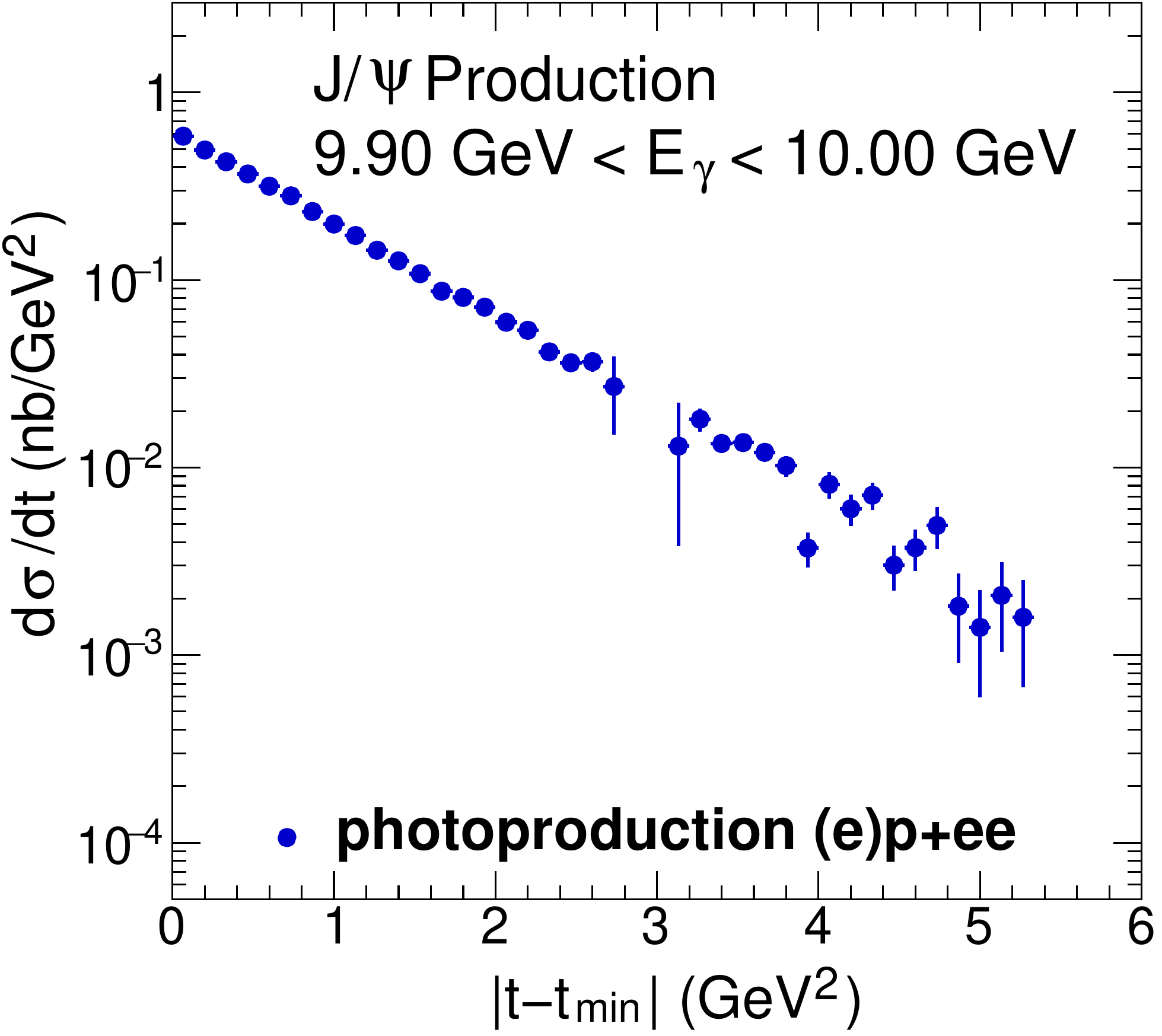}\\
    \includegraphics[width=0.23\textwidth]{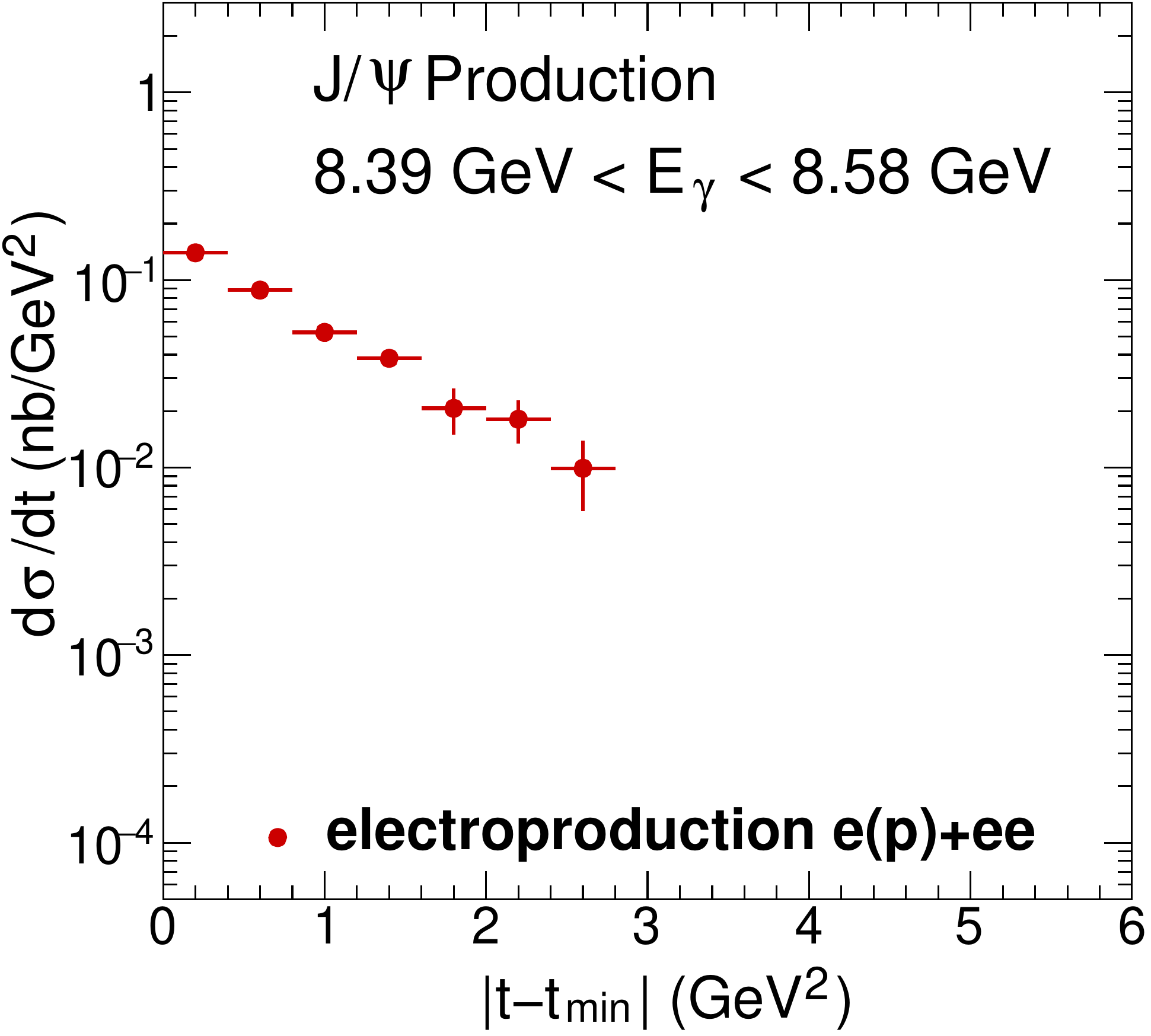}
    \includegraphics[width=0.23\textwidth]{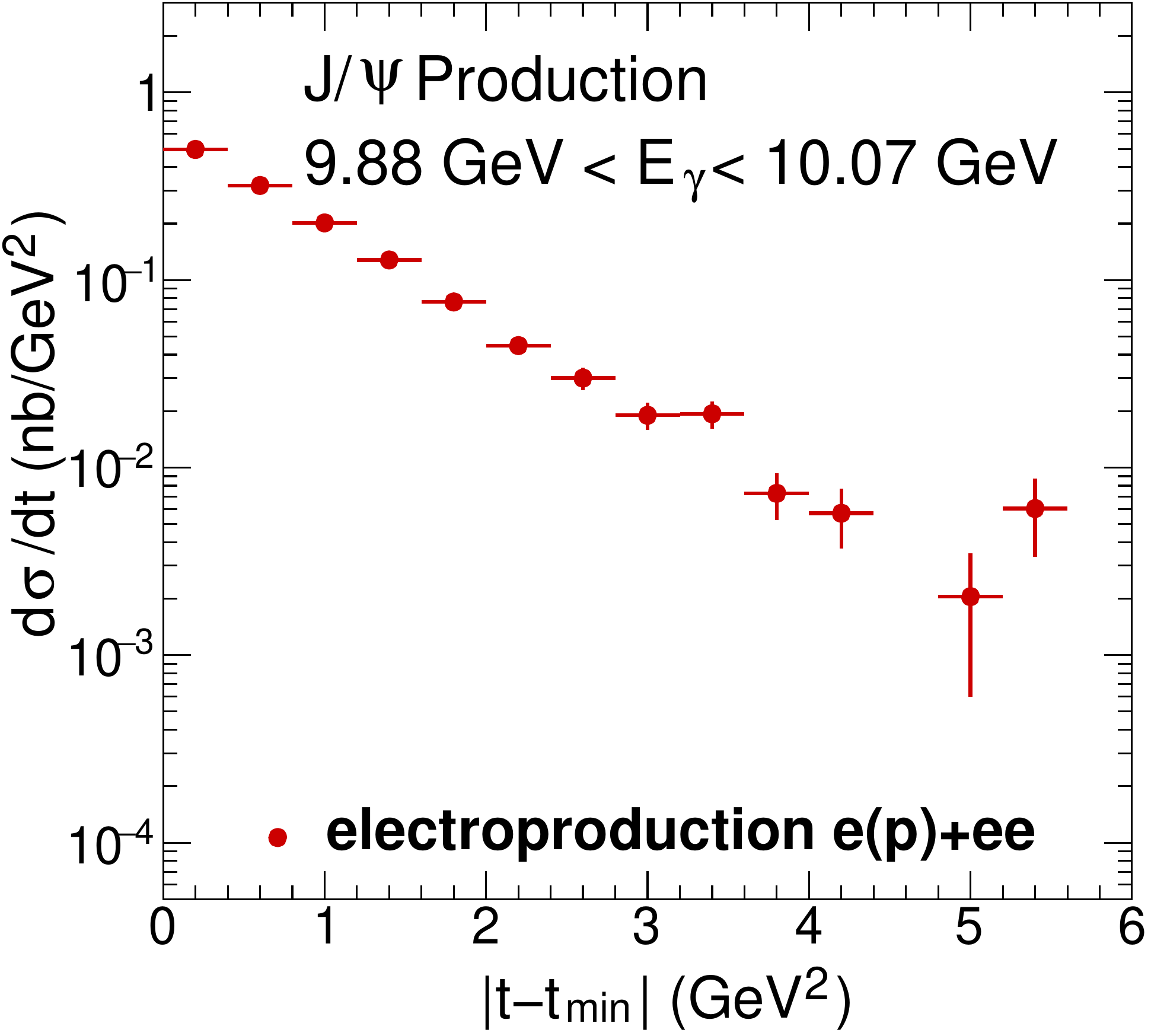}\\
    \caption{Top row: The projected differential cross section for a photoproduction bin at low (left) and high (right) photon energy from Fig.~\ref{fig:jpsi:1d}, assuming the nominal luminosity for SoLID-$J/\psi$. Bottom row: Same for two electroproduction bins. Precise measurements of these $t$-dependence over the full near-threshold phase space will hold the key to constrain the GFFs. }
    \label{fig:jpsi:2d}
\end{figure}

\lrpbf{
The $t$-dependent differential cross sections measurements that can be achieved by SoLID are shown in Fig.~\ref{fig:jpsi:2d}. 
The process to determine gluonic GFFs from the near-threshold $J/\psi$ differential cross section is currently under active discussion. One common theme to all proposed approaches\cite{Kharzeev:2021qkd,Hatta:2018ina,Hatta:2019lxo,Mamo:2022eui,Mamo:2019mka,Guo:2021ibg,Sun:2021gmi} is the need to precisely measure the $J/\psi$ differential cross section at larger values of $t$ as a function of the photon energy $E_\gamma$. A precise determination of the cross section at larger values in $t$ will help constrain extrapolation uncertainties, while enabling theoretical approaches that depend on a factorization at larger values of $t$. This measurement can only be accomplished with SoLID, due to the unique combination of large luminosity and large acceptance for this process. 
}

\subsection{Other Quarkonium Production Experiments at JLab and EIC} 
The increased profile of the physics topics that can be studied through near-threshold quarkonium production has spurred many experimental efforts at JLab and is an important component of the EIC scientific program~\cite{NASRep:2018}. The first 1-D and 2-D $J/\psi$ cross section results near threshold have been published by respectively GlueX and the Hall C $J/\psi$-007 experiment. In the next years, GlueX and CLAS12 will precisely measure the differential $J/\psi$ cross section at lower values of~$t$. 
\lrpbf{
SoLID-$J/\psi$ will fulfil a unique role within the Jefferson Lab program for near-threshold $J/\psi$ production, by precisely measuring the differential cross section at larger values of $t$, and by enabling a precise measurement of near-threshold electroproduction. The JLab $J/\psi$ program is complementary with the near-threshold~$\Upsilon$ program at the EIC.
}
\section{Generalized Parton Distribution Program}\label{sec:gpd}
Generalized parton distributions (GPDs) are a theoretical tool, developed in the late 90s, which offer correlation information between the transverse location and the longitudinal momentum of partons in the nucleon. At leading twist, there are four chiral-odd GPDs ($H$, $\Tilde{H}$, $E$, $\Tilde{E}$) and four chiral-even GPDs ($H_T$, $\Tilde{H}_T$, $E_T$, $\Tilde{E}_T$). Each GPD is a function of $x$, $\xi$ and $t$, where $x$ denotes the average light-cone momentum fraction of the quark, $\xi \approx x_B / (2 - x_B)$ is the skewness representing the longitudinal momentum fraction transferred to the nucleon, and $t$ represents the total square momentum transferred to the nucleon. GPDs also depend on $Q^{2}$, which is usually dropped out from the expressions since the $Q^2$-variation follows the QCD evolution equations. 
GPDs provide a link between electromagnetic form factors and parton distributions~\cite{Diehl:2003ny,Belitsky:2005qn,Guidal:2013rya} and can further access the contribution of the orbital angular momentum of quarks (and gluons) to the nucleon spin through the Ji's sum rule~\cite{Ji:1996ek},
\begin{eqnarray}
  J^{q} & = & \frac{1}{2}\Delta\Sigma^{q}+L^{q} \nonumber \\
  & = & \frac{1}{2}\int_{-1}^{+1}dx \, x [H^{q}(x,\xi,0)+E^{q}(x,\xi,0)],
\end{eqnarray}
where $\Delta\Sigma^{q}$ is the quark spin contribution that has been measured in polarized deep inelastic scattering, and $L^q$ is the quark orbital angular momentum contribution to the nucleon spin. Note that the sum rule also applies to the gluon GPDs. Hence, Ji's sum rule provides an experimental way to decompose the nucleon spin in terms of the contributions from the spin polarization and orbital angular momentum of quarks and gluons. 

\subsection{
Deep Exclusive Meson Production}

A special kinematic regime is probed in Deep Exclusive Meson Production (DEMP) reactions, where the initial hadron emits a quark-antiquark or gluon pair.  This has no counterpart in the usual parton distributions, and carries information about $q\bar{q}$ and $gg$-components in the hadron wavefunction. %
Because quark helicity is conserved in the hard scattering regime, the produced meson acts as a helicity filter~\cite{Goeke:2001tz}.  In particular, leading order QCD predicts that vector meson production is sensitive only to the unpolarized GPDs, $H$ and $E$, whereas pseudoscalar meson production is sensitive only to the polarized GPDs, $\tilde{H}$ and $\tilde{E}$.  
In contrast, DVCS depends at the same time on both the polarized ($\tilde{H}$ and $\tilde{E}$) and the unpolarized ($H$ and $E$) GPDs.  Thus, DEMP reactions provide a tool to disentangle the different GPDs from experimental data~\cite{Goeke:2001tz}.

The $\tilde{E}$ is particularly poorly known~\cite{Cuic:2020iwt}.  It is related to the pseudoscalar nucleon form factor $G_P(t)$, which is itself highly uncertain, because it is negligible at the momentum transfer of nucleon $\beta$-decay.  $\tilde{E}$ is believed to contain an important pion pole contribution, and hence is optimally studied in DEMP.  $\tilde{E}$ cannot be related to any already known parton distribution, and so experimental information about it can provide new nucleon structure information unlikely to be available from any other source.

Frankfurt et al.~\cite{Frankfurt:1999fp} identified the single spin asymmetry for exclusive $\pi^{\pm}$ production from a transversely polarized nucleon target as the most sensitive observable to probe the spin-flip $\tilde{E}$.  The experimental access to $\tilde{E}$ is through the azimuthal variation of the emitted pions, where the relevant angles are $\phi$ between the scattering and reaction planes, and $\phi_s$ between the target polarization and the scattering plane. The $\sin(\phi-\phi_s)$ asymmetry, where $(\phi-\phi_s)$ is the angle between the target polarization vector and the reaction plane, is related to the parton-helicity-conserving part of the scattering process, and is sensitive to the interference between $\tilde{H}$ and $\tilde{E}$~\cite{Frankfurt:1999fp,Diehl2005}.  The asymmetry vanishes if $\tilde{E}$ is zero.  If $\tilde{E}$ is not zero, the asymmetry will display a $\sin(\phi-\phi_s)$ dependence.
Refs.~\cite{Frankfurt:1999fp,Belitsky2004} note that ``precocious scaling'' is likely to set in at moderate $Q^2\sim 2-4$ GeV$^2$ for this observable, as opposed to the absolute cross section, where scaling is not expected until $Q^2>10$ GeV$^2$.

SoLID, in conjunction with a polarized $^3$He target, can be used to
probe $\tilde{E}$.  
Since polarized $^3$He is an excellent proxy for a polarized neutron, the reaction of interest is essentially $\vec{n}(e,e'\pi^-)p$ (after nuclear corrections are applied).  
The only previous data are from HERMES~\cite{HERMES:2009gtv}, for average values $\langle x_B \rangle =0.13$, $\langle Q^2 \rangle =2.38$ GeV$^2$. 
Although the observed $\sin(\phi-\phi_s)$ asymmetry moment is small, the HERMES data are consistent with GPD models based on the dominance
of $\tilde{E}$ over $\tilde{H}$ at low $-t=-(q-p_{\pi})^2$~\cite{Goloskokov:2009ia}.  An improved measurement of the $\sin(\phi-\phi_s)$ modulation of the transverse target spin asymmetry, is clearly a high priority.
In comparison to HERMES, SoLID will probe higher $Q^2$ and $x_B$, with much smaller statistical errors over a wider range of $t$.
Thus, the measurements should be more readily interpretable than those from HERMES, providing the first clear
experimental signature of $\tilde{E}$.

In the DEMP reaction on a neutron, all three charged particles in the final
state, $e^{-}$, $\pi^{-}$ and $p$, can be cleanly measured by SoLID.  
Hence, contamination from other reactions, including DEMP from the other two protons in $^{3}$He, can be greatly eliminated.  The dominant background of the DEMP measurement comes from the SIDIS reactions of electron scattering on the neutron and two protons in $\mathrm{^{3}He}$.
Further reduction in the background can be accomplished by reconstructing the missing momentum and missing mass of the recoil protons, via $\vec{p}_{miss}=\vec{q}-\vec{p}_{\pi}$, $M_{miss}=\sqrt{(\nu-E_{\pi})^2-(\vec{q}-\vec{p}_{\pi})^2}$.
After applying a missing momentum cut to exclude events for which $p_{miss}>1.2$~GeV/c, the SIDIS background is largely suppressed.

\begin{figure}[htb]
\centering\includegraphics*[width=0.99\linewidth]{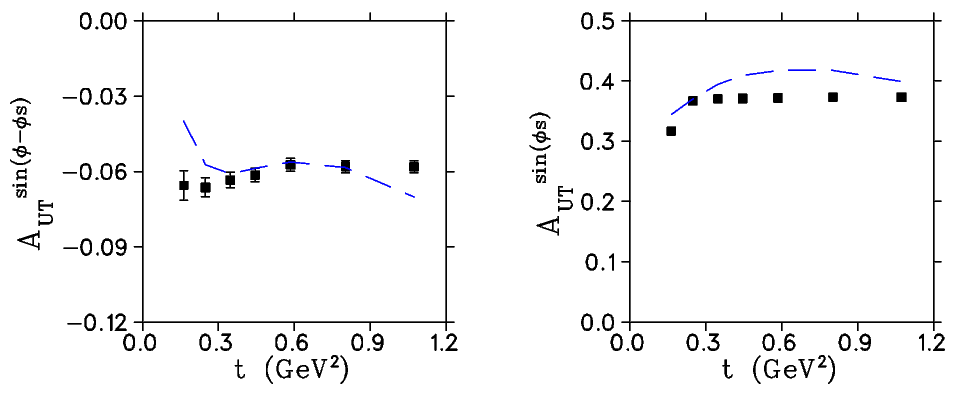}
\caption{\baselineskip13pt
Projected uncertainties for $A_{UT}^{\sin(\phi-\phi_s)}$ and $A_{UT}^{\sin(\phi_s)}$ in the $\vec{n}(e,e'\pi^-)p$ reaction from a transversely polarized $^3$He target and SoLID.  The dashed curve represents the input asymmetry into the simulation, and the data points represent the extracted asymmmetry moment values from an unbinned maximum likelihood (UML) analysis of simulated SoLID data.}
\label{fig:demp_proj}
\end{figure}

Figure~\ref{fig:demp_proj} shows E12-10-006B~\cite{JLabPR:demp} projections for the two most important transverse single spin asymmetry moments.  
The $\sin(\phi-\phi_s)$ moment (left) provides access to $\tilde{E}$ and is the primary motivation of the measurement.
There is growing theoretical interest in the $\sin(\phi_s)$ moment (right), as it provides access to the higher-twist transversity GPD $H_T$. 
The projected data points assume detection of triple-coincidence $\vec{^3He}(e,e'\pi^-p)pp$ events, after application of the $p_{miss}$ cut.  All scattering, energy loss, and detector resolution are included.  Fermi momentum has been turned off in the event generator, similar to where the recoil proton resolution is good enough to correct for Fermi momentum effects on an event-by-event basis. 
The agreement between the input and output fit values is very good, validating the unbinned maximum likelihood 
analysis procedure.

\lrpbf{
The high luminosity and full azimuthal coverage capabilities of SoLID make it well-suited for this measurement.
It is the only feasible manner to access to wide $t$ range needed to fully exploit the transverse target asymmetry information.  The projected SoLID data are expected to be a considerable advance over the HERMES data in terms of kinematic coverage and statistical precision.  The SoLID measurement is also important preparatory work for studies of the same asymmetries at the EIC, utilizing a transversely polarized proton or $^3$He beam.
}

\subsection{Deeply Virtual Compton Scattering with Polarized Targets}
 \begin{figure}[ht]
 \begin{center}
  \includegraphics[width=0.3\textwidth]{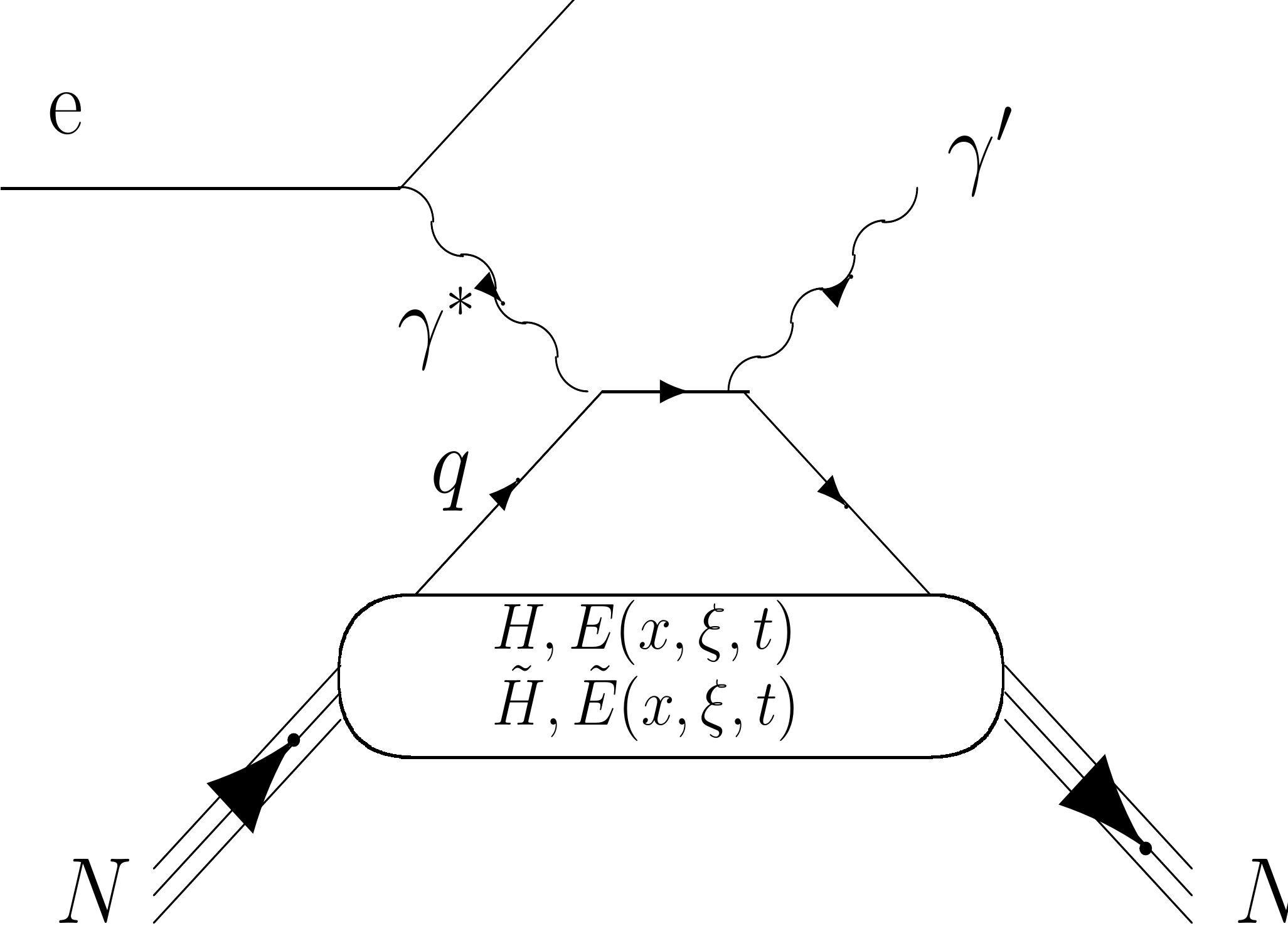}
   \caption[DVCS process in the $e+N\rightarrow eN\gamma$ reaction]{\footnotesize{DVCS process in the $e+N\rightarrow eN\gamma$ reaction. The cross section is composed of the amplitudes of DVCS and Bether-Heitler processes as well as their interference.}}
  \label{dvcs_bh}
 \end{center}
\end{figure}

Deeply Virtual Compton Scattering (DVCS) is the golden channel to experimentally study GPDs~\cite{Ji:1996nm,Belitsky:2001ns}. In electron scattering off nucleons with sufficiently large momentum transfer, a highly virtual photon scatters from a quark and excites the nucleon, which returns to its initial nucleon state by emitting a real photon so the nucleon remains intact. In this process, one measures the hard exclusive photons produced in the Bethe-Heitler (BH) and the DVCS processes, as well as their interference, i.e. $\sigma_{e+N\rightarrow eN\gamma}$ $\propto$ $\vert \mathcal{T}_{DVCS}\vert^2 $+$ \vert \mathcal{T}_{BH}\vert^2$+$ \mathcal{I}$, where the DVCS term and the interference term ($\mathcal{I} = \mathcal{T}_{DVCS}^{*}\mathcal{T}_{BH} + \mathcal{T}_{BH}^{*}\mathcal{T}_{DVCS}$) contain the information about the GPDs with the convolution integral, called Compton Form Factors (CFF). 

Several DVCS experiments with proton targets have been carried out in Halls A and B of Jefferson Lab with 6~GeV electron beam~\cite{JeffersonLabHallA:2006prd,JeffersonLabHallA:2015dwe,CLAS:2007clm,CLAS:2015uuo} as well as the HERMES experiment ~\cite{HERMES:2001bob, HERMES:2012gbh, HERMES:2010dsx, HERMES:2008abz, HERMES:2011bou, HERMES:2006pre, HERMES:2009cqe, HERMES:2009xsg}. With the 12~GeV upgrade, several experiments in Halls A and B have been approved to measure the beam-spin asymmetry and target-spin asymmetry with a longitudinally polarized proton target~\cite{halla:e12-06-114,clas12:e12-06-119}. The DVCS measurement on neutrons is more difficult, mainly due to lower production yields, smaller asymmetries, and bigger demands on the experimental techniques compared with the proton-DVCS case. The first neutron-DVCS measurement~\cite{JeffersonLabHallA:2007jdm} was performed in the E03-106 experiment in Hall A with polarized beam on a deuterium target. This pioneering work established the importance of the neutron-DVCS measurement, but was limited to a narrow phase space. An approved CLAS12 experiment~\cite{clas12:e12-11-003}, aims to measure the beam-spin asymmetry with an unpolarized neutron target. 

To allow for a full flavor decomposition to extract the GPDs of individual quarks, it is desired to collect precise neutron data over a more complete phase space and with more experimental observables. It is especially important to do measurements with a transversely polarized target, which is essential to access the poorly known GPD $E$. 
\lrpbf{
SoLID will enable the first measurement of DVCS on transversely polarized neutrons with 11~GeV longitudinally polarized electron beam, where the single-spin asymmetry ($A_{UT}$) and the double-spin asymmetry ($A_{LT}$) provide great sensitivities to decouple different CFFs in the neutron-DVCS reaction.
} 
A run-group measurement, in parallel with the already approved SIDIS experiment (E12-10-006), is under exploration. In combination with the DVCS measurement using polarized proton targets running parasitically with the approved SIDIS experiment (E12-11-108), one can perform flavor-decomposition to isolate the CFFs of $u$ and $d$ quarks. Possible detector upgrades, including a better energy resolution EM calorimeter or a recoil detector, will enable clean identification of the DVCS events and unlock the full power of the SoLID GPD program.

\subsection{
Timelike Compton Scattering}

\begin{figure}[ht]
 \begin{center}
  \includegraphics[width=0.48\textwidth]{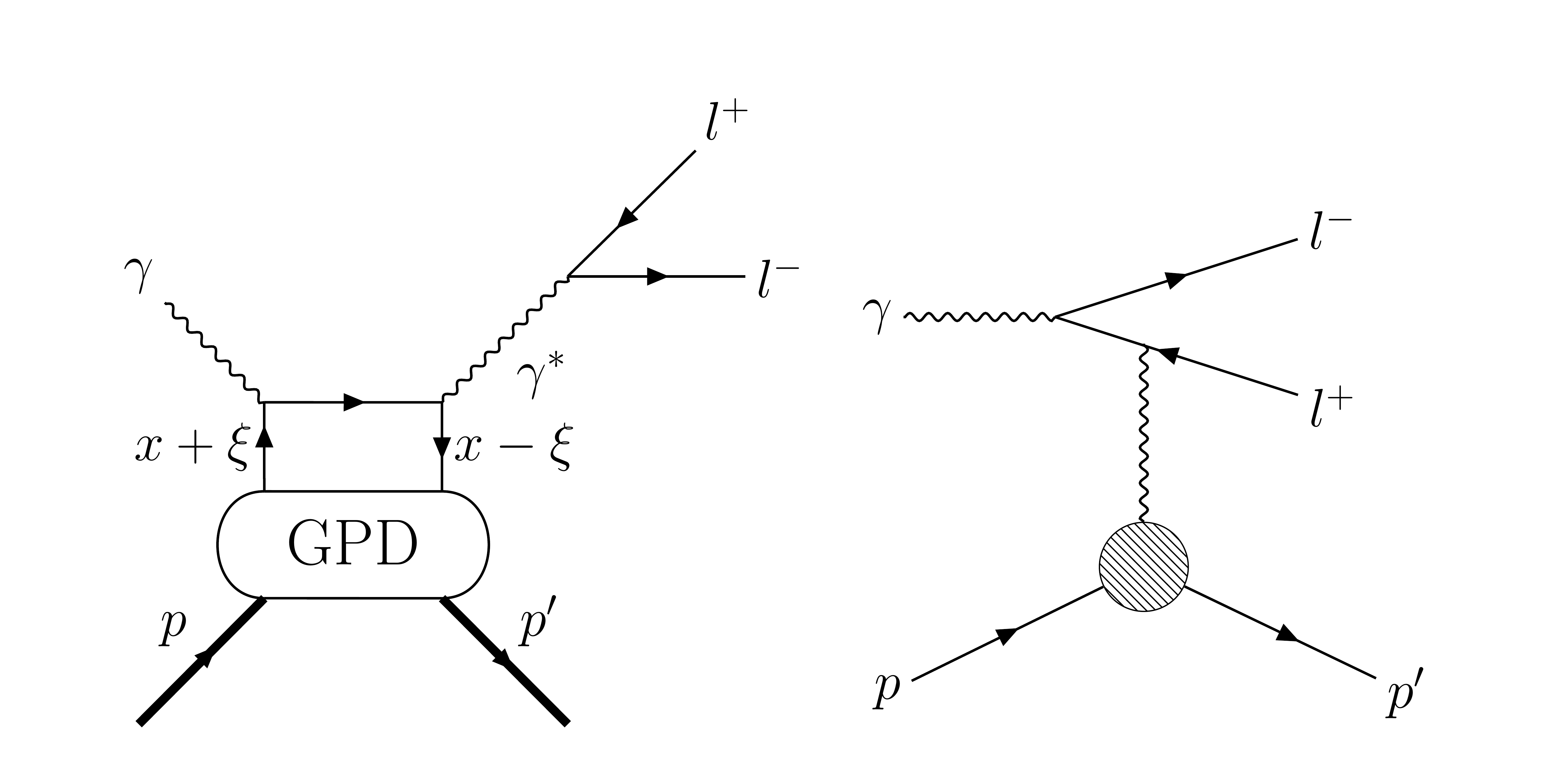}
   \caption{Left: handbag diagram of the TCS process. Right: diagram of the BH process.}
  \label{fig:TCS_diagram}
 \end{center}
\end{figure}

The most widely studied DVCS measurement is the electroproduction of a real spacelike photon on a nucleon. Correspondingly, Timelike Compton Scattering (TCS), is the photoproduction of a virtual timelike photon ($Q^{\prime 2} > 0$) on a nucleon, where the final-state virtual photon immediately decays into a lepton pair, as shown in Eq.~\ref{eq:TCS} and the left panel of Fig.~\ref{fig:TCS_diagram}~\cite{PhysRevLett.127.262501}. Like DVCS, TCS is also a direct process to access nucleon GPDs and can provide valuable information for GPD extraction. The study of both processes provides an upmost important test about the universality of GPDs and the QCD factorization approach.

\begin{equation}
\gamma + p \to \gamma^{\ast}+p^{\prime}  \to l^{-} + l^{+} + p^{\prime} 
\label{eq:TCS}
\end{equation}

\begin{figure}[ht]
 \begin{center}
  \includegraphics[width=0.45\textwidth]{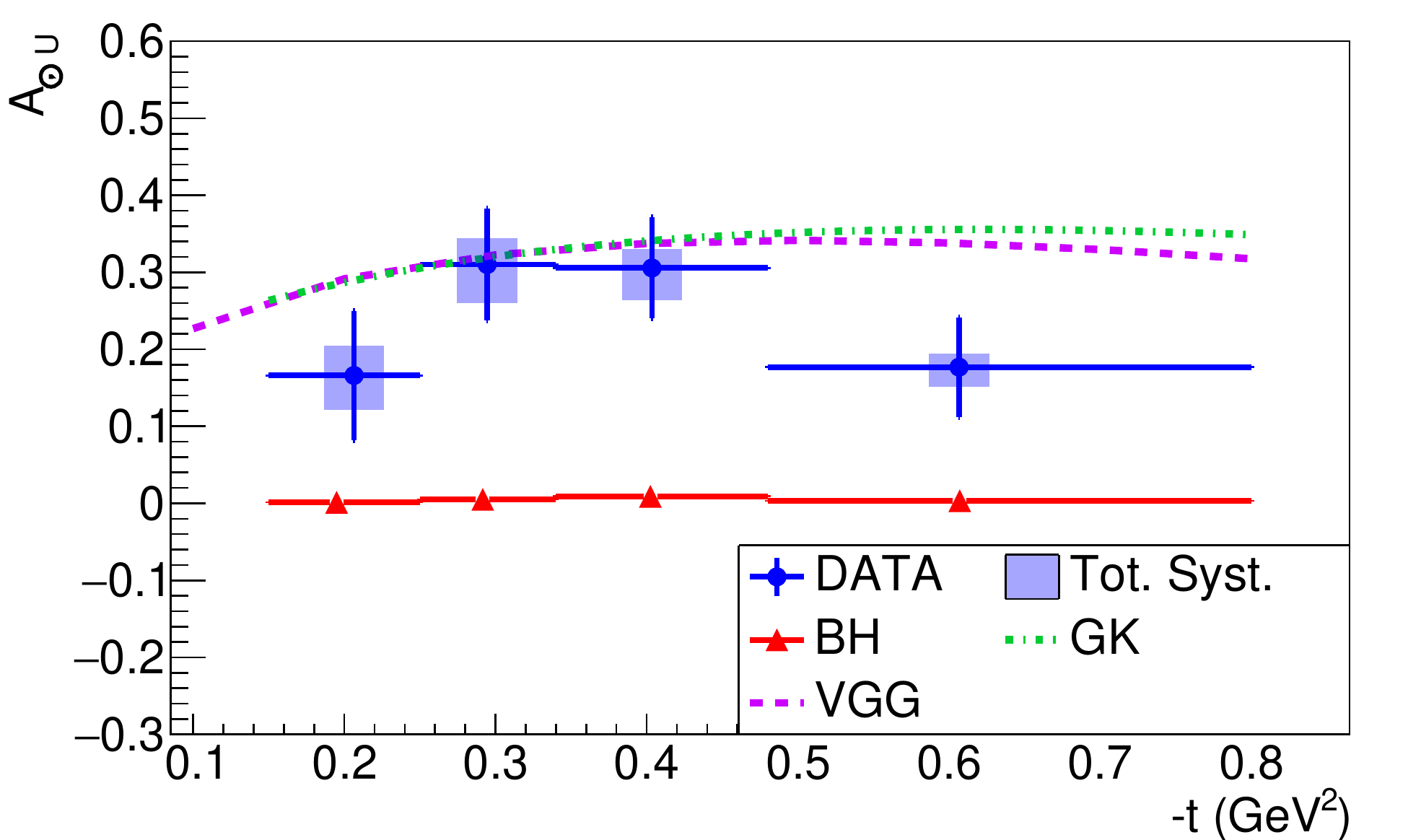}
    \includegraphics[width=0.45\textwidth]{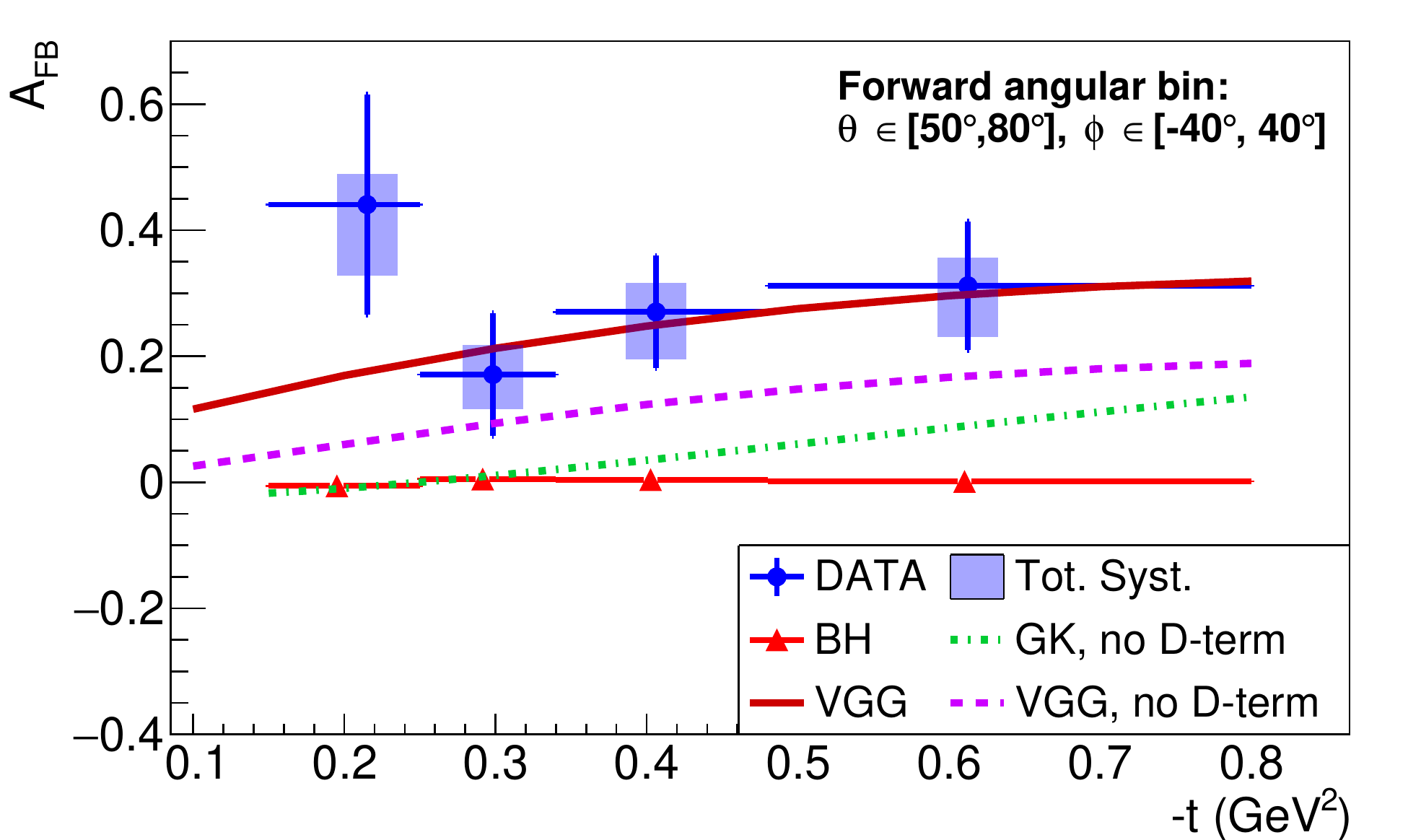}
   \caption{
   The photon polarization asymmetry $A_{\odot U}$ (top) and forward-backward (bottom) asymmetries as a function of $-t$ at the averaged kinematic point $E_\gamma = 7.29 \pm 1.55~{\rm GeV}$; $M=1.80\pm 0.26~{\rm GeV}$~\cite{PhysRevLett.127.262501}. The errors on the averaged kinematic point are the standard deviations of the corresponding distributions of events. The data points are represented in blue with statistical vertical error bars. The horizontal bars represent the bin widths. The shaded error bars show the total systematic uncertainty. The red triangles show the asymmetry computed for simulated BH events.  The dashed and dashed-dotted lines are the predictions of the VGG and GK models respectively. The solid line shows the model predictions of the VGG model with D-term.}
  \label{fig:TCS_result_clas12}
 \end{center}
\end{figure}

TCS is not the only physical process that can be observed in the exclusive photoproduction of lepton pairs, many resonance states decay into lepton pairs as well. In the resonance free region, the dominant background process with the same final state is the purely electromagnetic Bethe-Heitler (BH) reaction shown in the right panel of Fig.~\ref{fig:TCS_diagram}. Again like DVCS, the TCS and BH amplitudes interfere. Even though the BH cross section is significantly larger than the TCS cross section, we can take advantage of this interference to study TCS.

The JLab 12 GeV upgrade opens the door to access the TCS production in the resonance free region. The first TCS measurement on proton using the CLAS12 detector was recently published~\cite{PhysRevLett.127.262501} and the selection of results are shown in Fig.~\ref{fig:TCS_result_clas12}. The photon circular polarization and forward-backward asymmetries were measured to be nonzero, providing strong evidence for the contribution of the quark-level mechanisms parametrized by GPDs to this reaction. The comparison of the measured polarization asymmetry with DVCS-data-constrained GPD model predictions for the imaginary and real parts of $H$ points toward the interpretation of GPDs as universal functions. This is a great achievement, even with limited statistics. It is clear that more measurements are needed to expand the study of TCS.

Experiment E12-12-006A~\cite{JLabPR:E12-12-006A} will study the TCS reaction via exclusive $e^+e^-p$ production, using the SoLID detector with an 11 GeV polarized electron beam and a 15 cm LH$_2$ target. The experimental observables include the circularly polarized photon asymmetry and the forward-backward asymmetry just like CLAS12, but it can also study the moments of the weighted cross section with more available data. The kinematics can cover a wide range of squared four momentum transfer ($0.1 < -t < 0.7$ GeV$^2$), outgoing photon virtuality ($2.25 < Q^{\prime 2} < 9$ GeV$^2$) and skewness ($0.1 < \xi < 0.4$) with $\xi$=$Q^{\prime 2}/\left( (s-m_p^2)-Q^{\prime 2}\right)$ where $s$ is the center-of-mass energy and $m_p$ is the proton mass. As a run group experiment with the SoLID J/$\psi$ program E12-12-006, the two measurements would benefit each other on the normalization and systematic studies.

\lrpbf{
SoLID TCS is the perfect next stage experiment after the CLAS12 TCS measurement. It will provide an essential cross-check by using a different large acceptance detector to measure the same process. This is a safe approach, since TCS is still a new tool for GPD studies. The high luminosity $10^{37}$~cm$^{-2}\cdot$s$^{-1}$ of SoLID is 2 orders magnitude larger than CLAS12, making it possible to perform a mapping of the $t$, photon virtuality and skewness dependences at the same time. This is essential for understanding factorization, higher-twist effects, and NLO corrections. The experiment will collect unprecedented amount of high quality data. It will push the TCS study to a precision era, and together with DVCS, carry out global analyses to extract GPDs from the data.
}

\subsection{
Double Deeply Virtual Compton Scattering}
The dynamical properties of the nucleon that are expressed by the energy-momentum tensor~\cite{Ji:1996ek} involve integrals of GPDs over the average momentum fraction of partons at fixed skewness. Similarly, the tomography of the nucleon~\cite{Burkardt:2000za} involves integrals of GPDs over the transverse momentum transfer in the zero-skewness limit. Thus, it is of prime importance to obtain a separate knowledge of the $x$-and $\xi$-dependences of GPDs. Differently from the DVCS and TCS processes, which access GPDs along the line $x$=$\pm \xi$, the Double Deeply Virtual Compton Scattering (DDVCS) process~\cite{PhysRevLett.90.012001,PhysRevLett.90.022001}, where the initial and final photons are virtual, is the only known process allowing one to investigate independently the $(x,\xi)$-dependence of GPDs, i.e. at $x \neq \xi$.

\begin{figure}[!ht]
 \begin{center}
  \includegraphics[width=0.36\textwidth]{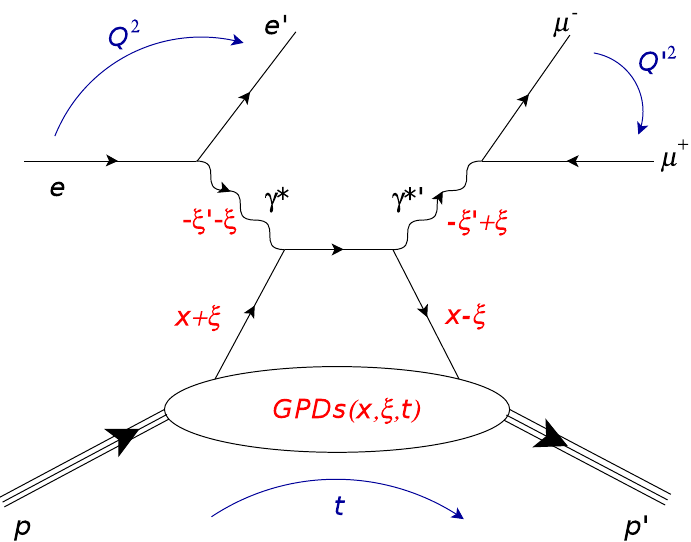}
   \caption[Leading twist and leading $\alpha_S$-order of the amplitude of the DDVCS process]{Schematic of the direct term of the DDVCS amplitude with a di-muon final state. The full amplitude contains also the crossed term, where the final time-like photon is emitted from the initial quark. $Q^2$=-$q^2$ is the virtuality of the space-like initial photon, and $Q'^2=q'^2$ is the virtuality of the final time-like photon.}
  \label{fig:ddvcs}
 \end{center}
\end{figure}

At leading twist and leading $\alpha_s$-order, the DDVCS process corresponds to the absorption of a space-like photon by a parton of the nucleon, followed by the emission from the same parton of a time-like photon decaying into a $l \bar{l}$-pair, see Fig.~\ref{fig:ddvcs}. The scaling variables attached to this process are defined as 
\begin{eqnarray}
\xi' & = & \frac{Q^2-Q'^2+t/2}{2Q^2/x_\text{B}-Q^2-Q'^2+t} \label{xip_sca} \\
\xi  & = & \frac{Q^2+Q'^2}{2Q^2/x_\text{B}-Q^2-Q'^2+t}, \label{xi_sca}
\end{eqnarray}
representing the Bjorken generalized variable ($\xi'$) and the skewness ($\xi$). When the final photon becomes real, the DDVCS process turns into DVCS, which corresponds to the restriction $\xi'$=$\xi$ in the Bjorken limit. When instead the initial photon becomes real, DDVCS turns into the TCS process, which corresponds to the restriction $\xi'=-\xi$ in the Bjorken limit. In these respects, the DDVCS process is a generalization of the DVCS and TCS processes. 

The DDVCS reaction amplitude is proportional to a combination of the Compton Form Factors (CFFs) ${\mathcal F}$ (with ${\mathcal F} \equiv \{ {\mathcal H}, {\mathcal E}, \widetilde{\mathcal H}, \widetilde{\mathcal E} \}$) defined from the GPDs $F$ (with $F \equiv \{ H,E,\widetilde{H},\widetilde{E} \}$) as
\begin{eqnarray}
\mathcal{F}(\xi',\xi,t) & = & \mathcal{P} \int_{-1}^1  F_+(x,\xi,t)\bigg[\frac{1}{x-\xi'} \pm \frac{1}{x+\xi'}\bigg] dx \nonumber \\
 & - & i \pi F_+(\xi',\xi,t), \label{eq1}
\end{eqnarray}
where ${\mathcal P}$ denotes the Cauchy's principal value integral, and  
\begin{equation}
F_+(x,\xi,t) = \sum_{q} \left(\frac{e_q}{e}\right)^2 {\left[ F^q(x,\xi,t) \mp F^q(-x,\xi,t) \right] }
\end{equation}
is the singlet GPD combination for the quark flavor $q$, where the upper sign holds for vector GPDs $(H^q,E^q)$ and the lower sign for axial vector GPDs $(\widetilde{H}^q,\widetilde{E}^q)$. In comparison to DVCS and TCS, the imaginary part of the DDVCS CFFs accesses the GPDs at $x$=$\pm \xi' \neq \xi$, and the real part of the DDVCS CFFs involves a convolution with different parton propagators. Varying the virtuality of both incoming and outgoing photons changes the scaling variables $\xi'$ and $\xi$, and maps out the GPDs as function of its three arguments independently. From Eq.~\ref{xip_sca}-\ref{xi_sca}, one obtains 
\begin{equation}
\xi' = \xi  \, \frac{Q^2-Q'^2+t/2}{Q^2+Q'^2},
\end{equation} 
indicating that $\xi'$, and thus the imaginary parts of the CFFs $\{ \mathcal{H}, \mathcal{E} \}$, changes sign around $Q^2$=$Q'^2$. This represents a strong testing ground of the universality of the GPD formalism~\cite{Anikin:2017fwu}. 

\begin{figure}[t]
 \begin{center}
\includegraphics[width=0.35\textwidth]{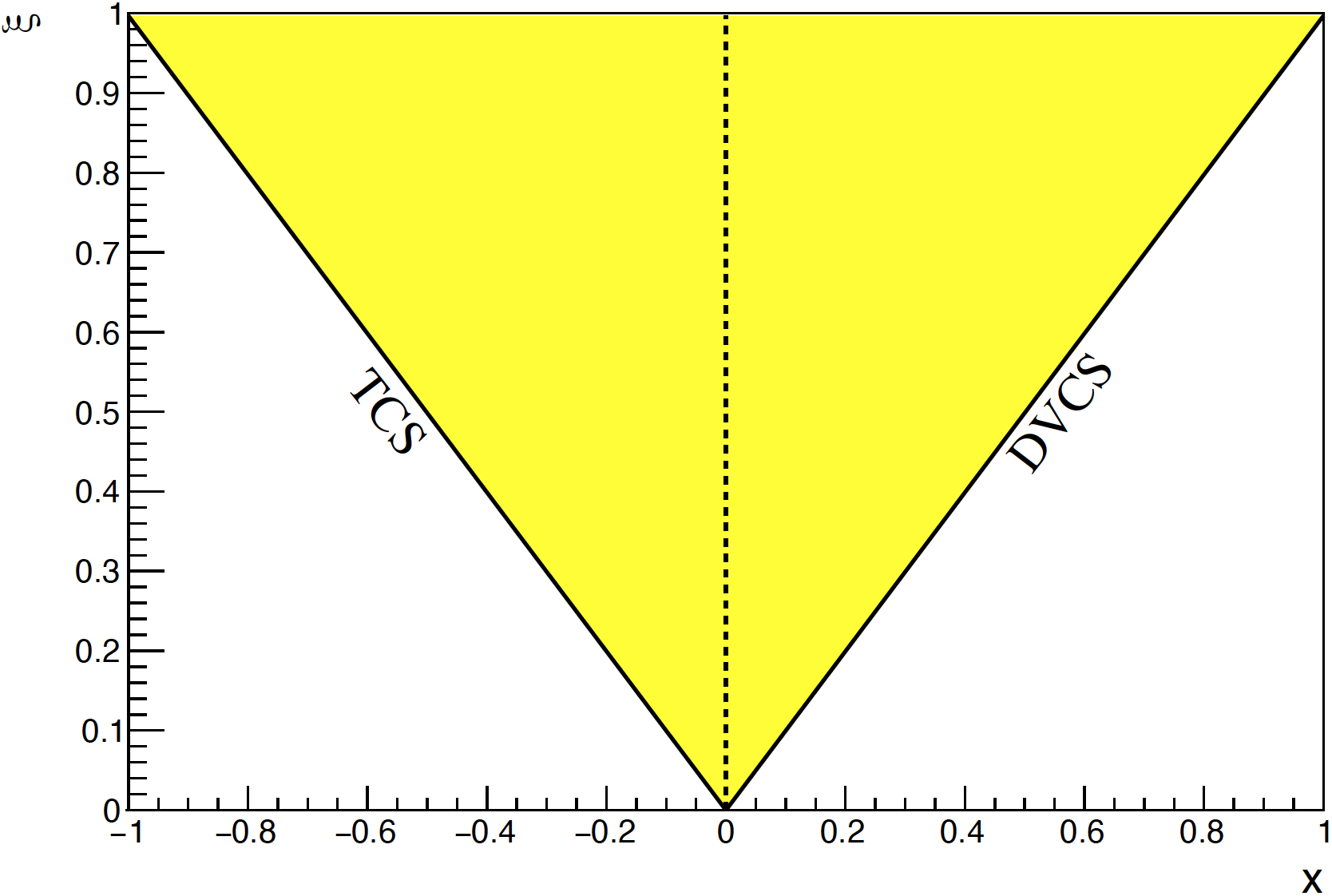}
\caption{Singlet GPD $F_+(x,\xi,0)$ coverage of the physics phase-space of the imaginary part of the CFFs: the yellow area represents the DDVCS reach, bounded on the one side by the TCS, and on the other side by DVCS lines. The $x$-axis corresponds to  the PDFs (Parton Distribution Functions) domain measured in inclusive Deep Inelastic Scattering.}
\label{Phy_Spa}
\end{center}
\end{figure}

\begin{figure}[t]
 \begin{center}
\includegraphics[width=0.45\textwidth]{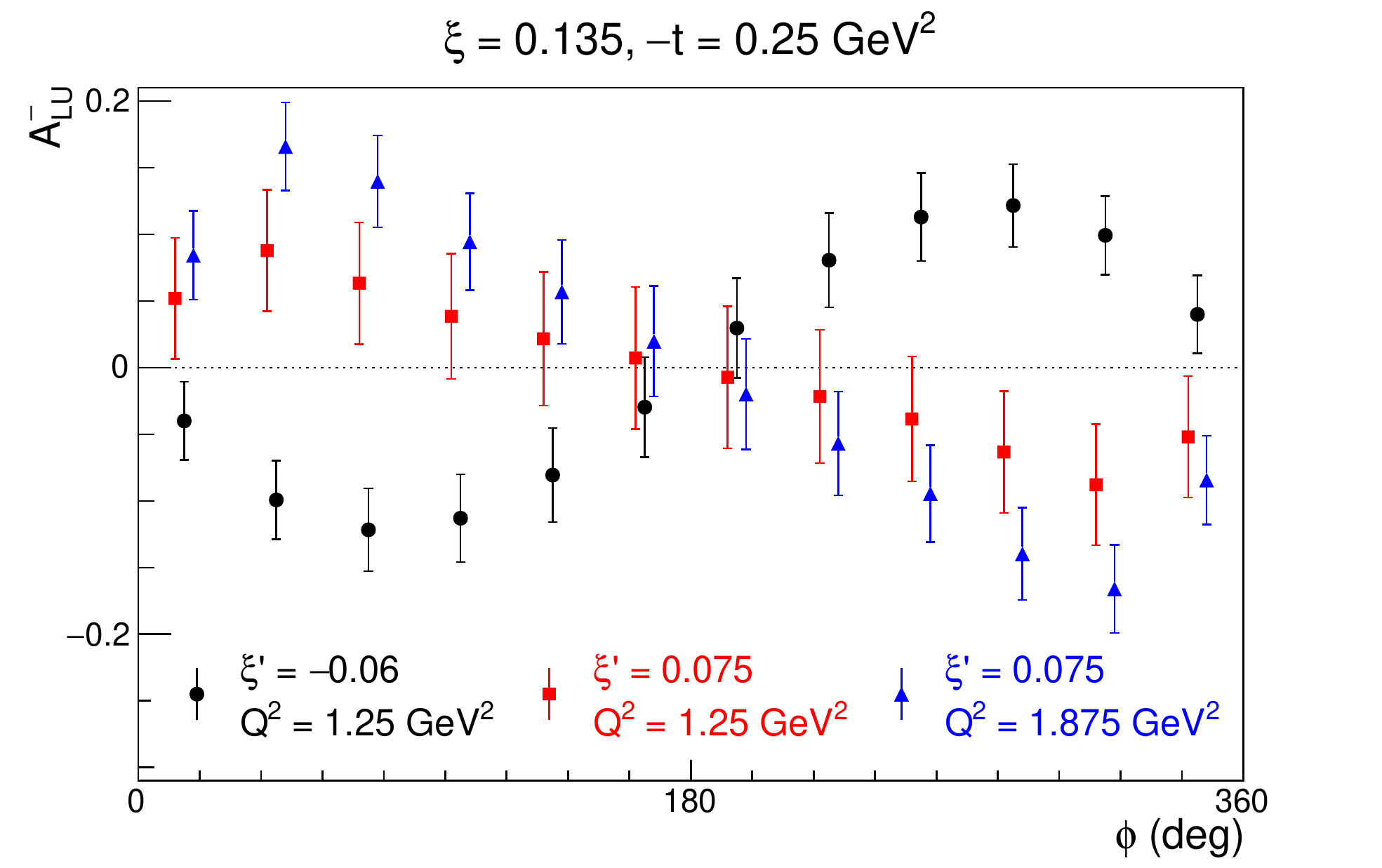}
\caption{Projections of selected DDVCS beam spin asymmetry measurements with SoLID, assuming 50 days of data taking on a liquid hydrogen target and a luminosity of 1.2$\times 10^{37}$cm$^{-2} \cdot$s$^{-1}$~\cite{Zhao:2020th}.}
\label{DDVCS_BSA}
\end{center}
\end{figure}

Similarly to DVCS, the imaginary part of the CFFs can be accessed by comparing experimental cross sections measured with polarized electron or positron beams of opposite helicities, and the real part of the CFFs is best measured by comparing experimental cross sections measured with unpolarized electron and positron beams~\cite{Zhao:2021zsm}. In order to achieve these measurements, the SoLID spectrometer is to be completed with specific devices dedicated for muon detection~\cite{JLabPR:DVCS_dimuon}. The Large Angle Muon Detector takes advantage of the material of the Large Angle Electromagnetic Calorimeter and the iron flux return to serve as shielding, and two layers of GEMs at the outer radius of the downstream encap ensure the detection of particles. The Forward Angle Muon Detector, placed after the downstream endcap, consists of three layers of iron slabs instrumented with GEMs. This configuration provides a significant coverage of the DDVCS muons and allows the efficient investigation of the $(\xi,\xi')$ space for $Q^2 \le 3.5$~GeV$^2$ and $-t < 1$~GeV$^2$. An unprecedented quantity of data will be collected and can be used to measure cross section and Beam Spin Asymmetry (BSA) observables of the DDVCS process. The experiment would operate over a period of 50 days with a 15 cm long unpolarized liquid hydrogen target and a 3~$\mu$A beam intensity. Selected BSA projections are shown in Fig.~\ref{DDVCS_BSA}. Particularly, it is worth noting that the quality of expected data would permit observation of the predicted sign change of the imaginary part of the CFFs, supporting GPD universality.

\lrpbf{
Both TCS and DDVCS measurements require detection of dilepton decay of virtual photons with high luminosity and large acceptance. The SoLID spectrometer uniquely meets such a demand. The SoLID TCS and DDVCS programs were featured in the 1st ``International Workshop on the Nucleon and Nuclear Structure through dilepton Production'' at ECT* in Trento, Italy in Oct 2016 and included in the resulting whitepaper~\cite{Anikin:2017fwu}.
}
\section{Other Physics Topics}\label{sec:others}
\lrpbf{
The multi-purpose feature of SoLID will allow many other physics topics to be studied, either as rungroup or stand-alone experiments. 
}
These physics topics are summarized in this section. 

\subsection{Inclusive Transverse Spin Structure Functions}
The transverse polarized structure function $g_{2}(x,Q^{2})$ probes transversely and also longitudinally polarized parton distributions inside the nucleon. It carries the information of quark\textendash gluon interactions inside the nucleon.  By neglecting quark masses, $g_{2}(x,Q^{2})$ can be decoded by a leading twist\textendash2 term and a higher twist term as follows:
\begin{equation}
g_{2}(x,Q^{2})=g^{WW}_{2}(x,Q^{2})+\bar{g_{2}}(x,Q^2),
\label{eq:g2}
\end{equation}
where twist\textendash2 term $g^{WW}_{2}$ was derived by Wandzura and Wilczek~\cite{Wandzura:1977qf} and it only depends on well\textendash measured $g_{1}$~\cite{SpinMuon:1998eqa,E143:1998hbs}.

The matrix element $d_{2}$ is the $x^2$ moment of $\bar{g_{2}}(x,Q^{2})$. This quantity measures deviations of $g_{2}(x,Q^{2})$ from the twist\textendash 2 term $g^{WW}_{2}$. At large Q2, where the operator product expansion (OPE)~\cite{Wilson:1969zs} becomes valid, one can access the twist\textendash3 effects of quark\textendash gluon correlations via the third moment of a linear combination of $g_{1}(x,Q^{2})$ and $g_{2}(x,Q^{2})$, presented as
\begin{eqnarray}
d_{2}(Q^2)&=&3\int_{0}^{1}x^{2}[g_{2}(x,Q^{2})-g^{WW}_{2}(x,Q^{2})]dx\nonumber\\
&=&\int_{0}^{1}x^{2}[2g_{1}(x,Q^{2})+3g_{2}(x,Q^{2})]dx.
\label{eq:d2}
\end{eqnarray}
Due to the $x^{2}$\textendash weighting, $d_{2}(Q^{2})$ is particularly sensitive to the large\textendash $x$ behavior of $\bar{g_{2}}$ and provides us a clean way to access twist\textendash3 contribution. 

A precision measurement of neutron spin structure function $g_{2} (x,Q^{2})$, running in parallel with this experiment and experiment E12-11-007~\cite{JLabPR:E12-11-007}, has been approved as a run group proposal~\cite{JLabPR:g2nd2n_solid} by PAC48. High statistics data will be collected within a large kinematic coverage of Bjorken scaling $x > 0.1$ and four momentum transfer $1.5<Q^{2}<10\;\rm{GeV}^{2}$ from inclusive scatterings of longitudinally polarized electrons off transversely and longitudinally polarized $^{3}$He targets, at incident beam energies of 11 GeV and 8.8 GeV. In addition to mapping out the $x$ and $Q^{2}$ evolution of $g_{2}$, the moment $d_{2}(Q^{2})$, which is connected to the quark-gluon correlations within the nucleon, will be extracted with $ 1.5<Q^{2}<6.5\;\rm{GeV}^{2}$. $d_{2}(Q^{2})$ is one of the cleanest observables that can be used to test the theoretical calculations from lattice QCD and various nucleon structure models.

\subsection{\label{sec:rg_kaon}SIDIS with Kaon Production}
While the JLab TMD program mostly focuses on measuring the pion production in SIDIS, the kaon production data are crucial to successfully decouple all light quark flavors. There are only limited kaon-SIDIS data from HERMES~\cite{HERMES:2010mmo}, COMPASS~\cite{COMPASS:2014bze}, and JLab Hall A collaborations~\cite{JeffersonLabHallA:2014yxb}, all of which are with poor precision and narrow kinematic coverage.  In the run-group proposal~\cite{JLabPR:sidis_kaon}, we will perform an offline analysis to extract the kaon-SIDIS events out from all the already approved SoLID pion-SIDIS measurements. The kaon events will be identified using the time-of-flight (TOF) information from the MRPC. A 20~ps time resolution of a new generation MRPC, which has been achieved with cosmic ray test by several groups~\cite{eic_mrpc, Wang:2013vha}, should be able to perform  $\pi^{\pm}/K^{\pm}$ separation up to a high hadron momentum (e.g. $P_{h}<7.0~GeV/c$), while the veto-signal from heavy-gas \v{C}erenkov detector can also effectively isolate $K^{\pm}$ from $\pi^{\pm}$. 

\begin{figure}[h]
\begin{center}
\includegraphics[width=0.48\textwidth]{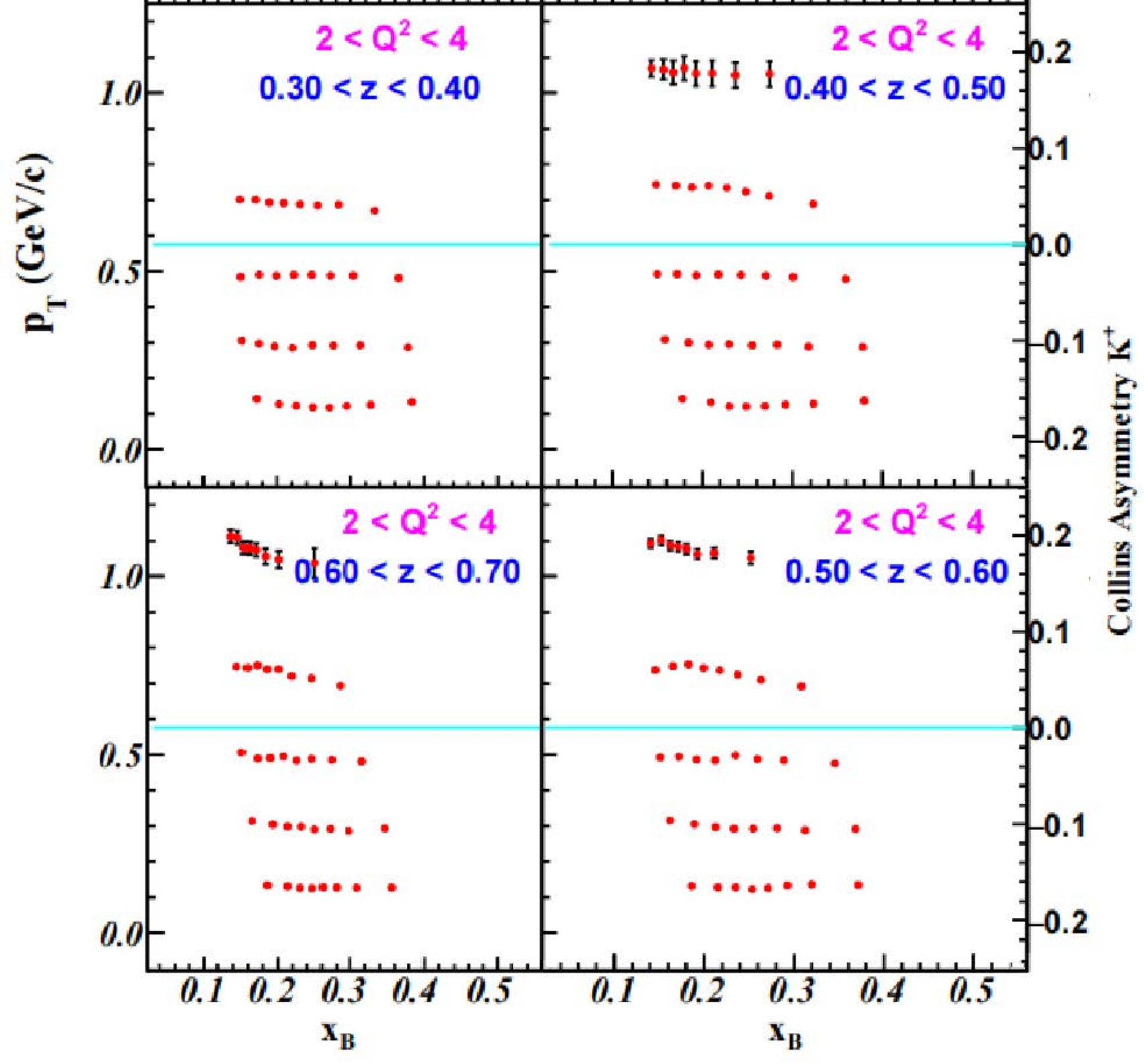} 
\end{center}
\vspace*{-20pt}
\caption{\label{fig:he3_project_collins}
One $Q^2$ bin of the 4D ($Q^2$, $z$, $p_{T}$, $x_{B}$) binning projection and statistical uncertainties of the Collins asymmetry ($A_{UT}^{\sin(\phi+\phi_{S})}$) in $\vec{n}(e,e'K^{+}$)X with transversely polarized $^3$He. The sizes of the uncertainties are indicated by the Y axis on the right. See the original proposal for all projection results.}
\end{figure}

\lrpbf{
Thanks to the high intensity and large acceptance features of the SoLID detector system, the new measurement will generate a large set of kaon data with great precision and a wide kinematic coverage in multiple dimensions as shown in Fig.~\ref{fig:he3_project_collins}. The combined analysis of both the pion and kaon SIDIS-data from both proton and neutron ($^3$He) targets on SoLID will allow us to systematically separate contributions from all light quarks, especially to isolate the sea-quark contributions.  The systematic uncertainties can also be largely reduce since the pion and kaon SIDIS events are measured all together. Model estimation shows that at the SoLID kinematics about 20\% of the kaon SIDIS events come from the current fragmentation region where the TMD factorization can be applied. The high-quality kaon data from SoLID are crucial for the validation of the model calculation. Our new measurement will provide high quality data for the continuous theoretical development of the TMD physics, and more importantly, provide strong guidance to future measurements on electron-ion collider (EIC), which will fully study the TMD of sea-quarks and gluons in a wider kinematic coverage and provide a more complete image of nucleon structures.
}

\subsection{\label{sec:rg_dh} SIDIS with Di-hadron Production}

Di-hadron SIDIS is an important part of the 12 GeV JLab physics program. Di-hadron single target spin asymmetry provided a wide range of insights into nucleon structure and hadronization. It is one of the easy channels to access the leading-twist PDF $h_1(x)$, the so-called transversity distribution function. The combination of the proton and neutron measurements on the transversity distribution function will also allow to operate a flavour separation.

In the process of $\ell(l) + N(P) \to \ell(l') + H_1(P_1)+H_2(P_2) + X $, the transversity distribution function $h_1(x)$ is combined with a chiral-odd Di-hadron Fragmentation Function ($DiFF$), denoted as $H_1^{<\kern -0.3 em{\scriptscriptstyle)}\,q}$, which describes the correlation between the transverse polarization of the fragmenting quark with flavor $q$ and the azimuthal orientation of the plane containing the momenta of the detected hadron pair. 
$DiFF$ can be extracted from electron-positron annihilation where two back-to-back jets are produced and a pair of hadrons are detected in each jet.
They also appear in the observable describing the semi-inclusive production of two hadrons in deep-inelastic scattering of leptons off nucleons or in hadron-hadron collisions. 
The $DiFFs$ also play a role in extending the knowledge of the nucleon col-linear picture beyond the leading twist.
 Contrary to the Collins mechanism, this effect survives after integration over quark transverse momenta and can be analyzed in a col-linear factorization framework. Thus this analysis will be much simpler compared to the traditional analysis of single-hadron fragmentation. 

Since the di-hadron proposal~\cite{JLabPR:sidis_dihadron} was accepted in 2014, research has been continued on improving $DiFF$~\cite{Pisano:2015wnq,Luo:2019frz}. The first extraction of transversity from a global analysis of electron-proton and proton-proton data was published in 2018 by M. Radici et al~\cite{Radici:2018iag}.
A measurement beam-spin asymmetry of di-hadron has also been published by the CLAS collaboration~\cite{CLAS:2020igs}, which leads to the extraction of the nucleon twist-3 parton distribution function $e(x)$~\cite{Courtoy:2022kca}.  
Recent measurements at CLAS12 showed the first empirical evidence of nonzero $G^{\perp}_1$, the parton helicity-dependent di-hadron fragmentation function ($DiFF$) encoding spin-momentum correlations in hadronization~\cite{Dilks:2021nry}. 
All these researches bring more and more attention to the di-hadron beam spin asymmetries in our field.

\subsection{Beam-Normal and Target-Normal SSAs}\label{sec:bnssa}
Our understanding and description of the internal structure of both nuclei and nucleons have seen a steady improvement over the past several decades. These improvements are sometimes brought on by inconsistent or unexplained experimental results, revealing limitations of our underlying assumptions. One such example is that of the discrepancy in the extraction of $\mathrm{G}^p_{E}/\mathrm{G}^p_{M}$, the ratio of the proton form factors of elastic scattering from either Rosenbluth or polarization transfer measurements at large $Q^{2}$, see e.g.~\cite{Puckett:2011xg} and references therein. At present, this discrepancy is attributed to two-photon exchange (TPE) and the size of the discrepancy is used to quantify TPE~\cite{Christy:2021snt}. Conversely, a reliable quantification of the TPE effects is needed to interpret electron scattering data in order to fully understand the structure of the nucleon. 

One way that TPE effects have been investigated is through a comparison of electron and positron elastic scattering off the proton, i.e. elastic lepton-charge asymmetry. Such measurements have been made at the VEPP-3 Storage Ring~\cite{Rachek:2014fam}, using CLAS~\cite{CLAS:2016fvy} at JLab, and by the OLYMPUS experiment at DESY~\cite{OLYMPUS:2016gso}. Studies of TPE also form part of the main thrust of a potential positron program at JLab~\cite{Accardi:2020swt}. However, a precision comparison between electron and positron scattering has its own challenges with one of the main systematic uncertainties being the relative luminosity control between the two beams. 

An alternate method to study TPE is through measurements of single spin asymmetries (SSA) where either the lepton (incoming or outgoing) or the target spin is polarized normal to the scattering plane, i.e., polarized along $\vec k\times\vec k'$ with $\vec k$ and $\vec k'$ the incoming and scattered electron's momentum, respectively. Experimentally, the most accessible would be the beam-normal SSA (BNSSA) or the target-normal SSA (TNSSA). 
At the Born level, in which a single photon is exchanged, both asymmetries are forbidden due to time-reversal invariance as well as parity conservation~\cite{Christ:1966zz}. 
Going beyond the Born approximation, they are no longer restricted and can provide direct access and insight into the imaginary part of the TPE amplitude. 

The initial theoretical predictions of both the TNSSA and BNSSA described the interaction in which the two photons couple to the same quark \cite{Metz:2006ssa}, and later, for the TNSSA, one in which the two photons couple to different quarks \cite{Metz:2012tnssa}. Finally, a complete partonic description of the TNSSA now exists \cite{Schlegel:2012tnssa}, in which all contributions have been taken into account.

Experimentally, the TNSSA has been measured for deep-inelastic $ep$ scattering~\cite{HERMES:2009hsi} and quasi-elastic and deep-inelastic $e-^3$He scattering~\cite{Katich:2013atq,Zhang:2015kna}, and comparison with available theory predictions is inconclusive as predictions vary up to two orders of magnitude depending on whether the two photons are assumed to couple with a single quark or two different quarks~\cite{Afanasev:2007ii,Metz:2012tnssa}, calling for experimental support to help distinguishing these model predictions. 
A run-group proposal~\cite{JLabPR:Ay} has been approved to measure the proton and the neutron TNSSA as part of the SoLID SIDIS running using transversely polarized NH$_3$ and $^3$He targets, at the level of $10^{-4} \sim 10^{-2}$.

The BNSSA data, on the other hand, mostly existed for elastic scattering as it is a typical background of elastic PVES experiments. A compilation of elastic BNSSA data can be found in~\cite{QWeak:2021jew}, along with new data from CREX/PREX-2~\cite{PREX:2021uwt}. In contrast, BNSSA data for DIS is nearly non-existent, except for the previous 6 GeV PVDIS experiment~\cite{Wang:2014guo} that measured this asymmetry to 20~ppm level. 
\lrpbf{
A new proposal~\cite{JLabPR:bnssa_dis} has recently been approved 
to measure BNSSA for the proton in the DIS region to ppm level for the first time. The measurement will utilize a transversely polarized electron beam and SoLID in its PVDIS configuration. The value of BNSSA $A_n$ will be extracted by fitting the measured asymmetry in the full azimuthal range to 2~ppm and 4~ppm level for the 6.6 and 11 GeV beam, respectively. It will add a new, high-precision observable to the landscape of TPE study and its impact on the understanding of the nucleon structure. 
}

\subsection{PVDIS with a Polarized \texorpdfstring{$^3$He}{3He} Target}
All existing PVES, elastic or DIS, focused on measurements of the cross section asymmetries with the electron spin flip on an unpolarized target. On the other hand, parity violation would cause a cross section difference in unpolarized electron scattering off right- and left-handed, longitudinally polarized hadrons. Such new observable, referred to as ``polarized parity-violating asymmetry'', can be written for the low to medium energy range as
\begin{eqnarray}
 A_{\mathrm{pvdis}}^{(h)} &\approx& \left(\frac{G_F Q^2}{2\sqrt{2}\pi\alpha}\right)
 \frac{g_V^e g^{\gamma Z}_5 + Y g_A^e g_1^{\gamma Z}}
  {F_1^\gamma}~, \label{eq:apolpv3}
\end{eqnarray}
where $Y$ is given by Eq.~(\ref{eq:y13_prd}), and we introduce polarized electroweak $\gamma Z$ interference structure functions: 
\begin{eqnarray}
 g_1^{\gamma Z} &=& \sum_i Q_{q_i} g_V^{i}(\Delta q_i +\Delta\bar{q}_i)\\
 g_5^{\gamma Z} &=& \sum_f Q_{q_i} g_A^{i}(\Delta q_i -\Delta\bar{q}_i)~. 
\end{eqnarray}
The $g_5^{\gamma Z}$ contribution to the asymmetry, however, is suppressed since $g_V^e\ll g_A^e$. Thus our main focus will be on the first determination of the $g_1^{\gamma Z}$, which provides information on new flavor combination of polarized PDFs. 

The flavor combination of the $g_1^{\gamma Z}$ is rather unique and provides connections to our understanding of the nucleon spin. 
This can be seen if we take the approximation  $\sin^2\theta_W \approx 0.25$), giving  
\begin{eqnarray}
 g_1^{p,\gamma Z} \approx g_1^{n,\gamma Z} 
 &\approx&\frac{1}{9}\left(\Delta u^+ + \Delta c^+ + \Delta d^+ + \Delta s^+\right)
\end{eqnarray}
where $\Delta q^+\equiv \Delta q + \Delta \bar q$. 
Therefore, the moment of $g_1^{\gamma Z}$ provides approximately the total quark spin contribution to the proton spin, which is believed to be 30\%~\cite{Deur:2018roz}. 

The measurement of such asymmetry is more difficult than the PVDIS asymmetry of Eq.~(\ref{eq:Apvdis1}) (often referred to as ``unpolarized PV asymmetry''), both because of the relatively small size of $A_{PV}^{(h)}$ and because of the lower luminosity of polarized than unpolarized targets. 
A letter-of-intent~\cite{JLabLOI:polpv_2016} was submitted to JLab PAC in 2016 with the goal to measure the $A_{PV}^{(h)}$ using a polarized $^3$He target and SoLID in the SIDIS configuration. To reach a high precision within a reasonable amount of beam time, the FOM of the polarized $^3$He target will need to be improved by factor 16 beyond its best projected performance of the 12 GeV era, 
via the use of higher fill pressure of $^3$He and cryo-cooling to increase the in-beam density. The projected relative uncertainty is $<10\%$ on $A_{PV}^{(^3\mathrm{He})}$ for $x=(0.2,0.4)$, using 180~days of production beam time at 100\% efficiency. While technically challenging, it will be the first measurement of the $g_1^{\gamma Z}$ structure functions. 
Similar measurements with the polarized proton could also be explored at the EIC. 
\section{SoLID Instrumentation}\label{sec:instrum}
\subsection{Overview of SoLID Setup}
SoLID is a large acceptance spectrometer designed to handle a very high luminosity to exploit the full potential of the 12 GeV beam of CEBAF. The equipment of SoLID is designed to satisfy the physics requirements of the five approved experiments. It has the capacity to handle very high signal and background rates, and it can sustain the high radiation environment with the very high luminosity in JLab’s experimental hall A.

\begin{figure}[!ht]
  \begin{center}
    \includegraphics[width=0.48\textwidth]{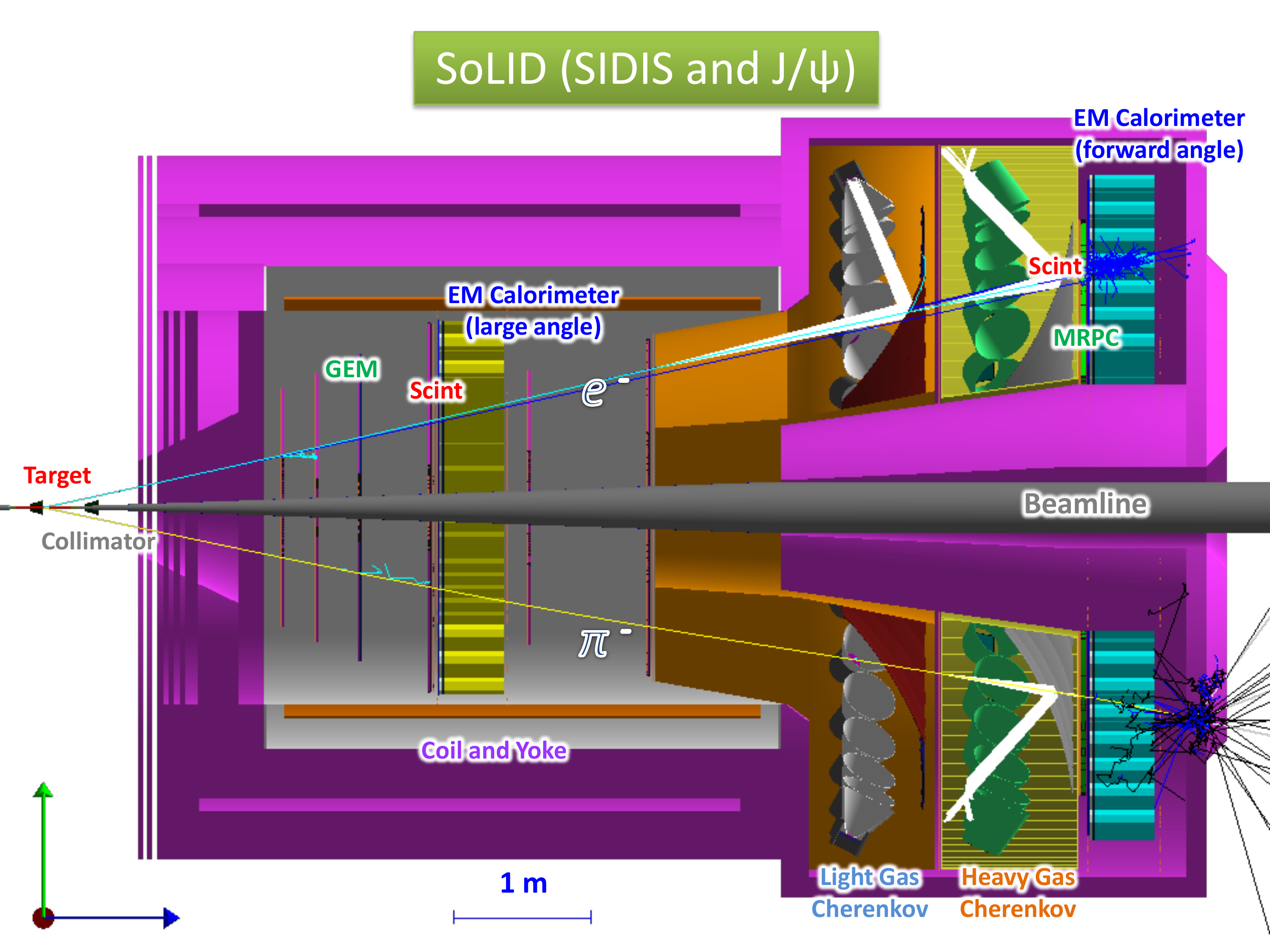}
    \includegraphics[width=0.48\textwidth]{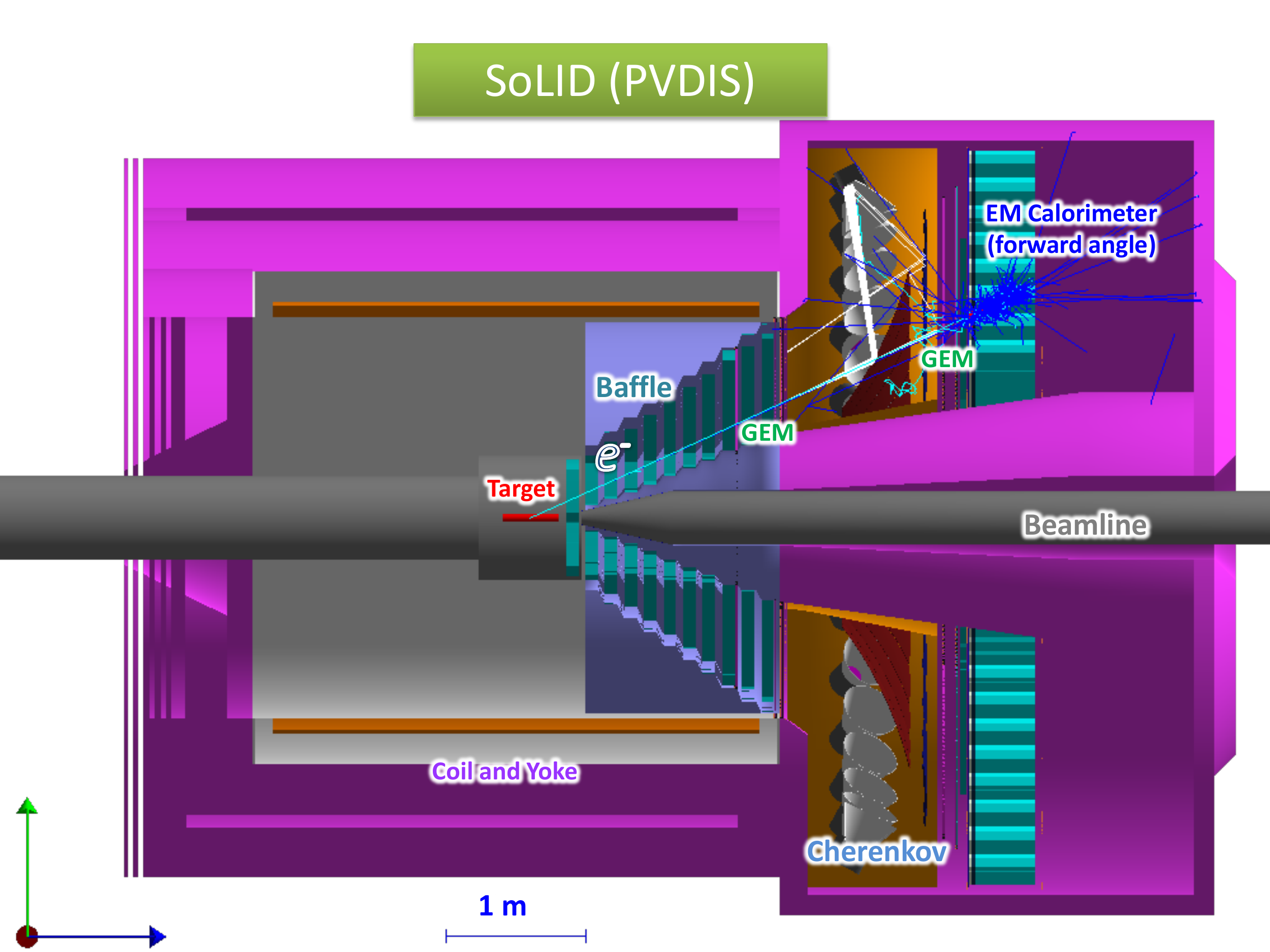}
    \caption{The two configurations of SoLID setup: SIDIS and $J/\psi$ (top) and PVDIS (bottom). }
    \label{fig:setup}
  \end{center}
\end{figure}

A large solenoid magnet sweeps away low-energy background charged particles, which makes it possible to operate at very high luminosities in an open geometry with full azimuthal coverage. The solenoid field is also necessary for tracking and momentum measurement. The CLEO-II magnet has been selected with modifications to its iron flux return. The detector system of SoLID includes two configurations: the ``SIDIS\&J/$\psi$'' configuration and the ``PVDIS'' configuration, as shown in Fig.~\ref{fig:setup}. 

The ``SIDIS\&J/$\psi$'' configuration consists of two groups of sub-detectors: the Forward Angle Detector group (FAD), and the Large Angle Detector group (LAD). The FAD group covers the nominal $8^\circ$-$15^\circ$ polar angle range and constitutes of five planes of Gas Electron Multipliers (GEM) for tracking, a light gas \v{C}erenkov (LGC) for $e/\pi$ separation, a heavy gas \v{C}erenkov (HGC) for $\pi/K$ and $\pi/p$ separation, a Multi-gap Resistive Plate Chamber (MRPC) for time-of-flight measurement and for kaon and proton particle identifications, a Scintillator Pad Detector (SPD) for photon rejection and a Forward Angle Electromagnetic Calorimeter (FAEC) for electron particle identification. The LAD group covers the nominal $15^\circ$-$24^\circ$ polar angle range and constitutes of four planes of GEM for tracking, a SPD for photon rejection and a Large Angle Electromagnetic Calorimeter (LAEC) for electron particle identification. This configuration can work with luminosity of 1$\times$10$^{37}$~cm$^{-2}\cdot$s$^{-1}$.

The ``PVDIS'' configuration uses five GEM planes for tracking and LGC and EC for $e/\pi$ separation to cover nominal $22^\circ$-$35^\circ$ polar angle range. It utilizes a set of baffles to reduce backgrounds while keeping a reasonable fraction of DIS electron events and can reach the luminosity of 1$\times$10$^{39}$~cm$^{-2}\cdot$s$^{-1}$.

The two configurations share three major detector components: GEMs, LGC and EC. They also share similar data acquisition (DAQ) system, supporting structure for the magnet and the detectors, and software tools for simulations and data analysis. 

There are additional components which are standard and existing at JLab that requires only slight modification, such as polarized NH$_3$ and polarized $^3$He targets, and the standard cryogenic hydrogen target. There are other additional components which are required by the MOLLER experiment and will become available before SoLID is operational, such as a high-precision Compton polarimeter, a super-conducting Moller polarimeter, and an upgraded End Station Refrigerator (ESR2) that is needed by the higher-power cryogenic target of PVDIS.

The SoLID spectrometer can handle high rates with high background by using the latest detector, data acquisition and computing technologies. The following subsections will describe those detector components and technologies in details. 

\subsection{The CLEO-II Magnet}
A solenoid magnet is a natural choice to meet the needs of SoLID's physics programs that require large acceptance in polar and azimuthal angles, and particle momentum. We have chosen the CLEO II's solenoidal magnet, that has a uniform axial central field of 1.5~T, a large inner space with a clear bore diameter of 2.9~m and a coil of 3.1~m diameter. With a coil length of 3.5 m, its magnetic field uniformity is $\pm0.2\%$. It was built in the 1980s by Oxford in England and installed for CLEO II in 1989~\cite{CLEO:1991qyy,Coffman:1990jt}.
After completion of experimental runs at Cornell, the coils and cryostat of the CLEO II magnet were moved to JLab in 2016 and the return steel moved in 2019. JLab is currently performing minor refurbishment of the magnet and preparing for a cold test to establish the magnet's operational condition. The cold test is scheduled to be completed before the end of 2022.

\begin{figure}[!ht]
\includegraphics[width=0.45\textwidth, angle = 0]{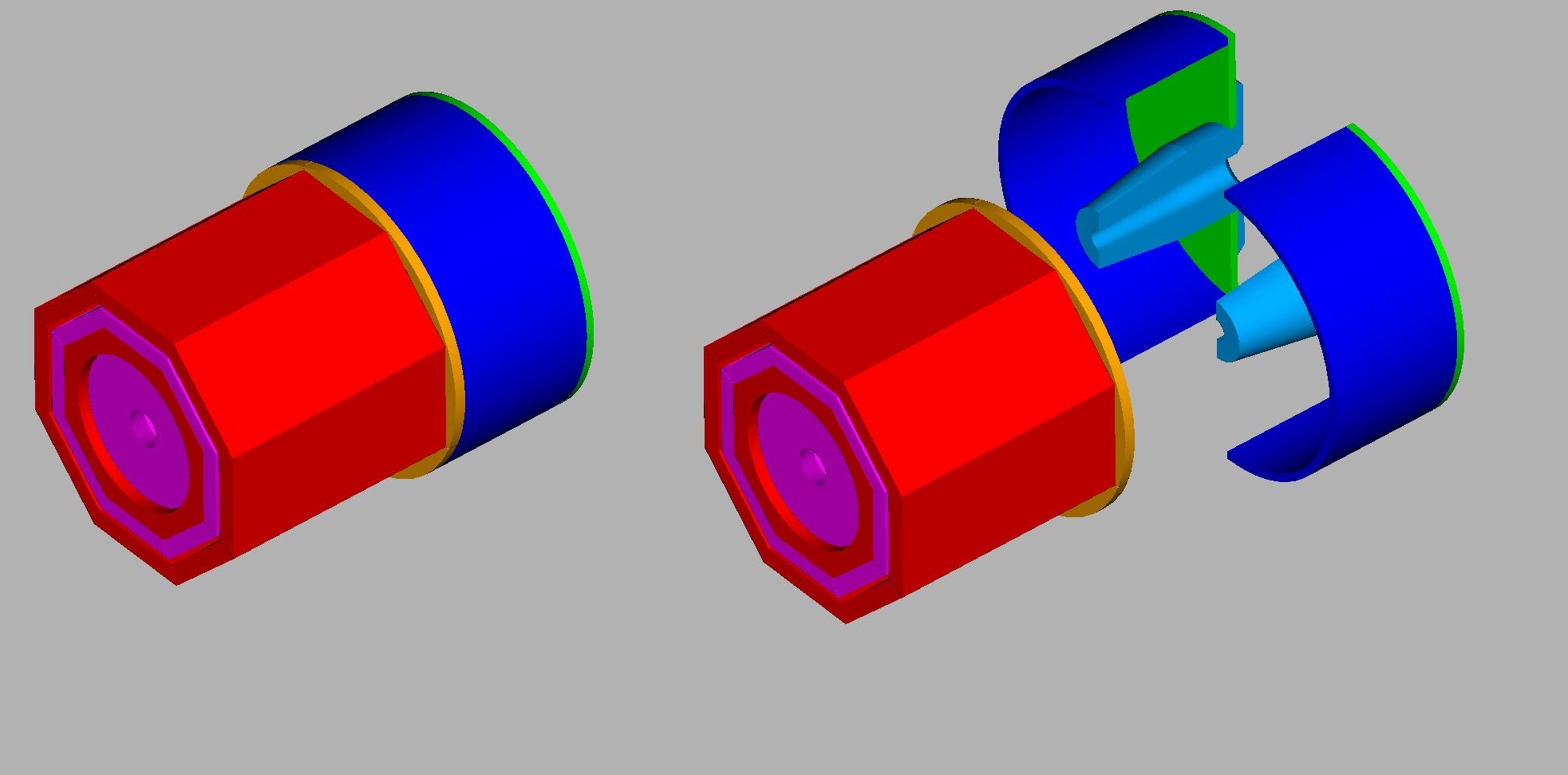}
\caption{The endcap will be split vertically and also have the capability of separating in the lateral direction.}
\label{fig:sequence}
\end{figure}

To use the CLEO magnet for SoLID, the coil and upstream coil collar will be reused as-is but the downstream coil collar and return yoke will be modified. 
A new detector endcap and front pieces will be fabricated that allow housing and installation of the detectors, see Fig.~\ref{fig:sequence}.

\subsection{
Gas Electron Multiplier Trackers}

Particle tracking for SoLID will be performed by Gas Electron Multiplier (GEM) trackers~\cite{Sauli:1997qp}. The GEM trackers are ideal for the SoLID detector because they provide for high resolution tracking and can operate in high-rate environments over a large area. More specifically, we expect the GEMs to provide a position resolution of 70 $\mu$m with rates over 100 MHz per cm$^{2}$. The current design of SoLID GEM chambers call for a triple design: each chamber is made of three GEM foils sandwiched between a drift area and a readout plane. 
Such triple GEM chambers have been successfully used in the COMPASS experiment at CERN~\cite{Ketzer:2004jk}, and  in the PRad experiment at JLab~\cite{Xiong:2019umf}. A large set of triple GEM detectors of the size comparable to those needed for SoLID is  currently used for the SBS program in Jlab Hall A. These SBS GEMs have performed very well in beam yielding highly stable operation.  In SBS experiments, these GEMs will be exposed to rates comparable to those expected in SoLID experiments. 

For the PVDIS configuration, five layers of GEMs will be used, each layer consisting of 30 sectors in the azimuthal direction that match the baffle design. This layout will allow for a 1~mrad polar angle and a 2\% momentum resolutions.

The SIDIS configuration of SoLID calls for six layers of GEM modules.  The SIDIS GEM will be assembled using the same GEM modules used in the PVDIS configuration. Because of the different coverage area required by SIDIS compared with PVDIS, this re-arrangement will allow small overlapping between GEM chambers, minimizing the acceptance loss due to inactive area caused by GEM chamber frames.  In the PVDIS configuration these frames sit in the shadows of the baffle-ribs and do not contribute to any loss of acceptance.

\subsection{Light Gas Cherenkov Detector}
The LGC detector provides electron identification in both SIDIS+$J/\psi$ and PVDIS configurations. The LGC is comprised of a tank of CO$_{2}$ gas as radiator, is divided into 30 sectors, each consisting of a pair of mirrors and one readout assembly onto which light is reflected. Each readout assembly is made of 9 Hamamatsu flat panel multianode photomultipliers (MAPMT) H12700-03 in a 3x3 array. Those MAPMT will be coated with a p-terphenyl wavelength shifter to enhance the efficiency of UV light detection. The MAPMTs have 64 pixels, each of which is sensitive down to single photon detection. Their signals can be read out individually or as sum of 16 pixels (quad-sum) or as the sum of all 64 pixels (total-sum). With these design features, the LGC is expected to have a nominal pion rejection on the order of $10^{3}$ while maintaining an electron efficiency close to 95\%. It will be part of electron trigger system. 

A parasitic beam test was conducted on an Cherenkov prototype at JLab Hall-C in 2020. The prototype telescopic Cherenkov device (TCD) was built with the same electronic components expected for use in the SoLID LGC. It used a UV mirror to collect light from 1m long CO$_{2}$ gas onto a 4x4 WLS coated MAPMT array. The device was tested at high rates that reached about twice the max rate expected during SoLID production running. The TCD performed within expectations at these large rates and the trigger capability using either quad-sum or total-sum were verified.

\subsection{Heavy Gas Cherenkov Detector}

For the SIDIS experiments, the HGC detector will identify charged pion and suppress charged kaon for a momentum range from 2.5 GeV/c to 7.5 GeV/c at the forward angle. Its radiator will be 1 m length of the heavy gas C$_{4}$F$_{8}$ at 1.7 atm absolute pressure at the room temperature of 20 C. Matching LGC and covering the full azimuthal angle, it will have 30 sectors. Each sector has one spherical mirror to collect lights onto a 4x4 MaPMT arrays which are surrounded by a light collection cone and magnetic shielding cone. The HGC mirror, MaPMT and readout electronics are similar to the components of LGC, but HGC will not be part of the trigger system. The detector is expected to have a pion detection efficiency of 90\% and a kaon rejection of 10. During the Cherenkov beam test at JLab Hall-C in 2020, the Cherenkov prototype was tested with C$_{4}$F$_{8}$ gas at 1 atm and it performed within expectations. Additionally, a full-size 4-sector HGC prototype was designed and constructed with an Aluminium thin front window to test the operating pressure of 1.7 atm.  This test showed the current design maintains mechanical stability with negligible leakage. 

\subsection{Electromagnetic Calorimeter}
The segmented electromagnetic calorimeter (ECal) consists of a preshower and a shower section, and will be used as the primary electron trigger and identification during all experiments. The preshower portion consists of a $2X_0$ pre-radiator and a 2-cm thick scintillator with wave-length shifting (WLS) fibers embedded for light readout. The shower portion is $18X_0$ long, based on the Shashlyk-type sampling~\cite{Atoian:2007up} with alternating layers of 1.5-mm thick scintillator and 0.5-mm thick lead absorber layers. The choice of the sampling-type design was mostly driven by a balance between cost and the required radiation hardness. The layout of ECal models will be different between the two SoLID configurations: The SIDIS+$J/\psi$ configuration will have the ECal at both forward and large-angle regions for electron detection with the large-angle ECal also providing MIP triggers for pions. The PVDIS configuration will have all ECal modules at the forward direction to detect DIS electrons. 
There will be approximately 1800 modules, each with a transverse size $100$~cm$^2$ in a hexagon shape such that they can be rearranged between the two configurations. A unique aspect of SoLID’s ECal is its light readout: because of the high radiation nature of SoLID’s operation, all WLS fibers will be connected to clear fibers and light will be routed outside of the solenoid magnet for readout by PMTs. Radiation hardness of a variety of WLS and clear fibers has been measured and is found to be sufficient to sustain the SoLID physics program. 

A number of prototypes have been constructed for the SoLID ECal preshower and shower modules and their light yield studied with both cosmic rays and particle beams. Using the Fermilab Test Beam Facility, the energy resolution of the ECal prototype was found to satisfy the needs of the SoLID physics program. 
Tests with the electron beam at JLab are ongoing to further study the ECal performance under the high-rate, high background environment. 

\subsection{Scintillator Pad Detector}
The Scintillator Pad Detector (SPD) will be used at both forward-angle and large-angle locations of the SIDIS configuration to provide photon rejection at the 5:1 and 10:1 level, respectively, and to reduce ECal-based trigger rates by requiring coincidence signals between the SPD and the ECal.  The forward-angle SPD (FASPD) will be made of 240 pieces of thin, large scintillator pads with WLS fibers embedded on the surface. Light from the WLF fibers  will be guided through clear fibers in a similar manner as the preshower ECal.  The large-angle SPD (LASPD) also provides time-of-flight (TOF) with a timing resolution goal of 150~ps, and as a result are made of 2-cm thick long, wedge shape scintillators with readout directly by field-resistant fine-mesh PMT on the edge of the solenoid field. The fine-mesh PMTs have been tested under a magnetic field up to 1.9~T and its gain and timing resolution characterized~\cite{Sulkosky:2016kad}. The SPD prototype modules have been tested with cosmic rays. We found that the fine-mesh PMTs combined with the LASPD can provide a 150~ps timing resolution as specified by the SoLID SIDIS program. Tests with the electron beam at JLab are ongoing to further study the SPD performance under the high-rate, high background environment.

\subsection{
Multi-Gap Resistive Plate Chamber}
The Multi-gap Resistive Plate Chamber (MRPC), which will be used as the TOF system in the SIDIS configuration, is located in front of the forward angle ECal. 
The unique advantage of the MRPC is that it not only can operate in a strong magnetic field, but also can handle extreme high rates. The new generation sealed MRPC developed by Tsinghua University can reach the rate capability as high as 50~kHz/cm$^2$ utilizing a new type of low resistivity glass (in the order of 10~$\Omega$cm)~\cite{Wang:2009bx,Wang:2012sa,Lyu:2018wmr,CMSRPC:2019uxi}. On top of that, the MRPC designed for SoLID has a thin gas gap of 104um with 8 gaps per stack and a total of 4 stacks~\cite{Wang:2013vha}. A cosmic ray test on two identical 4x8 gaps MRPCs conducted at Tsinghua University with a 5GS/s waveform digitizer shows a time resolution of 27~ps. Simulation shows that the intrinsic time resolution of such a MRPC can be as better as 14~ps using a much higher sampling rate ($\sim$10~GS/s) front-end electronics (FEE).  With a total path length of 8~meters in the SIDIS configuration, the MRPC with a time resolution of 20 (30)~ps can identify pions from kaons with momenta up to 7 (6)~GeV/c. The studies of the MRPC's realistic performance with several fast FEE candidates are ongoing using real high-energy beams at Fermilab and JLab.

\subsection{Baffles for PVDIS}
In order for the detectors in the PVDIS experiment to operate at the design luminosity, a set of baffles – 11 slitted plates made of an absorber material –  is designed such that a reasonable fraction of the DIS electrons pass through the slits. The slits in the multiple layers of the baffle system provide curved channels through which only spiraling high energy charge-negative particles can pass, while low energy and charge-neutral or charge-positive backgrounds are highly suppressed.

\begin{figure}[!ht]
    \centering
    \includegraphics[width=0.48\textwidth]{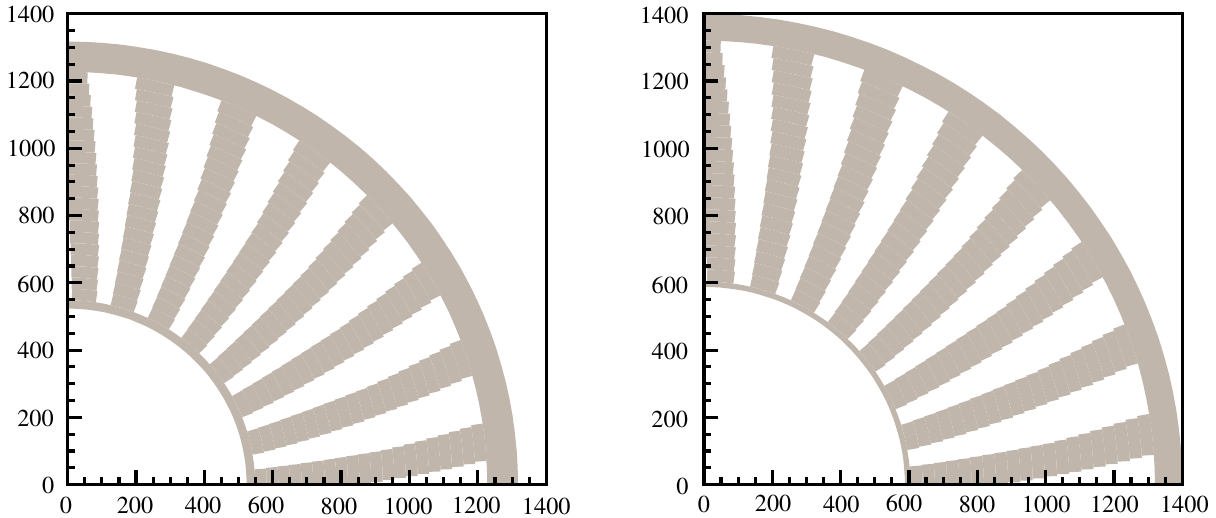}
    \caption{Face on view (first quadrant only) of the 10$^{th}$ and 11$^{th}$ plates in the PVDIS baffle system. Units are in mm.}
    \label{fig:baffle}
\end{figure}

To design the baffles for a specific magnetic field and detector configuration, ray-tracing of simulated DIS electrons is performed for the desired momentum range.  The number of sectors to be used for the PVDIS experiment is driven by the azimuthal angle $\phi$ traversed by the minimum momentum particles, which for the desired DIS kinematics is about $12^\circ$, hence the baffles are divided into 30 sectors. Segmentation of all detector system have the same number of sectors to match the baffle design. An illustration of the first 5 layers of the baffle system is given in Fig.~\ref{fig:baffle}. In practice, the simple ray-tracing model does not completely hold because the target has an extended length, allowing some fraction of background events to leak through. 

Three different material choices are being considered for the baffles. The baseline design is based on lead. Two other alternatives are tungsten power molded and glued to the desired shape, and copper. All will meet the requirement of PVDIS and with small differences in the resulting photon and hadron background rates. Additional care is taken to reduce secondary particles, such as those produced from photons hitting the baffle near the slits. Studies are being carried out on activation of the material and feasibility in construction will be taken into account. Overall, the baffles are expected to allow about 1/3 of DIS events to reach the detectors, while background events are suppressed by two orders of magnitude. 

\subsection{Support and Infrastructure}
The solenoid magnet will be supported on a stationary frame that will distribute the approximate 1000 ton load of the modified CLEO-II magnet section using eight 200-ton enerpac jacks.  Steel plates and large steel blocks and/or large I-beams will be used to distribute the load over a safe area.  The 200-ton jacks will be used for vertical alignment and have locking rings which allow for a full mechanical connection and not rely on hydraulic pressure for stationary support.

The endcap of the magnet will have a support structure that cradles each half the cylindrical ring.  The structure will be integrated into a track system that is mounted to steel plates resting upon the concrete floor of Hall A.  The track system will consist of a set of longitudinal tracks for moving the rear plate and nose unit of the endcap downstream away from the magnet.  A set of lateral tracks will separate the two endcap cylindrical halves that support the detectors and move each away from the beamline.  Motion can be achieved by using hydraulic or electric cylinders to push and pull the entire system into position.

Inside the magnet bore, the insertion of the SIDIS large angle detector packages that reside internal to the cryostat will be accomplished from the downstream side of the magnet using a supporting framework to roll the packages in and out. This will require the detector hut to be moved out of the way as described above to allow access to the cryostat.  In the inner bore region, an internal frame system is needed to mount the baffles in the PVDIS configuration and the large angle detectors for the SIDIS configuration. The frame cannot come into contact with the inside bore of the cryostat. This requires the frame to span the entire length of the cryostat and mount to the return yoke iron. A stainless steel support cylinder will be mounted between the two coil collars to bridge across the length of the cryostat.  Individual rails will bolt directly to the stainless cylinder to allow the internal detector packages to roll into place. The same rail system can be used for both configurations as well as the detectors in the endcap. A large universal installation fixture is envisioned to load each of the detector packages onto the rails of the magnet and endcap.

\subsection{Event Rates and Data Acquisition}\label{sec:daq}
The trigger rates were simulated with the full background. The SIDIS configuration, with an expected
trigger rate of 100 kHz and total data rate of over 3 GB/s, represents the greatest challenge for SoLID's data acquisition (DAQ) system. The PVDIS rates are also high, but are not as demanding as they are divided into 30 sectors with each equipped with individual DAQs. The SoLID DAQ is mostly based on JLab250 FADCs for the readout of PMTs of ECal and Cherenkov detectors. These electronics provide both readout and trigger capability on any detector fed into the FADCs.  The FADC readout so far has been shown to be able to operate up to 120 KHz of trigger rate at around 1\% of deadtime satisfying the SIDIS requirements. The GEM readout will use the VMM3 which has a minimum rate capability of 100 kHz at full occupancy.  So far, the SoLID DAQ system which can handle data rates of several GB/s is feasible using technology currently in use at JLab.

\subsection{Computing}
\label{sec:compreq}
Estimated computing needs for SoLID are summarized in Table~\ref{tab:compreq}. These are total resource requirements over the lifetime of the experiment, assuming that all simulation and production output is kept. Total overall resources needed are 188~PB storage and 233~M-core-hours CPU. This corresponds to 485 days of processing time on a 20,000-core cluster.

\begin{table}[ht]
	\begin{tabular}{l|c|c|c|c|c}
		\toprule
		Experiment   & SIDIS       & SIDIS      & SIDIS                 & J/$\psi$ & PVDIS \\
		& $^3$He (T)  & $^3$He (L) &  {\it NH$_{3}$} (T)   &          &       \\
		\midrule
		Storage (PB) &     26      &  10        &  35                   & 21       & 95    \\
		\midrule
		CPU time     &             &            &                       &          &       \\
		\; (M-core-hrs) &  30      &  12        &  40                   & 17       & 134   \\
		\bottomrule
	\end{tabular}
	\caption{Estimated SoLID computing requirements. CPU times are calculated
		assuming AMD EPYC 7502 processors.}
	\label{tab:compreq}
\end{table}

To arrive at the numbers in Table~\ref{tab:compreq}, 
average trigger rates of 100~kHz for the SIDIS experiments,
60 kHz for J/$\psi$, and 20~kHz per sector for PVDIS, are assumed 
({\it cf.}\ Section~\ref{sec:daq}). 
Event size estimates come from simulations and are 20~kB for SIDIS, 40~kB for J/$\psi$, and 6~kB per sector for PVDIS. The resulting instantaneous raw data rates range from 2.0 to 3.6 GB/s.

\subsection{Software}
\label{sec:software}
Software developed for SoLID to date comprises three main projects
\begin{enumerate}
	\item {\tt SOLID\_GEMC} \cite{GitHub:solid-gemc},
		 a simulation package built on GEMC \cite{Software:GEMC},
		 a generic simulation framework used by CLAS12 and other projects
		 at JLab. GEMC is based on Geant4 \cite{Software:Geant4}.
    \item \label{item:libsolgem}
    	{\tt libsolgem}, a digitization package
         for GEM detectors, which was developed by the SoLID collaboration
        \cite{GitHub:libsolgem}.
    \item \label{item:solidtracking}
        {\tt SoLIDTracking}, a library of experimental track reconstruction
        routines for the three main configurations of SoLID
        \cite{GitHub:SoLIDTracking}.
        This package employs a Kalman filter algorithm and is based
        in part on prior implementations for experiments at KEK and GSI.
    	
\end{enumerate}

A detailed description of packages \ref{item:libsolgem} and \ref{item:solidtracking}
can be found in Ref.~\cite{Weizhi:SoLID96v1},
which also includes a study of efficiency and accuracy of the track
reconstruction algorithm applied to simulated data from {\tt SOLID\_GEMC}.

The long-term goal for SoLID software development is to put in place
a unified end-to-end simulation and reconstruction framework, which will provide an integrated software environment for (almost) all parts of data processing. Implementing a software ecosystem for a new experiment requires considerable effort. In light of limited staffing, it will be necessary to adopt preexisting components wherever possible. At present, the most fruitful approach appears to be for SoLID software to be closely aligned with that of the Electron-Ion Collider (EIC) project. It is expected that EIC will converge on a unified software environment by the end of 2022, which would still be compatible with the timeline for SoLID.

\subsection{Advancing Detector Technology}

SoLID is designed to carry out experiments with high rate and high background. For many experiments, the luminosity achievable is limited by the detector occupancies. We are investigating new detector technologies with faster response time to improve the rate capability of SoLID. The Large Area Picosecond Photodetector (LAPPD) is being developed by INCOM and Argonne National Lab: it is a novel, affordable large area Microchannel Plate Photomultiplier (MCP PMT) and was tested in beam~\cite{Peng:2020iyl}. The pulse width of MCP PMT is of the order of 1~ns compared to about 20 ns for a regular PMT, possibly reducing greatly the pile-up for the Cherenkov detectors. This technology, when it becomes mature, would be a prime candidate as photosensor for the Cherenkov readouts.

Another technology being considered is the superconducting nano-wire technology~\cite{Natarajan:2012bw}. 
The detector exhibits excellent timing resolution and is likely to be more radiation hard than traditional technology. Such a detector could be used to complement the GEM tracking as a vertex tracker or provide additional tracking planes.

\section{Opportunities with Future Upgrades of CEBAF}\label{sec:upgrades}
\subsection{\texorpdfstring{$J/\psi$}{J/psi} and \texorpdfstring{$\psi^\prime$}{psi'} Production with 20\texorpdfstring{$^+$}{+} GeV}
A CEBAF energy upgrade to 20 GeV or higher would enable several additional topics to be pursued with the SoLID-$J/\psi$ setup. The electroproduction measurement at larger beam energies could operate without any changes to the 11 GeV setup, although further optimizations could be considered. A beam energy of 20 GeV or higher would access values of $Q^2$ up to 10 GeV$^2$ or larger, providing an additional large scale to aid with factorization. The photoproduction measurement would allow for a precision measurement of $J/\psi$ cross section at slightly larger energies, superseding the previous SLAC~\cite{Anderson:1976sd} and Cornell~\cite{Gittelman:1975ix} measurements. Furthermore, this would enable a small overlap region with the measurements at the EIC, where the SoLID measurement would have much a higher resolution in $W$ and a unique reach to high $t$ that cannot be done anywhere else. 
Finally, a measurement at higher energies allows a simultaneous measurement of $J/\psi$ and $\psi^\prime$ production, where the latter process provides for an independent knob to constrain the gluonic physics inside the proton, as it is a larger-size color dipole. 
\begin{figure}[!hb]
    \includegraphics[width=0.45\textwidth]{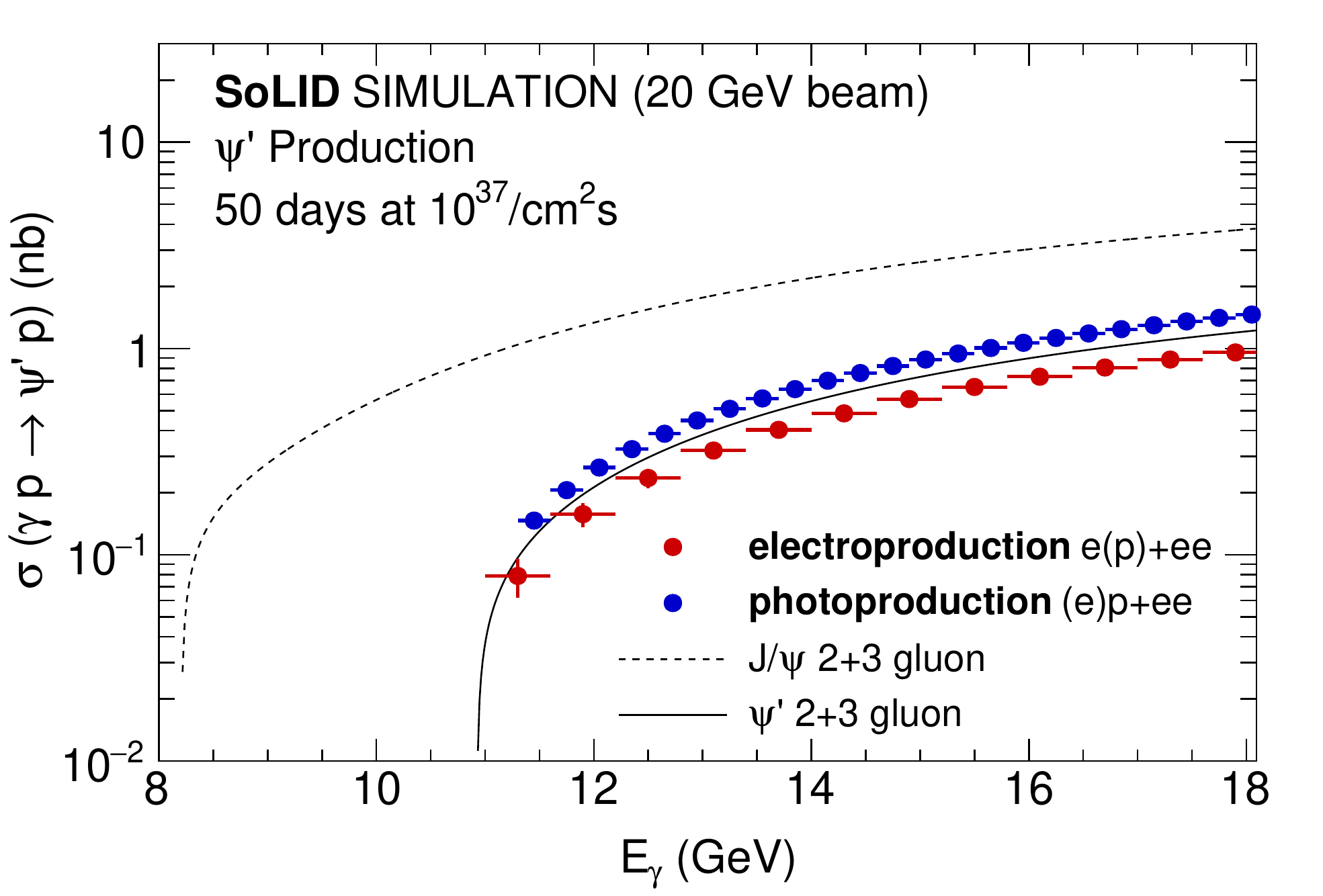}
    \caption{Projected 1-D cross section results for $\psi^\prime$ production assuming a 20 GeV beam energy and the nominal SoLID-$J/\psi$ experimental setup without any optimization for the higher beam energy, for 50 days at $10^{37}\text{cm}^{-2}$s$^{-1}$. The blue disks show the photoproduction results and the red disks the electroproduction results.}
    \label{fig:psiprime:1d}
\end{figure}
\begin{figure}[!ht]
    \includegraphics[width=0.23\textwidth]{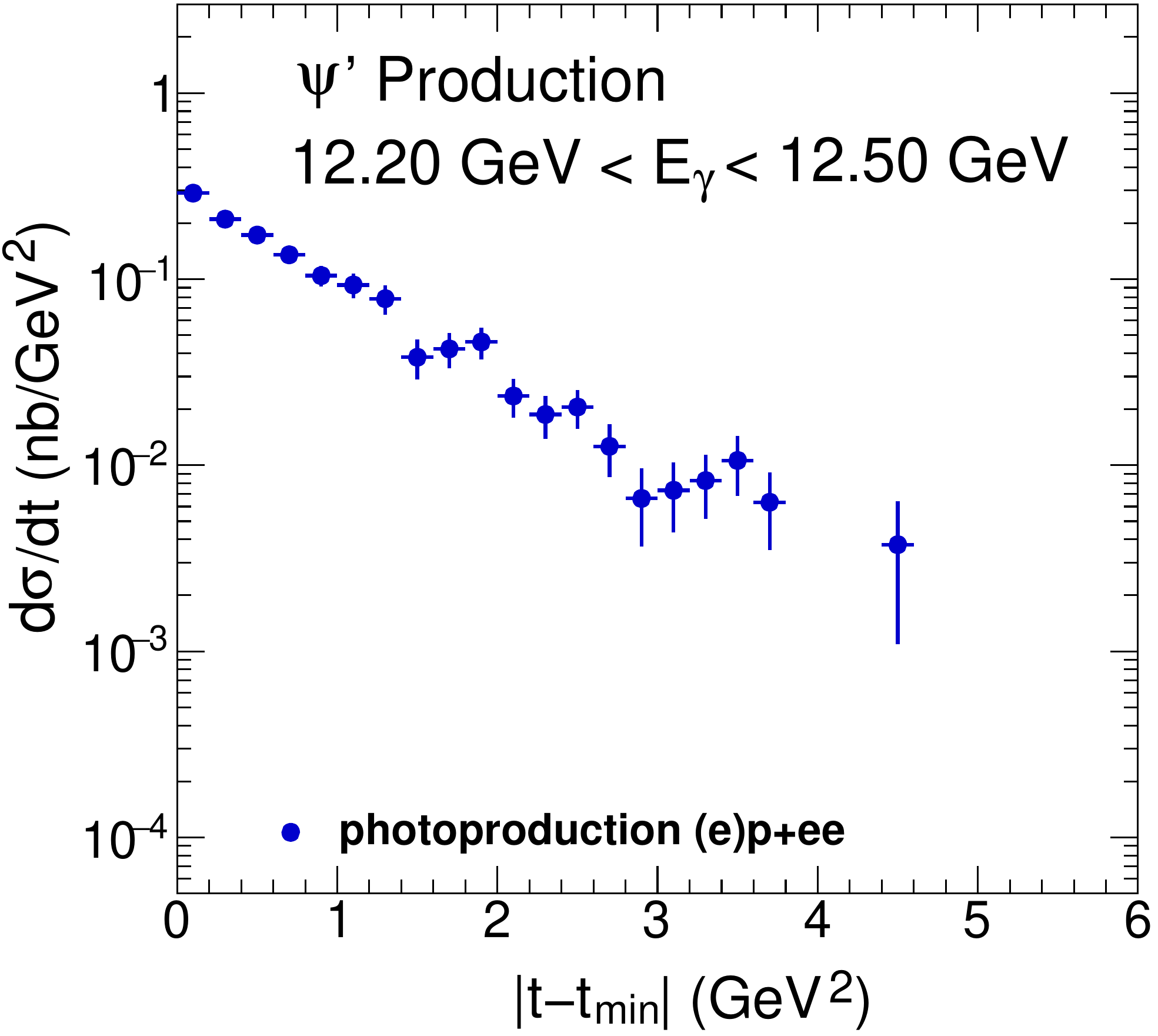}
    \includegraphics[width=0.23\textwidth]{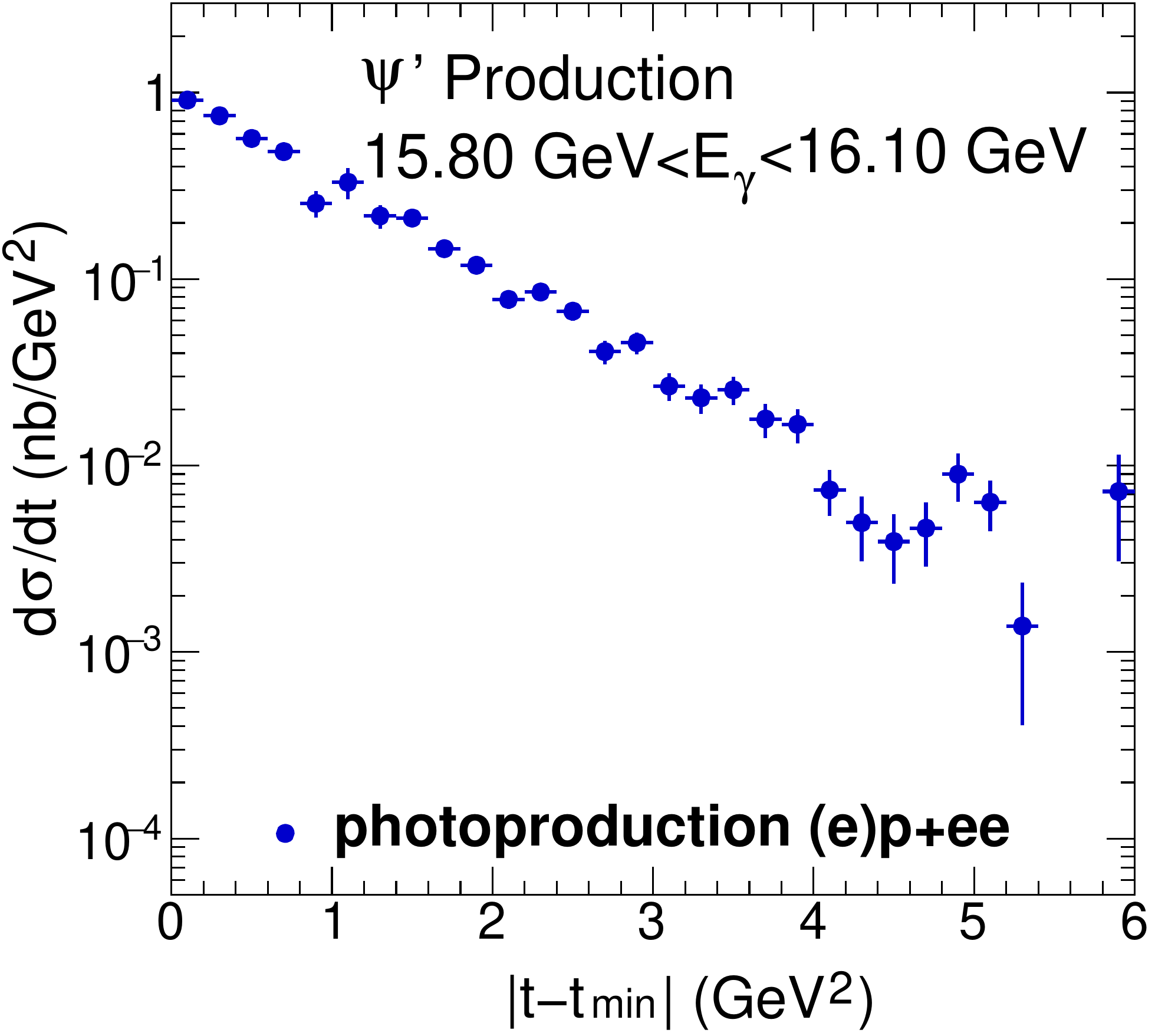}\\
    \includegraphics[width=0.23\textwidth]{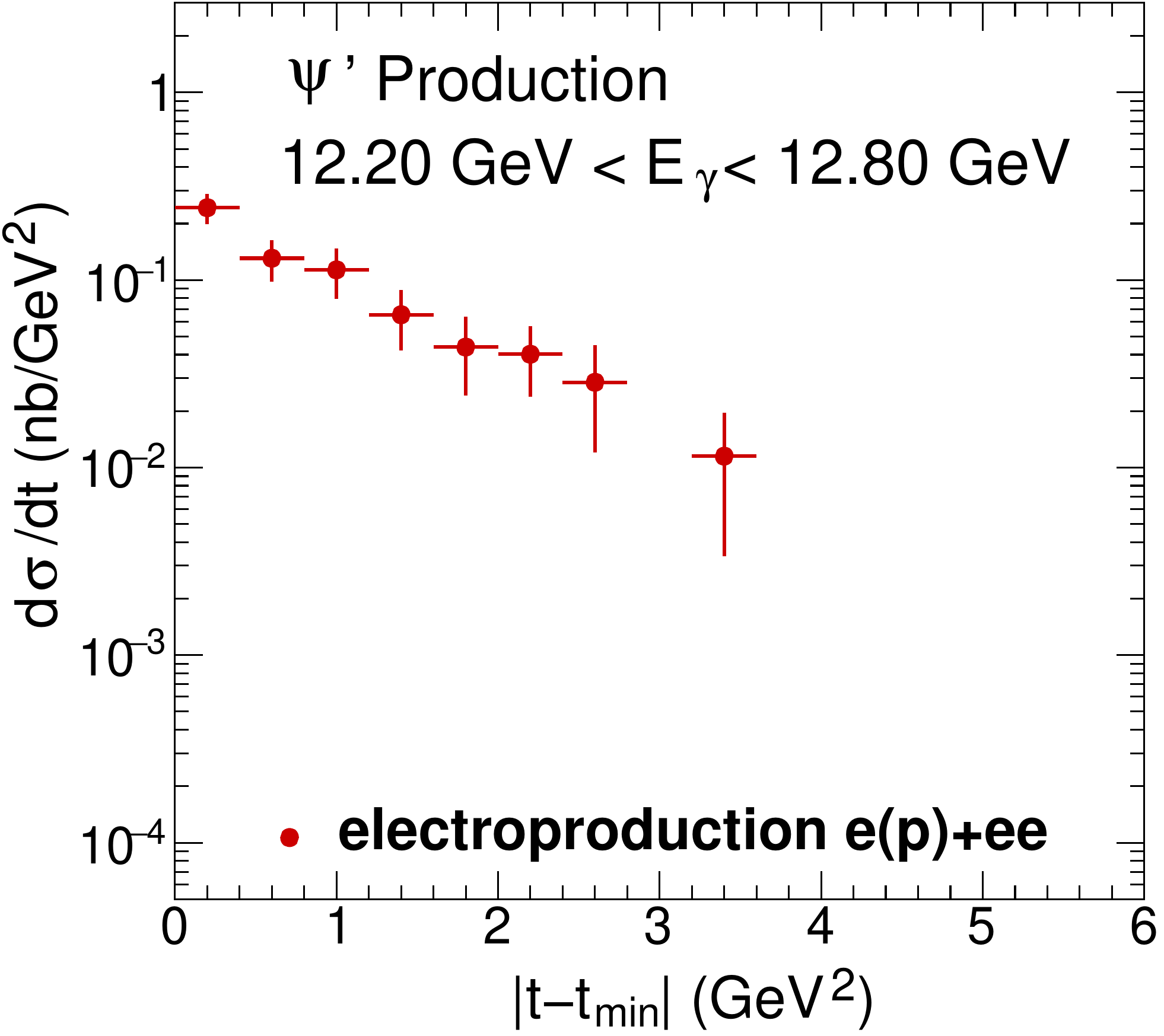}
    \includegraphics[width=0.23\textwidth]{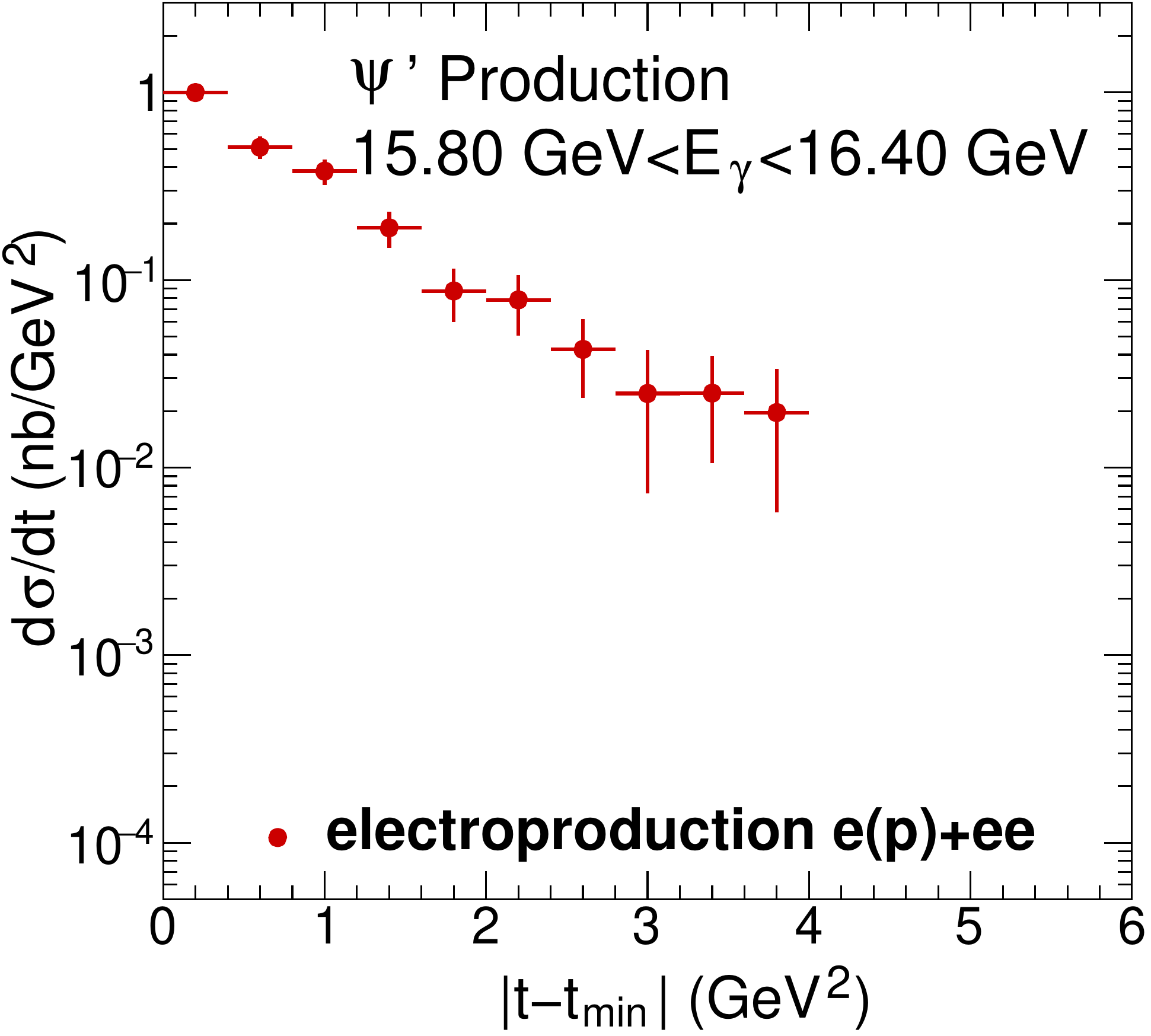}\\
    \caption{Top row: The projected differential cross section for a photoproduction bin at low (left) and high (right) photon energy from Fig.~\ref{fig:psiprime:1d}, assuming the nominal luminosity for SoLID-$J/\psi$ with a beam energy of 20 GeV. Bottom row: Same for two electroproduction bins. This figure illustrates that a precise measurement of the $t$-dependence for $\psi^\prime$ production is possible with the nominal SoLID-$J/\psi$ setup at higher energies.}
    \label{fig:psiprime:2d}
\end{figure}
Projected 1-D and 2-D cross section results for $\psi^\prime$ production with SoLID at 20 GeV are shown in Figs.~\ref{fig:psiprime:1d}~and~\ref{fig:psiprime:2d}.

\subsection{Nucleon 3D Structure with 20\texorpdfstring{$^+$}{+} GeV}
\begin{figure}[ht]
    \includegraphics[width=0.48\textwidth]{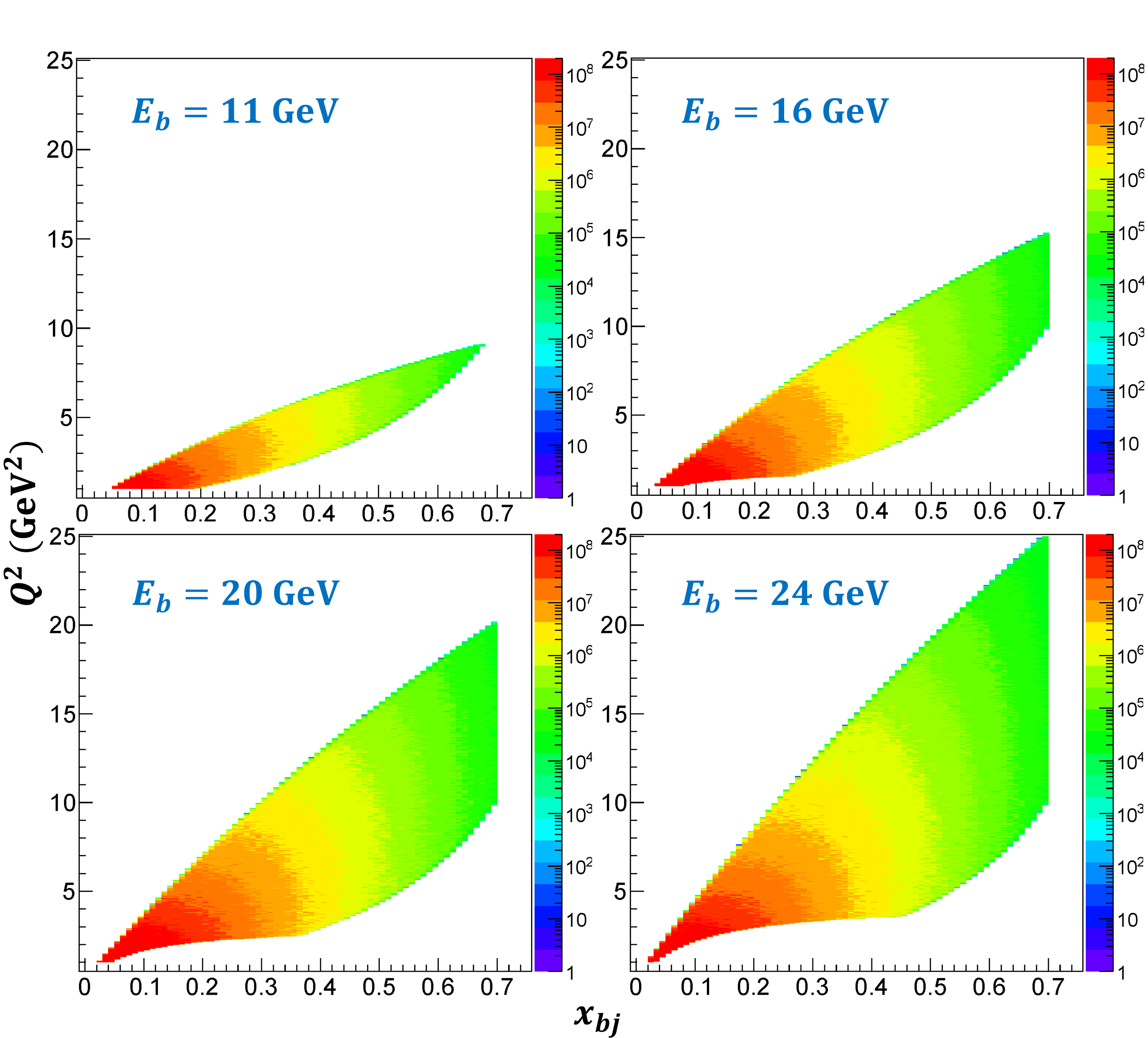}
    \caption{Projected kinematic coverage of SoLID-SIDIS with polar angles from $5^{\circ}$ to $27^{\circ}$ for various beam energies. The projections were simulated with a $^3$He target and the SoLID acceptance effects were turned off. The polar angle range, $5^{\circ}$-$27^{\circ}$, was optimized for the upgraded CEBAF energy.}
    \label{fig:kine-upgrade}
\end{figure}

\begin{figure}[ht]
    \includegraphics[width=0.48\textwidth]{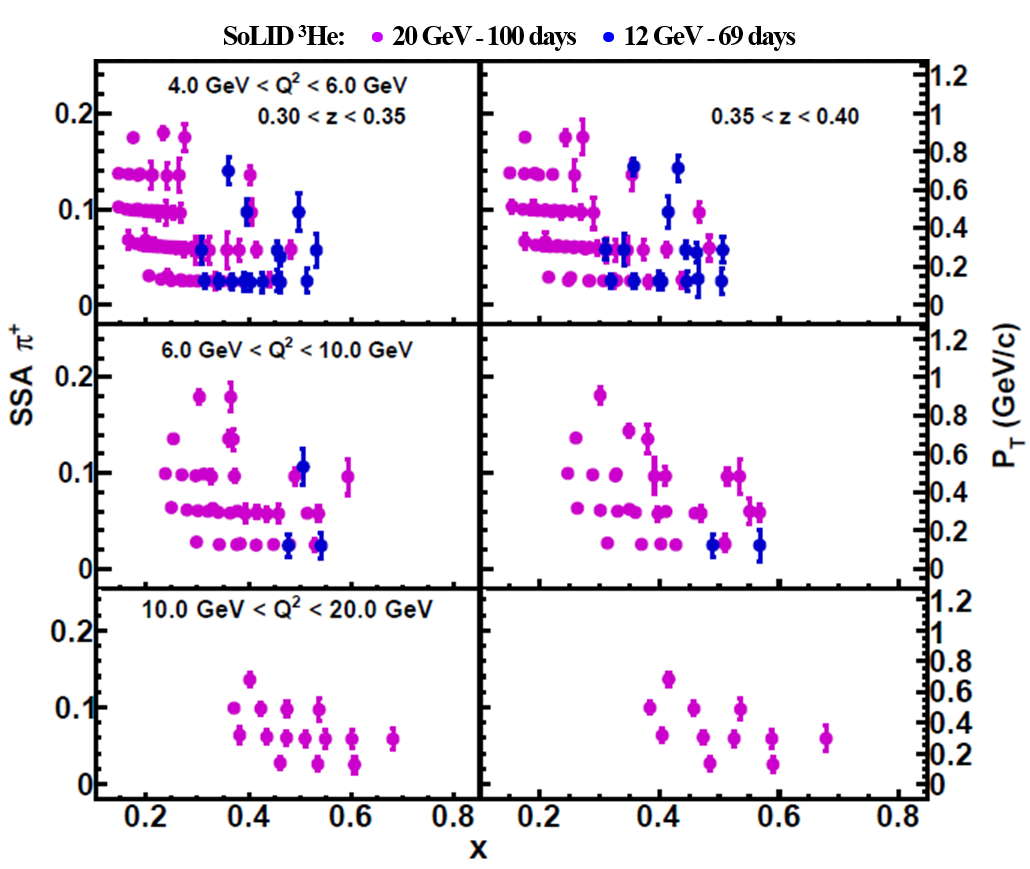}
    \caption{Selected $Q^2$-$z$ bins of projected Collins SSA with SoLID-SIDIS configuration and $^3$He target. Two different beam energies, 12 GeV and 20 GeV, are included to compare the their kinematic coverage.}
    \label{fig:collins-20GeV}
\end{figure}

The SIDIS and GPD programs of SoLID will also benefit from the CEBAF energy upgrade to 20 GeV or higher, resulting in significantly extended kinematic coverage of observables and potentially open up new physics channels for the nucleon 3D structure study. Figure~\ref{fig:kine-upgrade} shows the simulated $Q^2$-$x$ phase-space with various beam energies from 11 GeV to 24 GeV using the SIDIS configuration of SoLID with a polarized $^3$He target. 
Preliminary studies have been carried out for the Collins SSA, as shown in Fig.~\ref{fig:collins-20GeV}. A few $Q^2$-$z$ bins were selected from the full coverage of $2.0 < Q^2 < 20.0$ GeV$^2$ and $0.30 < z < 0.70$. As expected, SoLID with a higher energy beam of CEBAF will provide precision measurements of SIDIS and GPD in the higher $Q^2$ and lower $x$ region, that can not be charted with the 12 GeV beam.
More detailed studies, including those for the proton target and for other physics channels, will be carried out to optimize the potential physics programs of SoLID with a higher energy beam.

\subsection{Electroweak Physics with a Positron Beam}
With a higher beam energy of 20 GeV or above, the PVDIS measurements can be extended to higher $Q^2$, providing improved precision on the $\sin^2\theta_W$ and the $2g_{VA}^{eu}-g_{VA}^{ed}$ coupling, or the PDF ratio $d/u$ for higher~$x$. On the other hand, the addition of a positron beam at CEBAF will open up a wide range of physics topics not accessible with an electron beam alone~\cite{Accardi:2020swt}. One new observable that we can measure with SoLID and a positron beam is the lepton-charge asymmetry, defined as the cross section asymmetry between positron and electron DIS:
\begin{eqnarray}
A^{e^+e^-}&\equiv&\frac{\sigma^{e^+}-\sigma^{e^-}}{\sigma^{e^+}+\sigma^{e^-}}\ , 
\end{eqnarray}
and is related to the neutral current  coupling, $g_{AA}^{eq}$, 
predicted by the SM as $g_{AA}^{eq}=2g_A^e g_A^q$ and $g_{AA}^{eu}=-g_{AA}^{ed}=- 1/2$. 
More specifically, the asymmetry $A^{e^+e^-}$ between unpolarized $e^+$ and $e^-$ beams deep-inelastic-scattering off an isoscalar target has an electroweak contribution that is directly proportional to the combination $2g_{AA}^{eu}$ $-$ $g_{AA}^{ed}$~\cite{Zheng:2021hcf}: 
\begin{eqnarray}
A^{e^+e^-}&=& -\frac{3G_F Q^2}{2\sqrt{2}\pi\alpha}Y
\frac{R_V}{5}\left(2g_{AA}^{eu}-g_{AA}^{ed}\right)~,
\label{eq:Apm_unpol}
\end{eqnarray}
where $R_V$ was defined in Section~\ref{sec:pvdis} and the effect of sea quarks has been omitted for SoLID's kinematic coverage. 
Such measurement~\cite{JLabPR:C3q}, if successful, would provide the first measurement of this coupling for the electrons, superseding the previous measurement using muon beams at CERN~\cite{Argento:1982tq} that gave $2g_{AA}^{\mu u}$-$g_{AA}^{\mu d}$$=1.57$$\pm$0.38. 

The measurement of $A^{e^+e^-}$ faces both experimental and theoretical challenges. Experimentally, differences in beam energy, intensity, and the detection of the scattered particles between $e^+$ and $e^-$ runs will cause sizable contributions to $A^{e^+e^-}$, though these effects have a calculable kinematic-dependence and could be separated from electroweak contributions. Theoretically, electromagnetic interaction causes an asymmetry between $e^+$ and $e^-$ scatterings at the next-to-leading and higher orders, causing a contribution to $A^{e^+e^-}$ that are significantly larger than the electroweak contribution at the $Q^2$ values of JLab.  Progress in theory is needed in the coming decade to describe $A^{e^+e^-}$ at the level of precision required by the $g_{AA}^{eq}$ measurement.

\section{Summary}
The SoLID spectrometer is a multi-purpose device that can address many of the central issues in the studies of QCD and fundamental symmetries.  Three SIDIS experiments to perform precision measurements with transversely and longitudinally polarized $^3$He (effective polarized neutron) and transversely polarized proton will allow precision extractions of TMDs in the valance quark region to  
map out the 3D structure of the nucleon in momentum space. An experiment of electro- and photo-production of $J/\psi$ near threshold region probes the gluonic field and its contribution to the proton mass. A parity-violating DIS experiment will determine the effective electron-quark couplings of the Standard Model, pushing the limits in phase space in search for new physics, and will provide the PDF ratio $d/u$ at high $x$. A number of run-group experiments have been approved, including the exploration of GPDs with deep-exclusive reactions to study the 3D structure of the nucleon in coordinate space. 
\lrpbf{
At the latest JLab PAC meeting in 2022, all five SoLID experiments were re-approved with the highest rating ($A$) and two new experiments were added, including a measurement to study two photon exchange effects and a measurement to study isospin dependence of the EMC effect.
The key to the high impact of each of these experiments is the high luminosity combined with the large acceptance of SoLID, with orders of magnitudes higher figure-of-merit than all other devices at existing and future $ep$ (and $eA$) facilities. SoLID will thus exploit the full potential of the JLab 12 GeV beam, with a kinematic reach complementary to that of EIC.
}
The design of SoLID has been vetted by several JLab Director's reviews and a DOE Science Review. 
It shares significant synergy with EIC including detector technology, simulation, data acquisition capacity, software integration, data analysis aided by artificial intelligence and machine learning, radiative corrections and unfolding, and finally, training of the nuclear physics workforce for the QCD and fundamental symmetry frontier for the next decades. 
%


\acknowledgments{

This material is based upon work supported in part by the U.S. Department of Energy, Office of Science, Office of Nuclear Physics under contract numbers DE-AC05-06OR23177 (JLab), DE-AC02- 06CH11357 (ANL), DE-FG02-03ER41231 (Duke), DE-FG02-96ER40988 (Stony Brook U.), DE-FG02-84ER40146 (Syracuse U.), DE-FG02-94ER40844 and DE-SC0016577 (Temple U.), DE–SC0014434 (U. of Virginia); and Office of Science, Office of Workforce Development for Teachers and Scientists (WDTS) under the Science Undergraduate Laboratory Internships Program. We thank A. Accardi, S. Kuhn, and S. Mantry for the useful discussions that contributed to this manuscript. 
}


\bibliographystyle{JHEP}

\bibliography{solid_wp2022}

\end{document}